\documentclass[aps,pra,letterpaper,10pt,showpacs,superscriptaddress,longbibliography, twocolumn]{revtex4-2}

\usepackage{amsfonts,amssymb,amsmath,amsthm,array,mathtools}
\usepackage{graphicx,graphics}
\usepackage{booktabs}
\usepackage{gensymb}
\usepackage{dsfont}
\usepackage{bbm}
\usepackage[usenames,dvipsnames]{xcolor}
\usepackage{nicefrac,bigints}
\usepackage{url}
\usepackage{textcase}
\usepackage{bm}
\usepackage{setspace}
\usepackage[caption=false]{subfig}
\usepackage{comment}
\usepackage{enumitem}
\setlist[itemize]{noitemsep, nolistsep}

\usepackage{natbib}
\usepackage{physics}

\usepackage{tabularx}

\usepackage[colorinlistoftodos]{todonotes}
\usepackage[colorlinks=true, allcolors=blue]{hyperref}
\usepackage{color, colortbl}

\usepackage{times,todonotes}
\usepackage[normalem]{ulem} %strikethrough text, call with \sout{arg}

\usepackage{comment}

\definecolor{pinocol}{rgb}{.1,0.8,0.1}
\definecolor{pinocolbg}{rgb}{.1,0.8,0.2}

\definecolor{piccacol}{rgb}{.976,0.74,0.66}
\definecolor{piccacolbg}{rgb}{.976,0.74,0.66}

\definecolor{alecol}{rgb}{1, 0.5, 0}
\definecolor{alecolbg}{rgb}{1, 0.5, 0}

\definecolor{andreacol}{rgb}{0.5,0.1,1.0}
\definecolor{andreacolbg}{rgb}{0.5,0.5,1.0}

\definecolor{cescocol}{rgb}{1, 0.0, 1}
\definecolor{cescocolbg}{rgb}{1, 0.1, 1}

\definecolor{giuliocol}{rgb}{0, 0.8, 0.8}
\definecolor{giuliocolbg}{rgb}{0, 0.8, 0.8}

\graphicspath{{./}{./Figures/}}

\begin{abstract}
Free-space ground-to-ground links will be an integral part of future quantum communication networks.
The implementation of free-space and fiber links in daylight inter-modal configurations are however still hard to achieve, due to the impact of atmospheric turbulence, which strongly decreases the coupling efficiency into the fiber.
In this work, we present a comprehensive model of the performance of a free-space ground-to-ground quantum key distribution (QKD) system based on the efficient-BB84 protocol with active decoy states. Our model takes into account the atmospheric channel contribution, the transmitter and receiver telescope design constraints, the parameters of the quantum source and detectors, and the finite-key analysis to produce a set of requirements and optimal design choices for a QKD system operating under specific channel conditions.
The channel attenuation is calculated considering all effects deriving from the atmospheric propagation (absorption, beam broadening, beam wandering, scintillation, and wavefront distortions), as well as the effect of fiber-coupling in the presence of a partial adaptive correction with finite control bandwidth. We find that the channel fluctuation statistics must be considered to correctly estimate the effect of the saturation rate of the single-photon detectors, which may otherwise lead to an overestimation of the secret key rate.
We further present strategies to minimize the impact of diffuse atmospheric background in daylight operation by means of spectral and temporal filtering.
\end{abstract}

\begin{document}

\title{Optimal design and performance evaluation of free-space Quantum Key Distribution systems}
\author{Alessia~Scriminich}
\altaffiliation{These authors contributed equally to this work}
\affiliation{Dipartimento di Ingegneria dell'Informazione, Universit\`a degli Studi di Padova, via Gradenigo 6B, IT-35131 Padova, Italy}

\author{Giulio~Foletto}
\altaffiliation{These authors contributed equally to this work}
\affiliation{Dipartimento di Ingegneria dell'Informazione, Universit\`a degli Studi di Padova, via Gradenigo 6B, IT-35131 Padova, Italy}

\author{Francesco~Picciariello}
\affiliation{Dipartimento di Ingegneria dell'Informazione, Universit\`a degli Studi di Padova, via Gradenigo 6B, IT-35131 Padova, Italy}

\author{Andrea~Stanco}
\affiliation{Dipartimento di Ingegneria dell'Informazione, Universit\`a degli Studi di Padova, via Gradenigo 6B, IT-35131 Padova, Italy}

\author{Giuseppe~Vallone}
\affiliation{Dipartimento di Ingegneria dell'Informazione, Universit\`a degli Studi di Padova, via Gradenigo 6B, IT-35131 Padova, Italy}
\affiliation{Padua Quantum Technologies Research Center, Universit\`a degli Studi di Padova, via Gradenigo 6B, IT-35131 Padova, Italy}
\affiliation{Dipartimento di Fisica e Astronomia, Universit\`a degli Studi di Padova, via Marzolo 8, IT-35131 Padova, Italy}

\author{Paolo~Villoresi}
\affiliation{Dipartimento di Ingegneria dell'Informazione, Universit\`a degli Studi di Padova, via Gradenigo 6B, IT-35131 Padova, Italy}
\affiliation{Padua Quantum Technologies Research Center, Universit\`a degli Studi di Padova, via Gradenigo 6B, IT-35131 Padova, Italy}

\author{Francesco~Vedovato}
\email{Corresponding author: francesco.vedovato@unipd.it}
\affiliation{Dipartimento di Ingegneria dell'Informazione, Universit\`a degli Studi di Padova, via Gradenigo 6B, IT-35131 Padova, Italy}
\affiliation{Padua Quantum Technologies Research Center, Universit\`a degli Studi di Padova, via Gradenigo 6B, IT-35131 Padova, Italy}

\maketitle

\section{Introduction}

Quantum Key Distribution (QKD)~\cite{GisinQKD,ScaraniSecurityQKD,DiamantiQKD, pir2019advances} has the potential to allow secure communication between any two points on Earth. In a future continental-scale quantum network (or quantum internet)~\cite{Peev2009,Sasaki2011,Kimble08,Pirandolacomment,Wehnereaam9288} satellite, fiber, and free-space links will be required to operate jointly, in an inter-modal configuration.
While satellite-to-ground QKD has been demonstrated~\cite{Micius_Liao2017,Micius_BBM92_Yin2017,Bedington2017,Agnesi2018,Khan2018} and the development of fiber-based QKD is technologically mature~\cite{BoaronRecord,Yoshino2013,Islam2017,Yuan2018,MinderPittaluga2019,Optica_Agnesi2019,experimental_twin_field_Liu2019,centro_di_calcolo_Avesani2021}, the inter-modal operation of free-space and fiber links has only recently started to be investigated~\cite{Liao2017_daylight, Gong2018, qcosone_Avesani2021}. Free-space ground-to-ground links, although more lossy than a fiber equivalent over the same distance, require a lighter infrastructure investment, may exploit mobile stations and offer connectivity in remote locations.

An inter-modal QKD network must guarantee the compatibility of the free-space links with the fiber-based infrastructure, which is based on  the achievement of stable coupling of the free-space signal into a single-mode fiber (SMF) and on the use of a shared signal wavelength, typically in the telecom band. Coupling the received signal into a SMF brings in of itself several advantages, since the narrow field-of-view of the fiber limits the amount of background solar radiance that can reach the detector and the small mode-field-diameter of a standard SMF (typically $10~\mu$m) allows the use of detectors with a small active area, which are typically faster than larger detectors. This opens the way to daylight free-space QKD, thus enabling for continuous-time operation~\cite{qcosone_Avesani2021}.

However, the SMF coupling efficiency is strongly affected by the wavefront perturbations introduced by atmospheric turbulence, requiring the introduction of mitigation techniques such as Adaptive Optics (AO)~\cite{Jian2014}.

The performance of fiber-based QKD systems was studied in detail by Ref.~\cite{Rusca2018},  where the authors calculated the secret key rate (SKR), optimal decoy-state parameters, and key block length for the finite-key analysis, {considering} {the} channel loss as a fixed parameter. This approach is not appropriate for the case of free-space channels, since the statistics of atmospheric turbulence induces a random fading of the transmitted signal.

The statistics of the free-space channel transmission was derived in~\cite{Vasylyev2016, Vasylyev2018} to calculate the SKR of decoy-state QKD including the effect of collection losses due to beam-wander and scintillation. This treatment is however limited to the case of a QKD receiver with free-space detectors, and thus excludes single-mode fiber-coupled receivers. A similar approach was recently adopted in~\cite{Pirandola2021free,Pirandola2021sat}, for the specific case of continuous-variable (CV) QKD.

The effect of wavefront perturbations and atmospheric scintillation was calculated in~\cite{Canuet2018}, where the the single-mode fiber-coupling probability distribution was derived to extract the fading statistics of a satellite-to-ground link, considering the effect of a partial AO correction of the perturbed wavefront received.
This approach was found by~\cite{qcosone_Avesani2021} to be applicable also to ground-to-ground links.

In this article, we develop a comprehensive model of the performance of a free-space ground-to-ground QKD system. Differently from Ref.~\cite{Pirandola2021free}, we focus on the commonly used efficient-BB84 protocol with active decoy states, in the one-decoy variant of Ref.~\cite{Rusca2018}.
We generalize the approach of~\cite{Vasylyev2018} and~\cite{Canuet2018} to include both the collection losses at the receiver aperture, and the losses due to single-mode fiber-coupling.
Moreover the model of~\cite{Canuet2018} is further extended to include the effect of a finite control bandwidth of the AO system.

The model considers the effect of atmospheric absorption, receiver collection efficiency as a function of beam broadening, beam wandering and atmospheric scintillation, and SMF-coupling in the presence of atmospheric turbulence with partial AO correction of the wavefront deformations and finite AO control bandwidth to calculate the overall channel loss. The finite efficiency and saturation of the single-photon detectors are also included.

The expected error rate is calculated considering the intrinsic coding error caused by imperfect preparation and measurement of quantum states, the noise introduced by the detectors (dark counts and afterpulses), and the amount of diffuse atmospheric background coupled into the receiver in daylight operation.

\begin{figure}[b]
    \centering
    \includegraphics[width=0.9\columnwidth]{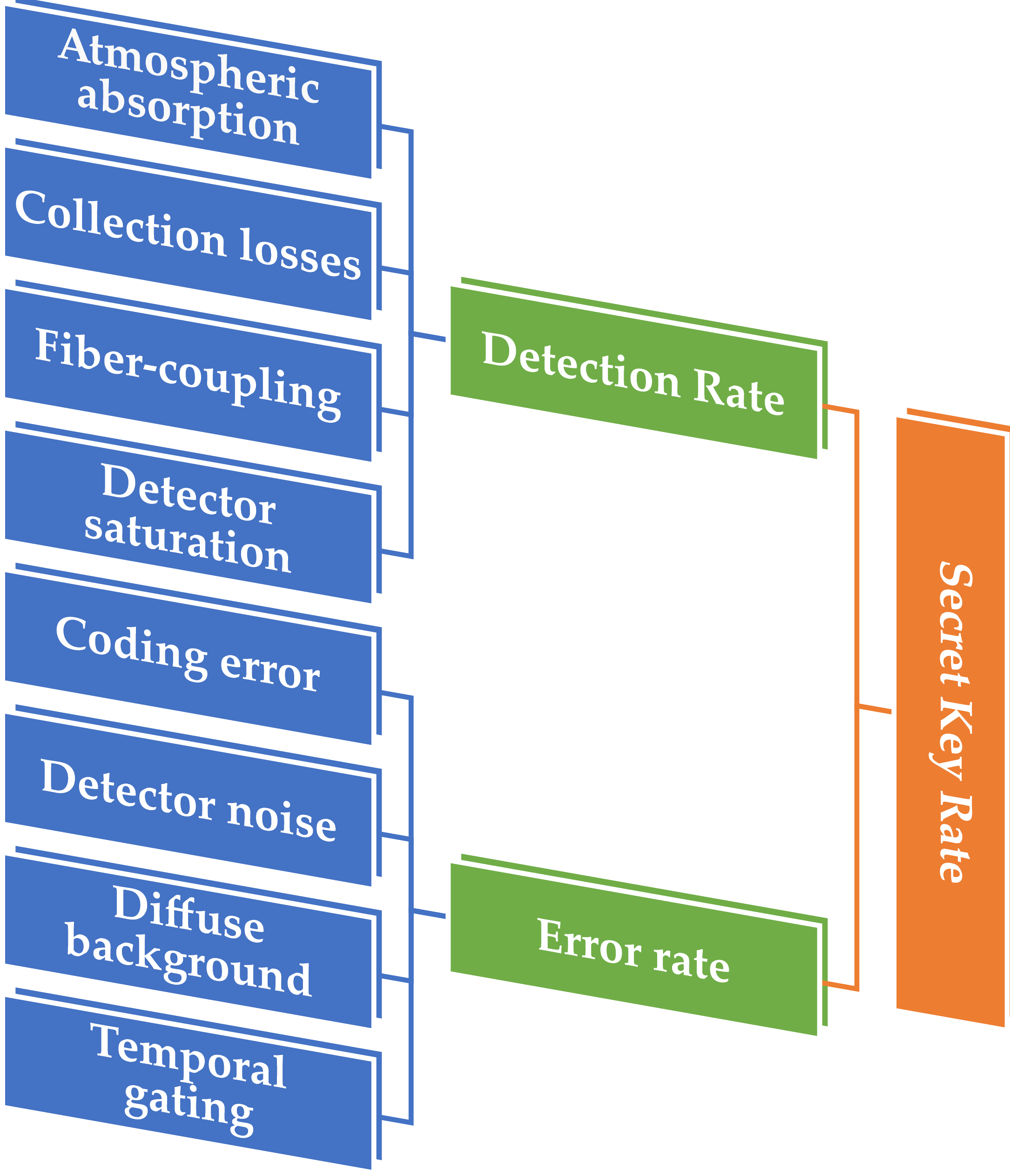}
    \caption{Sketch of the workflow of the model.}
    \label{fig:model_workflow}
\end{figure}

The present model gives as output the obtainable secret key rate (SKR), which takes into account the atmospheric channel contribution, the transmitter and receiver design constraints, the parameters of the quantum source and detectors, and the finite-key analysis to produce a set of requirements and optimal design choices for a QKD system operating under specific free-space channel conditions.
The workflow of our QKD model is sketched in Fig.~\ref{fig:model_workflow}.

In Sec.~\ref{sec:Efficiency}, we study the several contributions to the channel efficiency of ground-to-ground links, which reduce the signal detection rate.
In Sec.~\ref{sec:Errors}, we consider the effects which introduce errors in the exchange of qubits and reduce the secret key rate.
Finally, in Sec.~\ref{sec:Secret}, we combine the channel analysis with the decoy-state and finite-key analysis to estimate the final SKR.

\section{Channel efficiency and detection rate}
\label{sec:Efficiency}

The channel efficiency $\eta_{\rm CH}$ is given by the product of three terms and can be written as:
\begin{equation}
    \eta_{\rm CH} = \eta_\alpha \ \eta_{D_{\rm Rx}} \ \eta_{\rm SMF}
    \label{eqn:eta_tot}
\end{equation}
where $\eta_\alpha$ denotes the atmospheric absorption, $\eta_{D_{\rm Rx}}$ the receiver collection efficiency, and $\eta_{\rm SMF}$ the SMF coupling efficiency.

As a first order analysis, we will consider the effect that the atmospheric turbulence has on the \textit{average} value of the different terms composing the channel efficiency. However, when the receiver rate approaches the saturation limit of the single-photon detectors, the statistics of the collection efficiency and single-mode coupling efficiency can no longer be ignored, and the whole probability distribution $p(\eta_{\rm CH})$  has to be considered for the expected SKR to be estimated correctly (see Sec.~\ref{ss:detect_saturation}).

In our model, the probability distributions are numerically calculated and normalized as weight functions over a discretized array $\lbrace\eta_1,\dots,\eta_N\rbrace$, so that:
\begin{equation}
    \sum_{i=1}^N p(\eta_i) = 1~,
\end{equation}
where the probability $p(\eta_i)$ is the probability that the efficiency lies within the interval $[\eta_{i-1}, \eta_i]$ with $\delta\eta_i = \eta_i - \eta_{i-1}$ the spacing. 
Starting from the analytic probability density function (\textit{pdf}), for sufficiently fine binning we have
\begin{equation}
    p(\eta_i) = pdf(\eta_i)\delta\eta_i \ .
\end{equation}

\subsection{Atmospheric absorption}

The channel absorption $\eta_A$ for a link distance $z$ depends on the absorption coefficient $A(\lambda)$ for a specific wavelength $\lambda$ as: 
\begin{equation}
    \eta_A = 10^{-A(\lambda)\cdot z} ~,
\end{equation}
where the absorption coefficient is assumed constant for a horizontal link.
\begin{figure}
    \centering
    \includegraphics[width=\columnwidth]{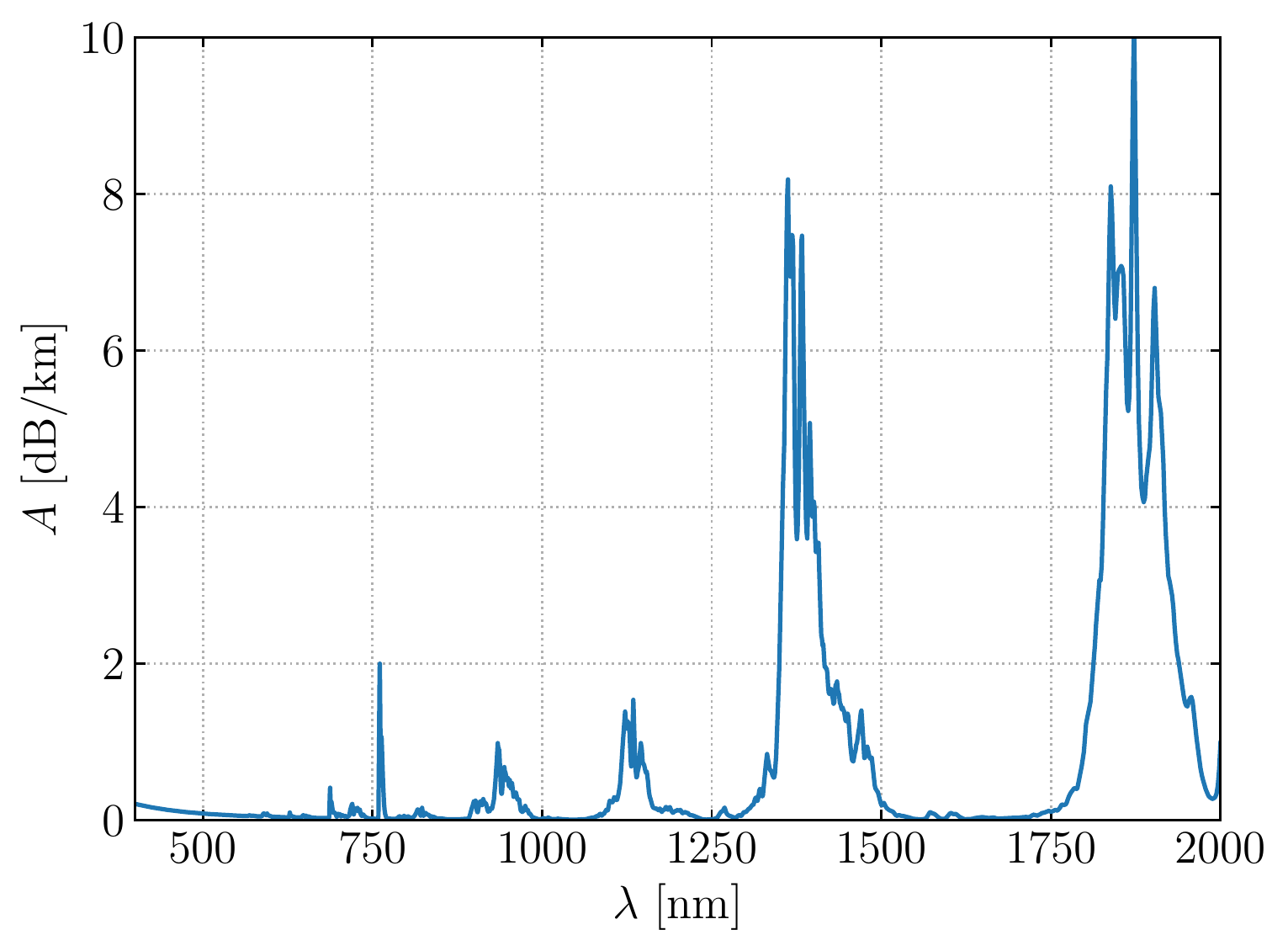}
    \caption{Atmospheric absorption coefficient computed by LOWTRAN for a horizontal path (sub-arctic winter atmospheric model).}
    \label{fig:lowtran_absorption_coefficient}
\end{figure}
An established tool for calculating the spectral properties of the atmosphere is the LOWTRAN software package~\cite{LOWTRAN}, which can be used to predict atmospheric absorption and scattering over a wide wavelength range, on horizontal or slanted paths, taking into account both geographical and seasonal atmospheric variations.
In Fig.\ref{fig:lowtran_absorption_coefficient}, we show the atmospheric absorption coefficient computed by LOWTRAN as a function of wavelength for a horizontal link, considering a sub-arctic winter atmosphere.

\subsection{Collection efficiency}

\subsubsection{Turbulence-induced beam broadening}

In vacuum, the beam size $W(z)$ of a collimated Gaussian beam of waist $W_0$ and wavelength $\lambda$ propagating for a distance $z$ is given by the formula for diffraction-limited propagation:
\begin{equation}
    W(z) = W_0\sqrt{1+\left(\frac{\lambda z}{\pi W_0^2}\right)^2} ~.
    \label{eqn:waist_z_diff_limit}
\end{equation}

{According to the Kolmogorov's theory of turbulence~\cite{Andrews_book}, when the propagation happens through the turbulent air, the index of refraction is treated as a fluctuating random field around a mean value which induces a perturbation of the wavefront, resulting in an overall loss of coherence of the optical wave. The atmospheric perturbation is captured by the so-called \emph{power spectral density} $\Phi(\kappa)$, which is the Fourier transform  of the refractive-index covariance function in terms of the spatial frequency $\kappa$. In our model, we use the power spectral density $\Phi(\kappa)$ for refractive-index fluctuations given by the well-known Kolmogorov spectrum of atmospheric turbulence }
\begin{equation}
    \Phi_n(\kappa) = 0.033 ~C_n^2~\kappa^{-11/3} \ ,
    \label{eqn:kolmogorov_spectrum}
\end{equation}
which is widely used in theoretical calculations.

The strength of the turbulence is parametrized by the refractive-index structure constant $C_n^2$, which may be considered constant along a horizontal link.
The effect of turbulence-induced coherence loss on Gaussian beam propagation has been studied by~\cite{Ricklin2002,Ricklin2003}, who found that the following formula holds:
\begin{equation}
    W(z) = W_0\sqrt{1+\left(1+\frac{2W_0^2}{\rho_0^2(z)}\right)\left(\frac{\lambda z}{\pi W_0^2}\right)^2} ~,
    \label{eqn:waist_z_turbulence}
\end{equation}
where 
\begin{equation}
    \rho_0(z) = (0.55~C_n^2k^2z)^{-3/5}
\end{equation}
is the spherical-wave atmospheric spatial coherence radius, with $k=2\pi/\lambda$ the wave-number.

\begin{figure}[b]
    \centering
    \includegraphics[width=\columnwidth]{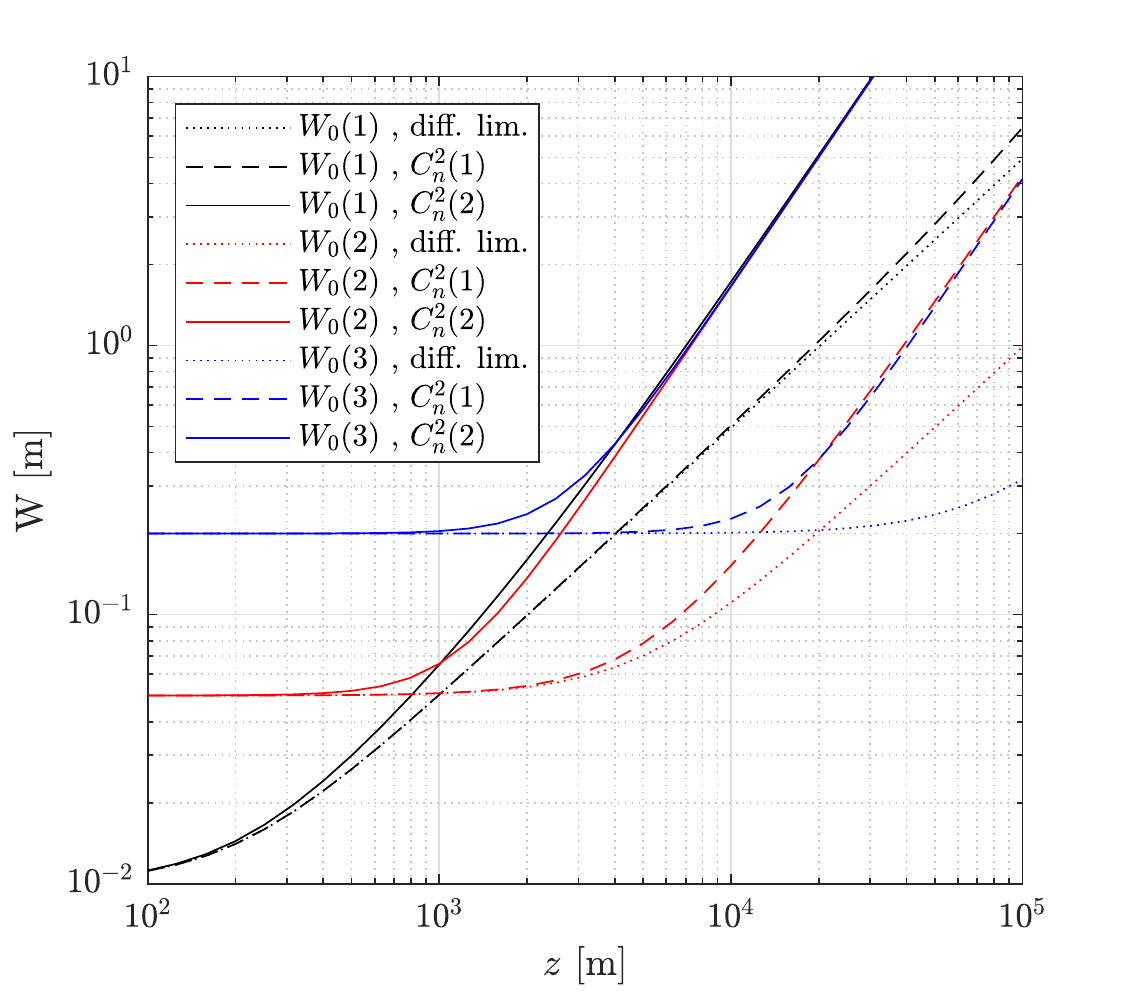}
    \caption{Beam size as a function of distance for different values of $C_n^2$ and $W_0$: $C_n^2(1)=10^{-15}~\rm m^{-2/3}$, $C_n^2(2)=10^{-13}~\rm m^{-2/3}$, $W_0(1)=10~\rm mm$, $W_0(2)=50~\rm mm$, $W_0(3)=200~\rm mm$.}
    \label{fig:beampropagation}
\end{figure}

While for the diffraction-limited case a larger beam waist $W_0$ implies a smaller intrinsic divergence $\theta_0 = \lambda/\pi W_0$, in the turbulence-affected case the larger the ratio $W_0/\rho_0$, the stronger is the loss of coherence.
%{While a larger beam waist implies a smaller intrinsic divergence (in both cases), the $\rho_0$ term introduces a loss of coherence in the turbulence-affected case.} 
Indeed, for $z\gg1$ we have that the two competing effects, diffraction-broadening and turbulence broadening, compensate each other and the beam size tends to a value that is independent of the initial waist $W_0$
\begin{equation}
 W(z) \overset{z\gg1}{\sim}\frac{\lambda\sqrt{2}}{\pi}\left(0.55~C_n^2k^2\right)^{3/10} z^{8/5} \ ,
\end{equation}
as shown in Fig.~\ref{fig:beampropagation} for different values of $W_0$ and $C_n^2$.

Assuming for simplicity that the pointing error is negligible, the average contribution to the collection efficiency $\eta_{D_{\rm Rx}}$ caused by diffraction and beam broadening for a receiver of finite aperture diameter $D_{\rm Rx}$ is given by
\begin{equation}
    \expval{\eta_{D_{\rm Rx}}} =  1 - \exp\left[-\frac{D_{\rm Rx}^2}{2W(z)^2}\right] \ ,
    \label{eqn:eta_diffraction}
\end{equation}
which is the integral of a Gaussian distribution of standard deviation $W(z)/2$ over a concentric circular area of diameter $D_{\rm Rx}$.
 
\subsubsection{Beam wander and scintillation}

The beam-size $W(z)$ appearing in Eq.~\eqref{eqn:eta_diffraction} is the so-called \textit{long-term} spot size, which represents the size of the beam averaged over a timescale much longer than the turbulence dynamic.
The instantaneous \textit{short-term} (ST) beam size $W_{\rm ST}(z)$ at a distance $z$ is given by:
\begin{equation}
    W_{\rm ST}(z) = \sqrt{W^2(z) - \expval{r_c^2}} ~,
    \label{eqn:waist_short_term}
\end{equation}
where $\expval{r_c^2}$ is the variance of beam-wander fluctuations at the receiver aperture plane. 

Beam wandering is caused by the larger-sized turbulence eddies, and results in a shift of the short-term spot on the receiver aperture. For a collimated beam transmitted along a horizontal channel one has~\cite{Andrews_book}:
\begin{equation}
    \expval{r_c^2} = 2.42~ C_n^2 z^3 W_0^{-1/3} ~.
    \label{eqn:beam_wander_variance}
\end{equation}

In addition to beam wander, one must also consider the effect of atmospheric scintillation, which introduces random fluctuations in the beam irradiance profile.
The probability density function $p_{D_{\rm Rx}}$ of $\eta_{D_{\rm Rx}}$ was derived analytically in~\cite{Vasylyev2016, Vasylyev2018} exploiting the law of total probability and separating the contributions from turbulence-induced beam wandering and atmospheric scintillation.
The distribution $p_{D_{\rm Rx}}$ varies depending on the strength of turbulence, which is parametrized by the Rytov variance $\sigma_R^2$, which is defined as:
\begin{equation}
    \sigma_R^2 = 1.23~C_n^2k^{7/6}z^{11/6} ~.
    \label{eqn:rytov_variance}
\end{equation}

\begin{figure}
    \centering
    \includegraphics[width=\columnwidth]{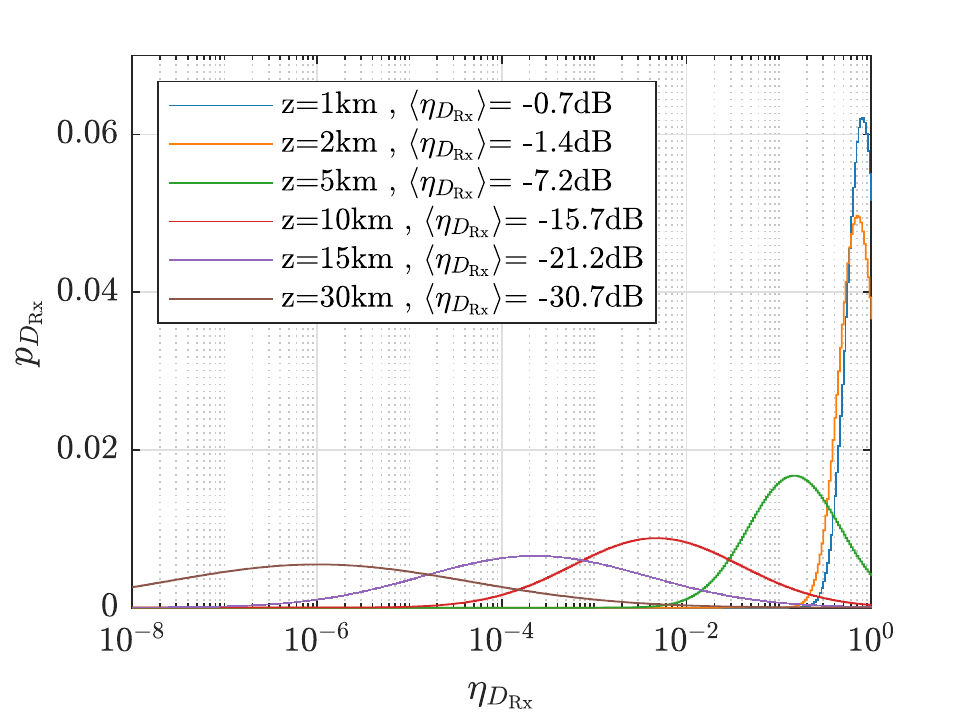}
    \caption{Probability distribution $p_{D_{\rm Rx }}$ of the collection efficiency $\eta_{D_{\rm Rx}}$ for a fixed receiver aperture $D_{\rm Rx}=100~\rm mm$, turbulence parameter $C_n^2=10^{-14}~\rm m^{-2/3}$, transmitter waist $W_0=50~\rm mm$, wavelength $\lambda=1550~\rm nm$, and different link distances: $z=1~\rm km$ ($\sigma_R=0.4$), $z=2~\rm km$ ($\sigma_R=0.8$), $z=5~\rm km$ ($\sigma_R=2$), $z=10~\rm km$ ($\sigma_R=3.7$), $z=15~\rm km$ ($\sigma_R=5.3$), $z=30~\rm km$ ($\sigma_R=10.1$). The probability is given as a weight function, given an array with logarithmic spacing: $\eta_{D_{\rm Rx}} = 10^{[-8~:~0.02~:~0]}$.}
    \label{fig:prob_wander_scintillation}
\end{figure}

For weak turbulence ($\sigma_R^2<1$) $p_{D_{\rm Rx}}$ resembles a log-negative Weibull distribution, while for stronger turbulence ($\sigma_R^2>1$) one finds a truncated log-normal distribution. 
The exact form and derivation of $p_{D_{\rm Rx}}$ can be found in~\cite{Vasylyev2018} and requires the knowledge of the following quantities: the average collection efficiency $\expval{\eta_{D_{\rm Rx}}}$ of Eq.~\eqref{eqn:eta_diffraction},the collection efficiency calculated using the short-term waist $W_{\rm ST}$ of Eq.~\eqref{eqn:waist_short_term}, the beam-wander variance $\expval{r_c^2}$, and the mean-squared efficiency $\expval{\eta_{D_{\rm Rx}}^2}$, that is:
\begin{equation}
    \expval{\eta_{D_{\rm Rx}}^2} = \expval{\eta_{D_{\rm Rx}}}(1 + \sigma_{\rm I}^2(D_{\rm Rx})) ~,
\end{equation}
where $\sigma_{\rm I}^2(D_{\rm Rx})$ is the aperture-averaged scintillation index (\textit{flux variance})~\cite{Andrews_book}:
\begin{align}
    \sigma_{\rm I}^2(D_{\rm Rx}) &=
    \exp\left[
        \frac{0.49\beta_0^2}
            {\left( 1 + 0.18d^2 + 0.56\beta_0^{12/5} \right)^{7/6}} \right. \nonumber \\
    &+\left.   \frac{ 0.51\beta_0^2\left( 1 + 0.69\beta_0^{12/5} \right)^{-5/6}}
            {1 + 0.90d^2 + 0.62d^2\beta_0^{12/5}} 
    \right] - 1 ~,
\label{eqn:scintillation_flux_variance}
\end{align}
with $d = \sqrt{\frac{kD_{\rm Rx}^2}{4z}}$ and $\beta_0^2=0.4065~\sigma_R^2$.
Some examples of probability distributions $p_{D_{\rm Rx}}$ calculated in this way are provided in Fig.~\ref{fig:prob_wander_scintillation}.

\subsection{Fiber coupling efficiency}

The fiber coupling efficiency is given by the normalized overlap integral between the fiber mode, and the incident optical field $U(\vec{r},t)$,
\begin{equation}
    U(\vec{r},t) = U_0(\vec{r},t)\exp\left[\chi(\vec{r},t)+i \Psi(\vec{r},t)\right] \ ,
\end{equation}
where $\chi$ is the log-amplitude perturbation term, and $\Psi$ is the wavefront phase term. The different origins of $\chi$ and $\Psi$ perturbations imply the statistical independence of scintillation and phase effects~\cite{Canuet2018,Fried1966}, and  the average coupling efficiency $\eta_{\rm SMF}$ can be factorized  into three terms:
\begin{equation}
    \eta_{\rm SMF} =\eta_0 \ \eta_{\rm AO} \  \eta_S \
     ,
\end{equation}
where $\eta_0$ is the optical  efficiency of the receiver telescope, $\eta_{\rm AO}$ is the coupling efficiency due to wavefront perturbations that may be partially corrected by AO, and $\eta_S$ is the coupling efficiency due to the spatial structure of atmospheric scintillation. We now discuss the three terms separately.

\subsubsection{Optical coupling efficiency}

\begin{figure}[t]
    \centering
    \includegraphics[width=\columnwidth]{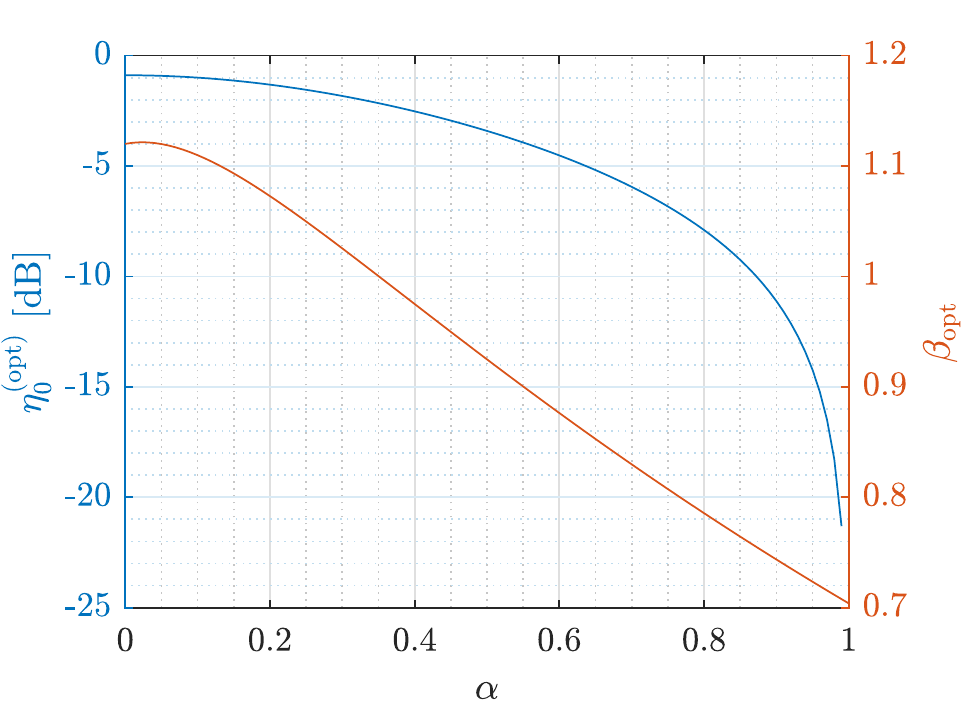}
    \caption{Maximum ideal SMF coupling efficiency and optimum $\beta$ parameter as a function of obscuration ratio $\alpha$. }
    \label{fig:eta0_beta_alpha}
\end{figure}

The \emph{optical coupling efficiency} $\eta_0$ of an optical system measures the matching between an unperturbed received beam and the mode-field diameter (MFD) of the SMF. $\eta_0$ is determined by the design optics of the receiving telescope, and particularly by the ratio $\alpha = D_{\rm Obs}/D_{\rm Rx}$ between the diameters of the central obscuration and of the telescope aperture~\cite{Ruilier2001}. The ideal coupling efficiency can be parametrized by
\begin{equation}
    \eta_0(\alpha,\beta) = 2\left[\frac{\exp(-\beta^2) -\exp(-\beta^2\alpha^2)}{\beta\sqrt{1-\alpha^2}} \right]^2 \ ,
    \label{eqn:eta0}
\end{equation}
where $\beta$ is given by
\begin{equation}
    \beta = \frac{\pi D_{\rm Rx} \rm}{2\lambda} \frac{\rm MFD}{f} \ ,
\end{equation}
with $D_{\rm Rx}$ the receiver diameter, $f$ the effective focal length of the optical system, and {\rm MFD} the mode field diameter of the SMF.
Given a particular $\alpha$, the value of $\beta$ can be optimized to achieve the optimal coupling efficiency $\eta_0^{\rm (opt)} = \eta_0(\beta_{\rm opt})$. Knowing the value of $\beta_{\rm opt}$ allows to choose the optimal design value of $f$, since typically the working wavelength $\lambda$, fiber ${\rm MFD}$ and receiver diameter are constrained.  Fig.~\ref{fig:eta0_beta_alpha} shows $\eta_0^{\rm (opt)}$ and $\beta_{\rm opt}$ as a function of the obscuration ratio $\alpha$.
For $\alpha=0$, we have $\beta_{\rm opt}=1.12$, that allows to achieve a maximum optical coupling efficiency of $81.5\% \approx -0.89$~dB.

\subsubsection{Effect of wavefront perturbations and adaptive optics correction}

As was first derived by~\cite{Fried1965} and~\cite{Noll1976}, 
the instantaneous wavefront aberration $\Psi(\vec{r},t)$ introduced by the turbulent channel at a point $\vec{r}$ on the receiver aperture can be decomposed in a superposition of Zernike polynomials defined over the normalised pupil coordinates $(r,\varphi)$, with $r = 2|\vec{r}|/D_{\rm Rx}$. In the decomposition of $\Psi(\vec{r},t)$, each polynomial term $Z_n^m(r,\varphi)$ of radial degree $n$ and azimuthal degree $m$ is weighted by a time-dependent coefficient $b_n^m(t)$, yielding
\begin{equation}
\Psi(r,\varphi,t) = \sum_{n,m} b_n^m(t) Z_n^m(r,\varphi) ~.
\end{equation}

The Zernike coefficient variances $\expval{b_n^{m2}}$  represent the statistical strength of a particular aberration order, and depend on the ratio of the receiver aperture $D_{\rm Rx}$ to the atmospheric coherence width $r_0=2.1\rho_0$ (also known as the Fried parameter),  with a modal term scaling with the radial order $n$ ~\cite{Noll1976, Boreman1996}:
\begin{equation}
\expval{b_n^{m2}} = \left( \frac{D_{\rm Rx}}{r_0} \right) ^{\frac{5}{3}} \frac{n+1}{\pi}  \frac{ \Gamma \left( n-\frac{5}{6} \right) \Gamma \left(\frac{23}{6}\right) \Gamma \left(\frac{11}{6}\right) \sin \left(  \frac{5}{6}\pi \right)}{ \Gamma \left(n+\frac{23}{6}\right)} ~.
\label{eqn:zer_variance}
\end{equation}

An expression of the instantaneous coupling efficiency in the presence of wavefront perturbations was derived by~\cite{Ma2015} and~\cite{Canuet2018} directly in terms of the Zernike coefficients $b_n^m$:
\begin{equation}
\eta_{\rm AO}(t) = \exp \left[ - \sum_{n,m} b_n^m(t)^2\right] ~.
\label{eqn:eta_smf_instant}
\end{equation}

From Eq.~\eqref{eqn:eta_smf_instant}, knowing that the coefficients are independent, Gaussian-distributed random variables with zero mean and variance given by Eq.~\eqref{eqn:zer_variance} -- i.e. $b_n^m\sim\mathcal{N}(0,\expval{b_n^{m2}})$ --  we derive the average coupling efficiency:
\begin{equation}
    \expval{\eta_{\rm AO}} = \prod_{n,m} \frac{1}{\sqrt{1+2\expval{b_n^{m2}}}} ~.
    \label{eqn:eta_smf_average}
\end{equation}

\begin{figure}[b]
    \centering
    \includegraphics[width=\columnwidth]{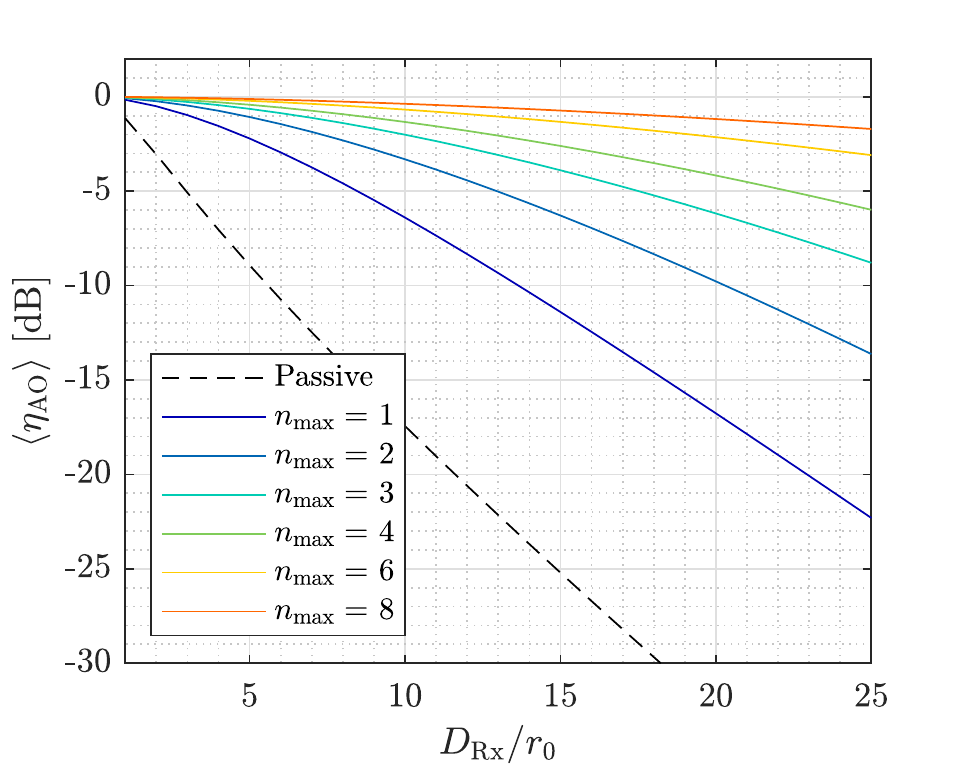}
    \caption{SMF coupling efficiency as a function of the ratio of receiver diameter to atmospheric coherence width for increasing order of AO correction, assuming infinite control bandwidth.}
    \label{fig:smf_n_inf_bw}
\end{figure}

Eq.~\eqref{eqn:eta_smf_average} makes the calculation of the average SMF coupling efficiency in the presence of a partial adaptive-optic compensation of turbulence up to an order $n_{\rm max}$  straightforward.
Assuming an ideal AO system with infinite control bandwidth, it is sufficient to completely suppress the coefficients in the productory corresponding to radial orders $n\leq n_{\rm max}$. Fig.~\ref{fig:smf_n_inf_bw}
shows the SMF coupling efficiency as a function of the ratio $D_{\rm Rx}/r_0$ of receiver diameter to atmospheric coherence width for increasing order of AO correction, assuming infinite control bandwidth.

\begin{figure}
    \centering
    \includegraphics[width=\columnwidth]{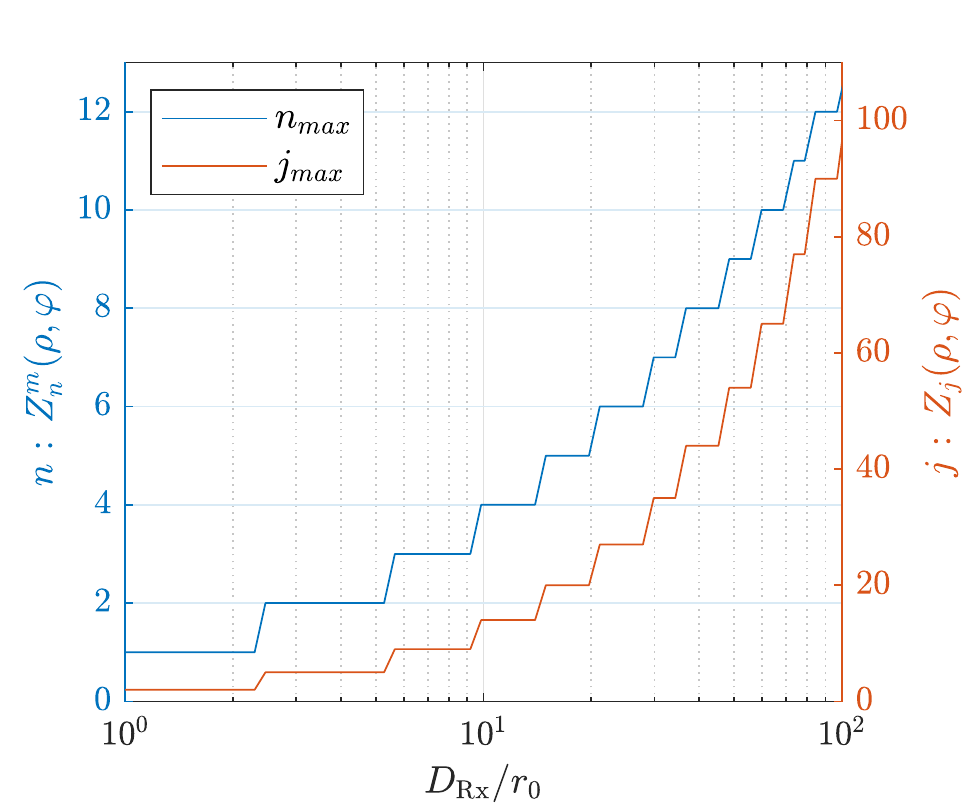}
    \caption{Required Zernike correction shown as the maximum order (radial $n$ and single OSA/ANSI index $j = [n(n+1)+m]/2$) that is above the threshold for diffraction-limited quality as a function of the ratio of receiver diameter to Fried parameter.}
    \label{fig:required_zernike_correction}
\end{figure}

From Eq.~\eqref{eqn:zer_variance}, we have that the strength of the aberration orders decreases for higher orders. As is shown in Fig.~\ref{fig:required_zernike_correction}, it is interesting then to evaluate how many Zernike modes should be compensated to regain near diffraction-limited wavefront quality. This can be done referring to the Rayleigh criterion~\cite{malacara_book}, from which we have that aberrations whose $\rm RMS\leq 0.05\lambda$ are below the threshold for diffraction-limited quality.

As discussed in Ref.~\cite{Roddier_AO}, the effect of control bandwidth limitations can be taken into account by introducing a mode attenuation factor $\gamma_n^2$ for the $n$-th order aberration coefficients: % Note: citation was Roddier2004
\begin{equation}
    \gamma_n^2 = \frac{\int |W_n(\nu)|^2|\varepsilon(\nu) |^2\rm d \nu}
                      {\int |W_n(\nu)|^2 \rm d \nu} ~,
\label{eqn:gamma_nu_bandwidth}
\end{equation}
where $|W_n(\nu)|^2$ represents the power spectral density of the temporal spectrum of the $n$-th order aberrations, and $\varepsilon(\nu)$ represents the transfer function between the
residual phase and the turbulent wavefront fluctuations, and depends on the AO system's open-loop transfer function $G(\nu)$:
\begin{equation}
    \varepsilon(\nu) = \frac{1}{1+G(\nu)} ~.
    \label{eqn:epsilon_nu}
\end{equation}
In this work, we consider a typical AO system with a pure-integrator control based on wavefront sensing with a Shack-Hartmann wavefront sensor, and correction with a deformable piezo-electric mirror. The open-loop transfer function is then given by
\begin{equation}
    G(\nu) = K_i\frac{e^{-\tau \nu}\left(1-e^{-T\nu} \right)}{(T\nu)^2} ~,
    \label{eqn:g_open}
\end{equation}
where $K_i$ is the gain of the integrator, $\tau$ is the overall latency of the control-actuator stage, and $T$ is the inverse of the wavefront sensor frame rate.

The temporal power spectra of the different aberration orders have been studied in Ref.~\cite{Conan1995}, where the authors find that the power spectral density scales polynomially with a cut-off frequency $\nu_c^{(n)}$ depending on the radial order $n$, average wind velocity $\bar v$ and receiver diameter $D_{\rm Rx}$:
\begin{equation}
    |W_n(\nu)|^2 \sim 
    \begin{cases}
    \nu^{-2/3}  & \nu\leq\nu_c, n=1\\
    \nu^{0}     & \nu\leq\nu_c, n\neq1\\
    \nu^{-17/3} & \nu>\nu_c
    \end{cases}
\end{equation}
with
\begin{equation}
    \nu_c^{(n)}=0.3(n+1)\bar{v}/D_{\rm Rx} \ .
\end{equation}
The SMF efficiency of Eq.~\eqref{eqn:eta_smf_average} is thus modified in the case of finite control bandwidth as:
\begin{align}
    \expval{\eta_{AO}} &= \prod_{\substack{n,m \\n\leq n_{\rm max}}} \frac{1}{\sqrt{1+2\gamma_n^2\expval{|b_n^{m2}}}} \nonumber \\
    & \quad + \prod_{\substack{n,m \\n>n_{\rm max}}} \frac{1}{\sqrt{1+2\expval{b_n^{m2}}}} \ ,
\end{align}
with $n_{\rm max}$ the maximum aberration order corrected.

\begin{figure}[b]
    \centering
    \includegraphics[width=\columnwidth]{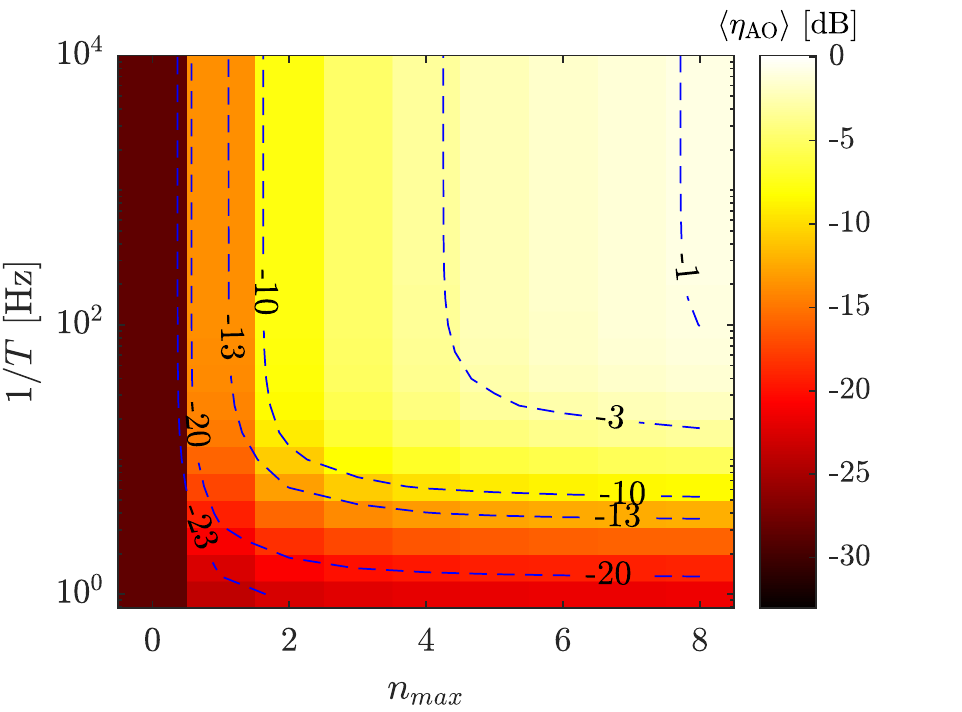}
    \caption{Coupling efficiency $\expval{\eta_{\rm AO}}$ as a function of maximum aberration order corrected $n_{\rm max}$, and integration time of the wavefront sensor $T$, for a scenario with $\lambda= 1550$~nm, $z= 20$~km, $D_{\rm Rx}= 400$~mm, $D_{\rm Rx}/r_0= 17$, $C_n^2= 10^{-14}{\rm m}^{-2/3}$, and $\bar{v}=3~\rm m/s$.}
    \label{fig:eta_SMF_vs_n_BW}
\end{figure}

Fig.~\ref{fig:eta_SMF_vs_n_BW} shows $\expval{\eta_{\rm AO}}$ as a function of the corrected aberration order and wavefront sensor integration time, in a scenario with $D_{\rm Rx}/r_0=17$ and an average wind velocity corresponding to a light breeze.

\subsubsection{Effect of atmospheric scintillation}

In addition to phase perturbations, we also take into account the irradiance fluctuations introduced by atmospheric scintillation, which result in a random apodization of the pupil transmittance function, which in turn affects the maximum SMF coupling efficiency. 
A rigorous calculation of the scintillation contribution $\eta_S$ to the SMF efficiency, which considers the optical system modulation transfer function and log-amplitude spatial covariance function $C_\chi(r)$ can be found in Ref.~\cite{Canuet2018} and is based on the result of Ref.~\cite{Fried1965,Fried1966}. Nonetheless, a good approximation for the average value of $\eta_S$ is only dependent on the on-axis pupil-plane scintillation index $\sigma_I^2$, which is given by Eq.~\eqref{eqn:scintillation_flux_variance} in the limit of an infinitesimal aperture $d=0$.
\begin{equation}
    \expval{\eta_S} \approx \exp[-C_\chi(0)] = \exp[-\sigma_\chi^2] = (1+\sigma_I^2)^{-\frac{1}{4}} ~,
\end{equation}
where we use the fact that the log-amplitude variance $\sigma_\chi^2$ is related to the scintillation index through $\sigma_I^2=\exp(4\sigma_\chi^2)-~1$.

\begin{figure}
    \centering
    \includegraphics[width=\columnwidth]{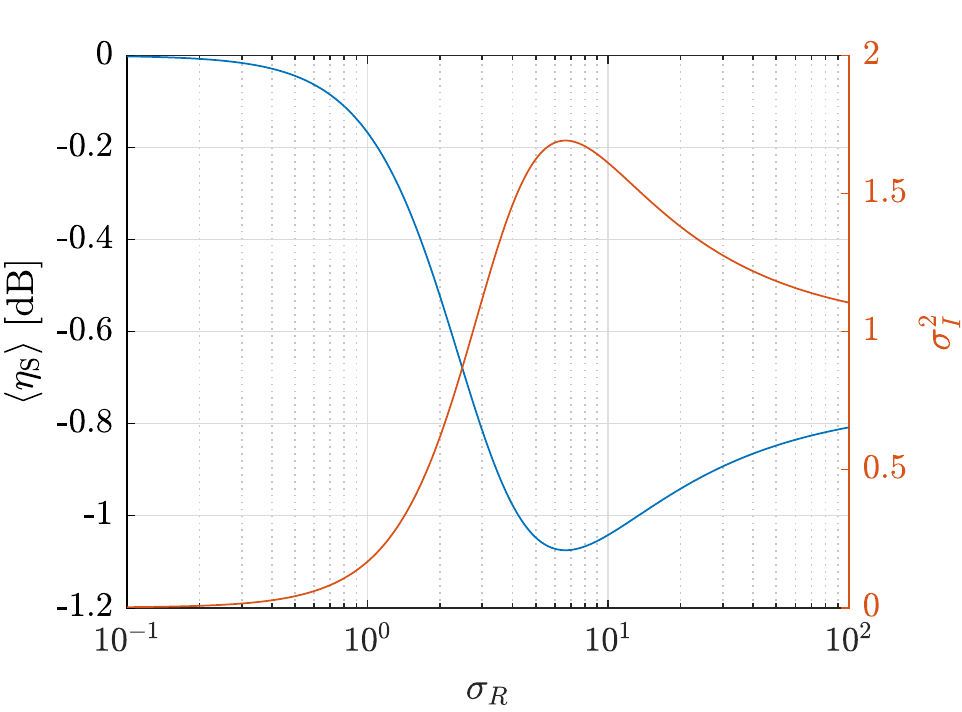}
    \caption{Scintillation contribution to the SMF coupling efficiency and scintillation index as a function of the Rytov variance $\sigma_R$ (see Eq.~\eqref{eqn:scintillation_flux_variance}).}
    \label{fig:eta_scintill}
\end{figure}

Fig.~\ref{fig:eta_scintill} shows a plot of $\expval{\eta_S}$ and $\sigma_I^2$ as a function of the Rytov variance \eqref{eqn:rytov_variance}, highlighting the behavior of $\sigma_I^2$, which increases for the \textit{weak fluctuation} regime, reaches a maximum value in the \textit{focusing} regime, and then tends to $\sigma_I^2\sim 1$ in the \textit{strong fluctuation} regime.

\subsubsection{Optimum receiver diameter}

\begin{figure}[t!]
    \centering
    \includegraphics[width=\columnwidth]{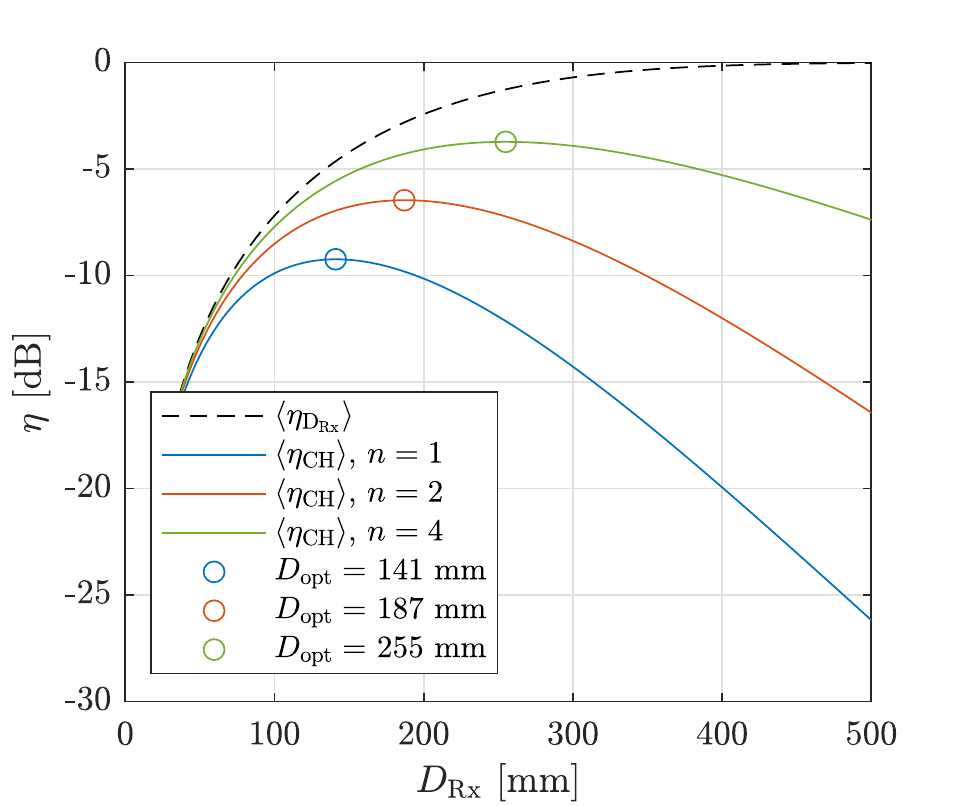}
    \caption{Trade-off between collection efficiency and fiber-coupling efficiency. Fixed link parameters: $\lambda=1550~\rm nm$, $z= 5~\rm km$, $W_0=50~\rm mm$, and $C_n^2= 10^{-14}{\rm m}^{-2/3}$.}
    \label{fig:efficiency_vs_D}
\end{figure}

We have seen that the average collection efficiency $\expval{\eta_{D_{\rm Rx}}}$ of Eq.~\eqref{eqn:eta_diffraction} increases as the receiver diameter increases.
Conversely, for a fixed AO correction order $n$, the SMF coupling efficiency of $\expval{\eta_{\rm AO}}$ Eq.~\eqref{eqn:eta_smf_average} decreases with increasing receiver diameter.
This leads to a trade-off between the collection efficiency and the fiber-coupling efficiency  (as represented in Fig.~\ref{fig:efficiency_vs_D}) in order to maximize the overall channel efficiency $\eta_{\rm CH}$, and we can find the optimum receiver diameter $D_{\rm opt}$, once the other link parameters, such as wavelength, link distance, transmitter waist, and turbulence strength are fixed.

\begin{figure}
    \centering
    \includegraphics[width=\columnwidth]{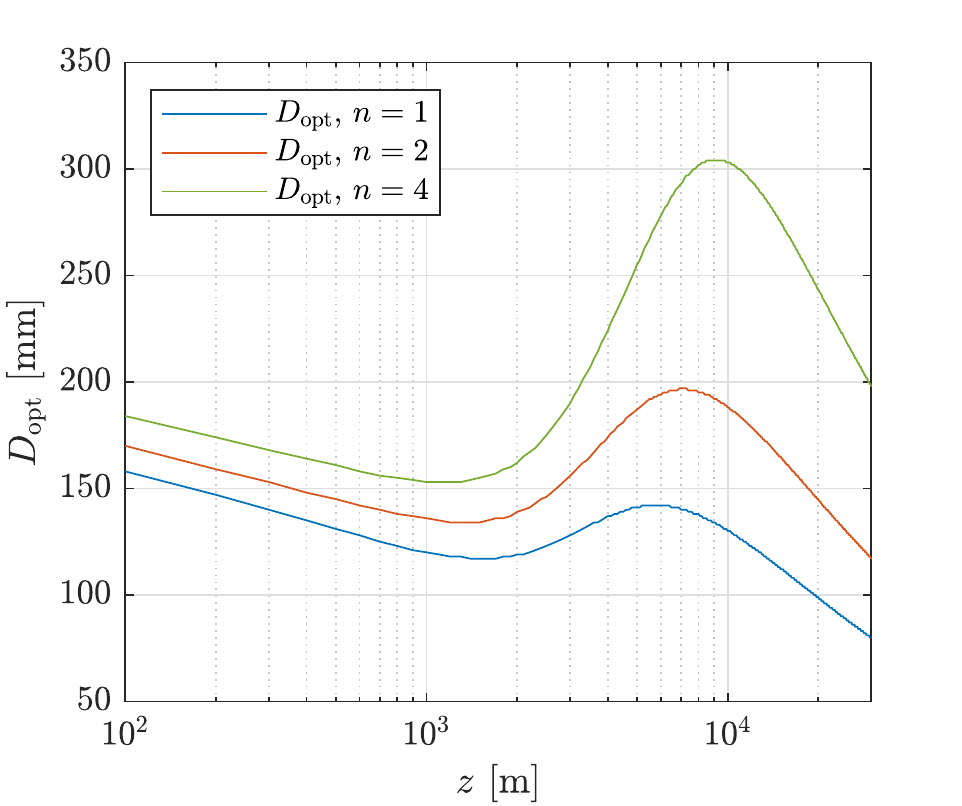}
    \caption{Optimum receiver beam diameter as a function of link distance. Fixed link parameters: $\lambda=1550~\rm nm$, $W_0=50~\rm mm$, $z_0=5~\rm km$, and $C_n^2= 10^{-14}{\rm m}^{-2/3}$.}
    \label{fig:optimum_diameter}
\end{figure}

Fig.~\ref{fig:optimum_diameter} shows the optimum receiver diameter as a function of link distance, for different orders of aberration correction and a moderate turbulence strength of $C_n^2=10^{-14}~\rm m^{-2/3}$.
For short link distances (shorter than the transmitter Rayleigh distance $z_0=\pi W_0^2/\lambda$) the beam size does not diverge much, and the fiber-coupling term dominates, causing the optimal diameter to decrease with increasing distance.
For link distances $z\sim z_0$, the beam size starts increasing, the collection efficiency term dominates, leading to an increasing optimum beam diameter.
For longer propagation distances ($z\gg z_0$), the decrease in spatial coherence of the beam is more severe and the turbulence term dominates again, leading to a smaller optimum diameter.

\subsubsection{Probability distribution of coupling efficiency}

As anticipated, when the receiver rate approaches the saturation limit of the single-photon detectors, the full  statistics of channel efficiency can no longer be ignored, and the whole probability distribution $p_{\rm CH}$  has to be considered for the expected SKR to be estimated correctly, as we will present in Sec.~\ref{ss:detect_saturation}).

\begin{figure}[t]
    \centering
    \includegraphics[width=\columnwidth]{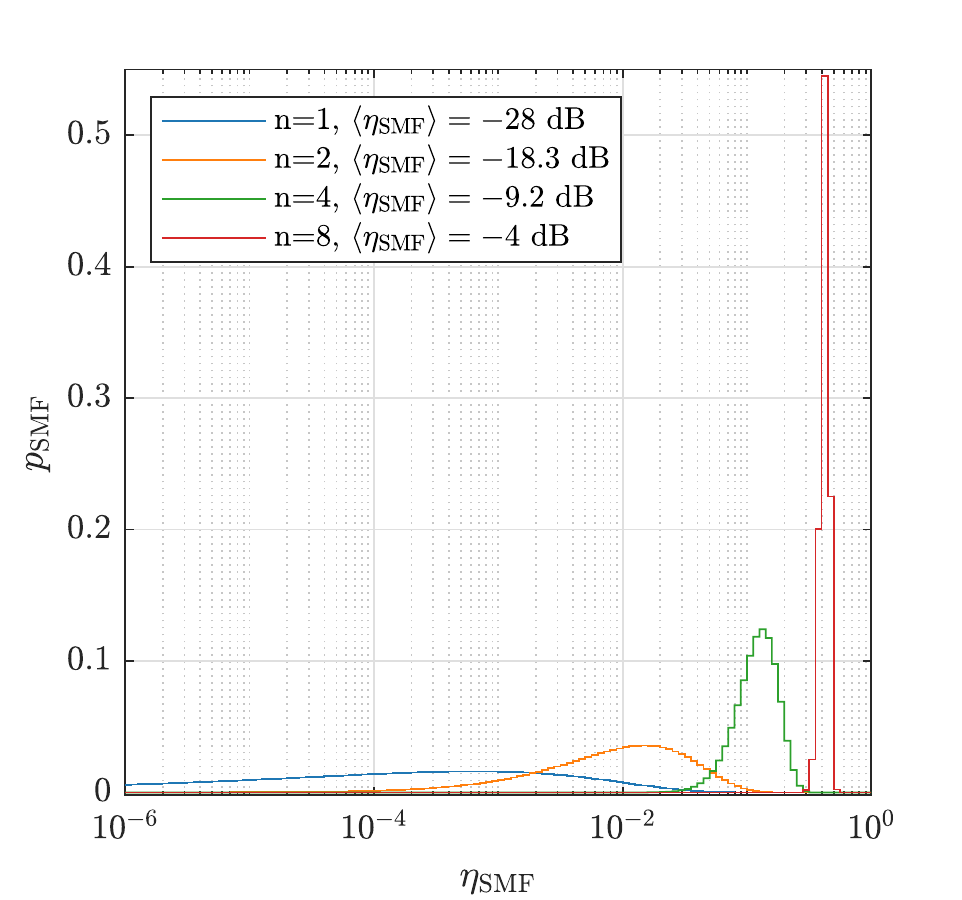}
    \caption{Probability distribution of the SMF efficiency for a fixed link distance $z=10~\rm km$, turbulence parameter $C_n^2=10^{-14}~\rm m^{-2/3}$, transmitter waist $W_0=50~\rm mm$, wavelength $\lambda=1550~\rm nm$, ideal design efficiency $\eta_0=0.8145$, receiver apertures  $D_{\rm Rx}=400~\rm mm$, for different orders of AO correction, assuming infinite correction bandwidth. The probability is given as a weight function, given an array with logarithmic spacing: $\eta= 10^{[-6 : 0.05 : 0]}$.}
    \label{fig:pdf_smf_n}
\end{figure}

\begin{figure}[t]
    \centering
    \includegraphics[width=\columnwidth]{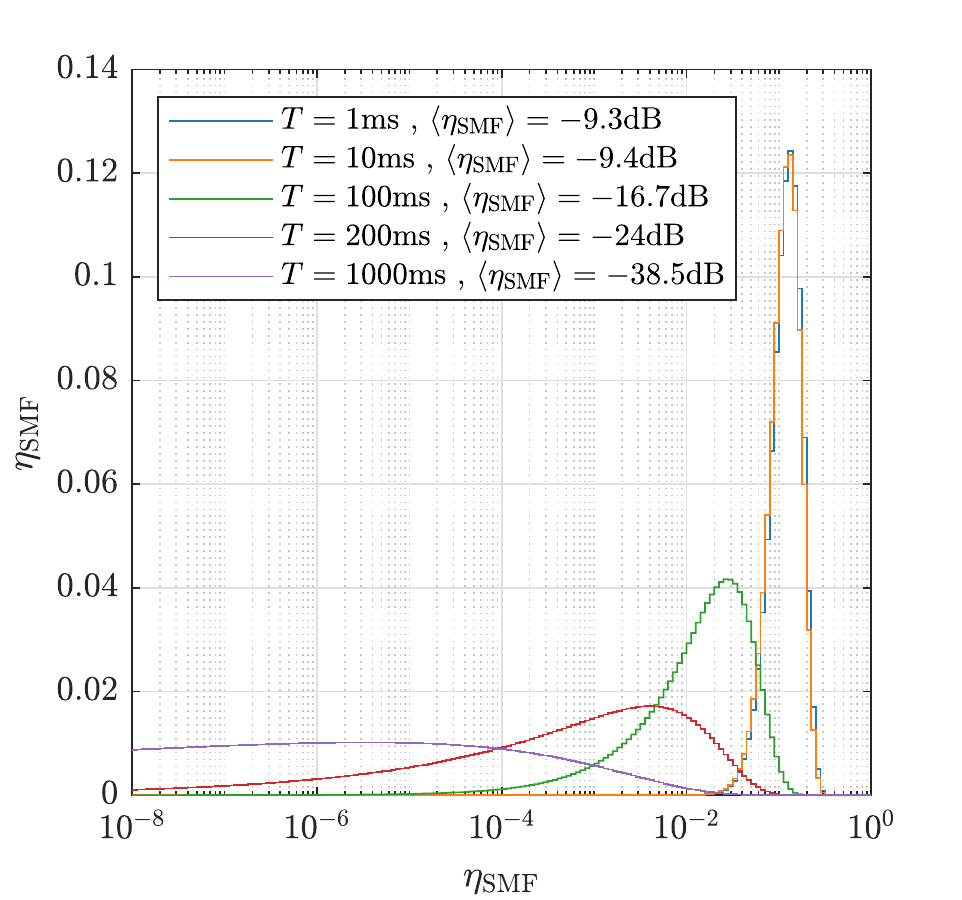}
    \caption{Probability distribution of the SMF efficiency for a fixed link distance $z=10~\rm km$, turbulence parameter $C_n^2=10^{-14}~\rm m^{-2/3}$, transmitter waist $W_0=50~\rm mm$, wavelength $\lambda=1550~\rm nm$, ideal design efficiency $\eta_0=0.8145$, receiver aperture  $D_{\rm Rx}=400~\rm mm$, AO correction order $n=4$, as a function of the wavefront sensor integration time. The probability is given as a weight function, given an array with logarithmic spacing: $\eta= 10^{[-8 : 0.05 : 0]}$.}
    \label{fig:pdf_smf_bw}
\end{figure}

The derivation of the probability distribution of the SMF coupling efficiency with partial adaptive optics correction can be found in Ref.~\cite{Canuet2018}, which, however, does not include the effect of finite control bandwidth.
Since the irradiance fluctuation statistical contribution is already taken into account for the collection efficiency, we restrict the calculation of the SMF coupling distribution to phase distortions only, and present some examples of probability distributions calculated as a function of the maximum corrected aberration order (Fig.~\ref{fig:pdf_smf_n}), and varying the bandwidth of the AO control loop (Fig.~\ref{fig:pdf_smf_bw}).

Given a set of Zernike coefficients $\lbrace b_j\rbrace$ with variances $\langle b_j^2\rangle$, which may be corrected by a compensation factor $\gamma_j^2$ -- where $\gamma_j^2=0$ for perfect compensation, and $\gamma_j^2=1$ for uncorrected coefficients~\cite{Canuet2018} -- we define the quantity $z(t)$ as the instantaneous sum of the squared Zernike coefficients:
\begin{equation}
    z(t) = \sum_j b_j^2(t) ~. 
\end{equation}
The probability distribution of $z$, including the effect of finite AO control bandwidth is then, using the result of Ref.~\cite{gil_pelaez}:
\begin{equation}
    p_z(z) = \frac{1}{\pi} \int_0^\infty
    \frac{\cos\left[\sum_j\frac{1}{2}\text{arctan}\left(2\gamma_j^2\langle b_j^2\rangle u\right)  -zu\right]}
         {\prod_j \left[1+u^2\left(\gamma_j\langle b_j^2\rangle\right)^2 \right]^{1/4}}{\rm d}u ~.
\end{equation}
The probability distribution of $\eta_{\rm SMF}$ is then:
\begin{equation}
    p_{\rm SMF}(\eta_{\rm SMF}|\eta_{\rm max}) = \frac{1}{\eta_{\rm SMF}}p_z\left[\log\left(\frac{\eta_{\rm max}}{\eta_{\rm SMF}}\right)\right] ~, \label{eq_32}
\end{equation}

where $\eta_{\rm max}=\eta_0\cdot\eta_S$ is the maximum normalized coupled flux, given by the product of the optical coupling efficiency of the system $\eta_0$, and the spatial scintillation term $\eta_S$.


\subsection{Channel probability distribution}\label{sec:prob_ch}

Based on the results of Refs.~\cite{Canuet2018}, which calculated the channel loss term due to SMF coupling, and Ref.~\cite{Vasylyev2018}, which calculated the channel loss due to the finite receiver aperture in the presence of beam broadening, beam wandering and scintillation, we calculate the overall channel transmittance probability distribution considering both contributions (pupil-plane and focal-plane losses) and exploiting the law of total probability to write $p_{\rm CH}(\eta_{\rm CH})$ as:
\begin{equation}
    p_{\rm CH}(\eta_{\rm CH}) = \int p_{\rm SMF}(\eta_{\rm CH}|\eta_0 \eta_S \eta_{D_{\rm Rx}})p_{D_{\rm Rx}}(\eta_{D_{\rm Rx}}) \rm d \eta_{D_{\rm Rx}} ~,
\end{equation}
where $p_{\rm SMF}(\eta_{\rm CH}|\eta_0\eta_S\eta_{D_{\rm Rx}})$ is the probability of obtaining a normalized flux $\eta_{\rm CH}$ in the SMF fiber, given a maximum input normalized flux $\eta_0 \ \eta_S \ \eta_{D_{\rm Rx}}$, through Eq.~\eqref{eq_32}. 

\begin{table*}[htb!]
    \centering
    \begin{tabular}{lccccccccc}
    \toprule
        & Case 1 & Case 2 & Case 3 & Case 4 & Case 5 & Case 6 & Case 7 & Case 8 &\\
       \midrule
       $\langle\eta_{\rm CH}\rangle$ & -7 & -15 & -17 & -23 & -25 & -38 & -43 & -48 & $[\rm dB]$\\
       \midrule
       $C_n^2$  & $10^{-13}$ & $10^{-14}$ & $10^{-13}$ &
       $10^{-14}$ & $10^{-14}$ & $10^{-14}$ & $10^{-14}$ & $10^{-14}$ & $\rm[m^{-2/3}]$\\
       $W_0$ & 25 & 60 & 25 & 60 & 60 & 60 & 60 & 25 & $[\rm mm]$\\
       $D_{\rm Rx}$ & 50.8 & 200 & 50.8 & 200 & 50 & 200  & 400 & 200 & $[\rm mm]$\\
       $z$ & 1 & 10 & 2 & 10 & 10 & 20 & 20 & 30 & $[\rm km]$\\
       $n_{\rm max}$ & 1 & 4 & 1 & 1 & 1 & 1 & 2 & 1\\
       $\eta$ & $10^{[-3:0.02:0]}$ & $10^{[-5:0.05:0]}$ & $10^{[-5:0.05:0]}$ & $10^{[-8:0.1:0]}$ & $10^{[-8:0.1:0]}$ & $10^{[-12:0.1:0]}$ & $10^{[-15:0.1:0]}$ & $10^{[-15:0.1:0]}$ \\
    \bottomrule
    \end{tabular}
    \caption{Input parameters for the simulation of the eight case studies.}
    \label{tab:case_studies}
\end{table*}

\begin{figure}[b]
    \centering
    \includegraphics[width=\columnwidth]{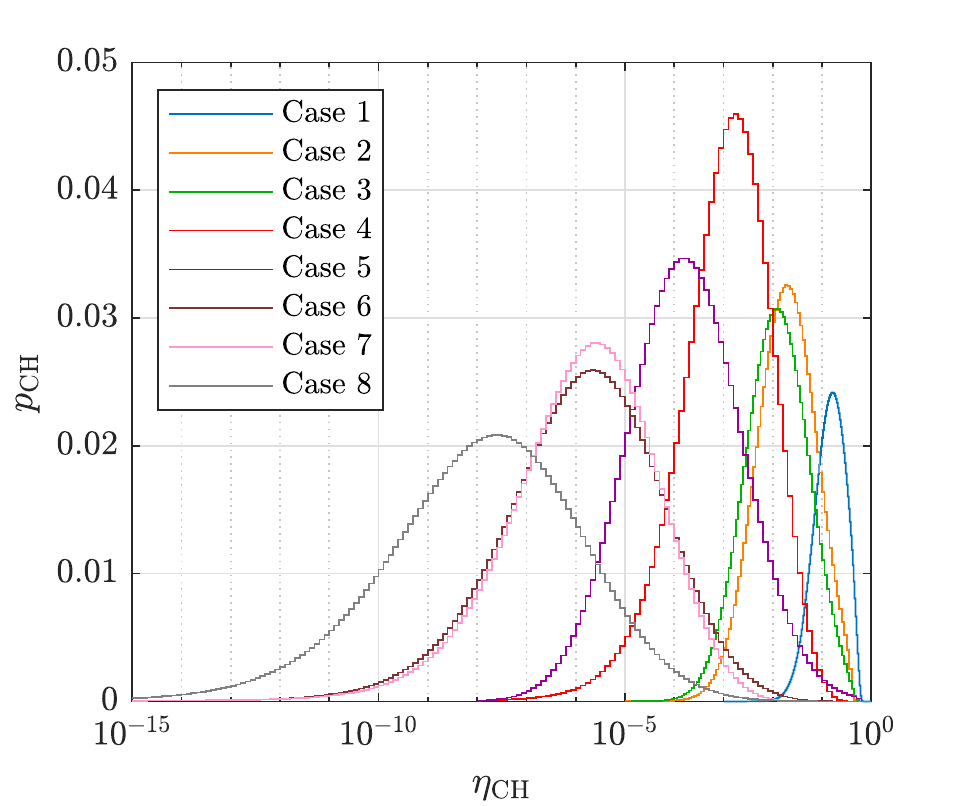}
    \caption{Overall channel probability distributions for the case studies described in Table~\ref{tab:case_studies}.}
    \label{fig:joint_distributions}
\end{figure}

In Fig.~\ref{fig:joint_distributions} we show some examples of channel probability distributions for eight case studies, with average link losses in the range $[-7:-48]~\rm dB$ and input parameters summarized in Table~\ref{tab:case_studies}. In all case studies, we assume an unobstructed receiver aperture, and maximum optical efficiency of $\eta_0=81.5\%$. Cases 1 and 3 correspond to a scenario with a short urban link with strong turbulence, small aperture Tx/Rx telescopes, and mere tip/tilt correction.
Cases 4, and 5 correspond to a scenario with moderate turbulence and longer link distance and highlight the effect of a smaller/larger receiver aperture, which leads to a trade-off between large collection efficiency and minimum aberration order corrected.
Cases 2 and 6 show the effect of AO on longer, moderately turbulent links with large aperture receivers.
Cases 7 and 8 show examples of highly lossy channels.

\subsection{Effect of detector saturation}
\label{ss:detect_saturation}

The key performance indicators of QKD, such as the SKR and error rate, are estimated using large samples of data acquired during a long experiment, averaging over the fluctuations of the channel efficiency.
This might suggest that these fluctuations can be neglected and only their mean value is relevant.
Yet, this would be equivalent to calculating the expected value of a function by applying it to the expected value of its parameters.
Since the functions that model the generation of a key are not all affine, i.e., they are not all compositions of a translation and a linear map, neglecting the fluctuations and using only the mean values is in general incorrect.

The most important non-affine effect is caused by the saturation of the detectors.
Single photon detectors are blinded just after an event and might be further kept off to combat the phenomenon of afterpulses, which increases noise.
This so-called dead time $T_d$ implies that there is maximum rate of output signals $R_{\rm sat} = 1/T_d$ that the detectors can produce.

If $R_0(\eta_{\rm CH})$ is the rate of photons reaching a detector multiplied by its finite efficiency, the output detection rate is~\cite{Muller1974}
\begin{equation}
    R_{\rm det} = \frac{R_0(\eta_{\rm CH})\cdot R_{\rm sat}}{R_0(\eta_{\rm CH}) + R_{\rm sat}} \ .
    \label{eq:Saturation}
\end{equation}
Because this is not an affine function of $R_0(\eta_{\rm CH})$, its fluctuations cannot be neglected.
Although this expression is derived for a continuous source, it is approximately valid also for a pulsed one if the repetition rate is much greater than $R_{\rm sat}$.

\begin{figure}
    \centering
    \includegraphics[width=\columnwidth]{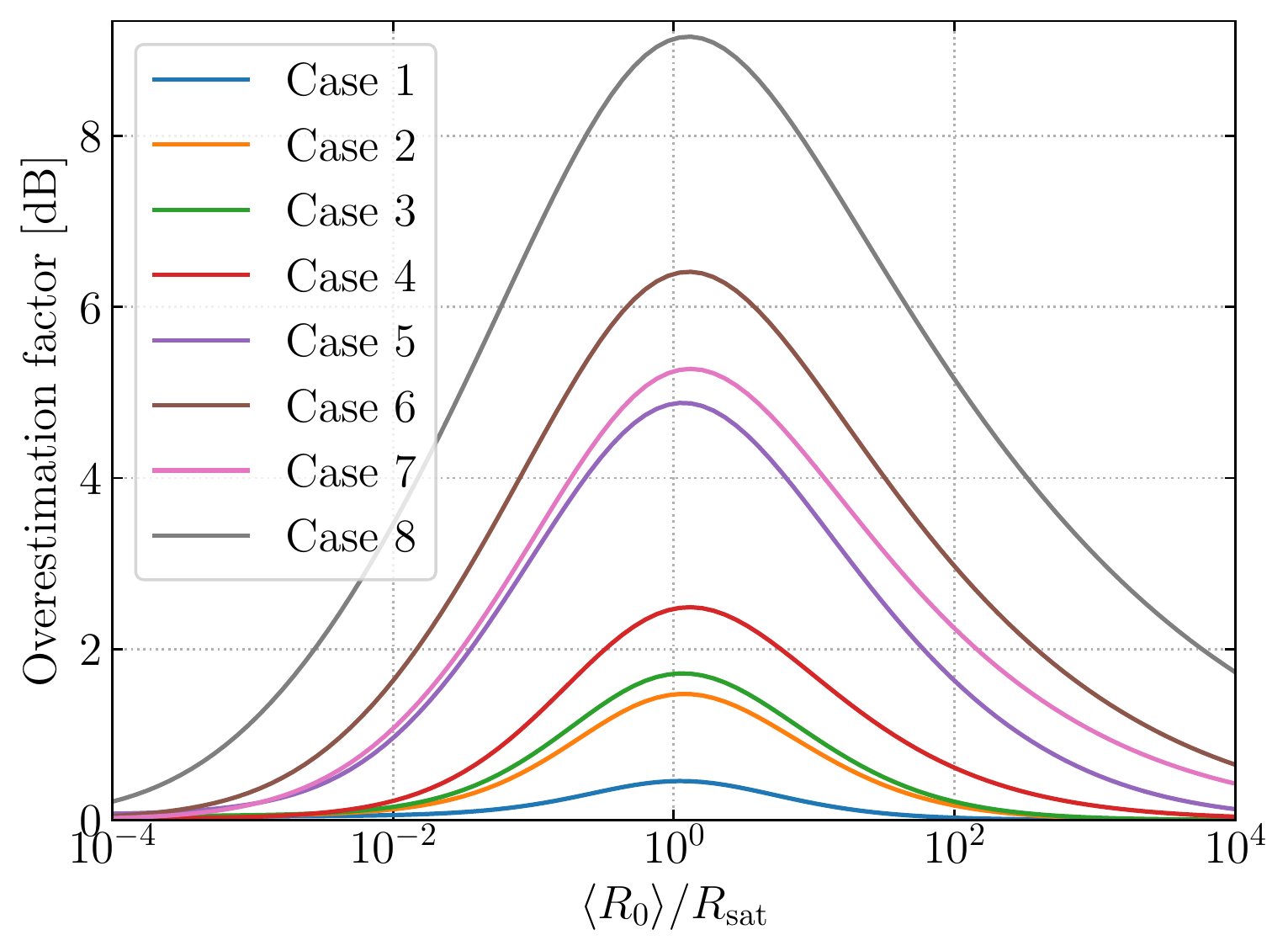}
    \caption{Overestimation caused by neglecting fluctuations for the different distributions of Fig.~\ref{fig:joint_distributions}.
    The effect is larger when $\langle R_0\rangle\approx R_{\rm sat}$ and for distributions of high kurtosis.}
    \label{fig:multiplot}
\end{figure}

To quantify the importance of these fluctuations, we estimate the raw key rate of a QKD system for the eight cases of Fig.~\ref{fig:joint_distributions}, first considering the entire distributions and then only their mean values.
In Fig.~\ref{fig:multiplot} we show the overestimation factor (converted to dB to visualize it better) caused by neglecting fluctuations.
This can reach a value of almost $9$ dB for the distribution of case $8$.
The effect is larger when $\langle R_0\rangle\approx R_{\rm sat}$ and vanishes for $\langle R_0\rangle\gg R_{\rm sat}$ or $\langle R_0\rangle\ll R_{\rm sat}$, when Eq.~\eqref{eq:Saturation} is well approximated by its linearization.
The distributions for which this error is greater are those which have stronger tails (i.e., a high kurtosis).
Indeed, neglecting fluctuations means neglecting the suppression of the tails caused by the saturation of the detectors: the greater the tails are, the graver is the error caused by neglecting them. 

\begin{figure}
    \centering
    \includegraphics[width=\columnwidth]{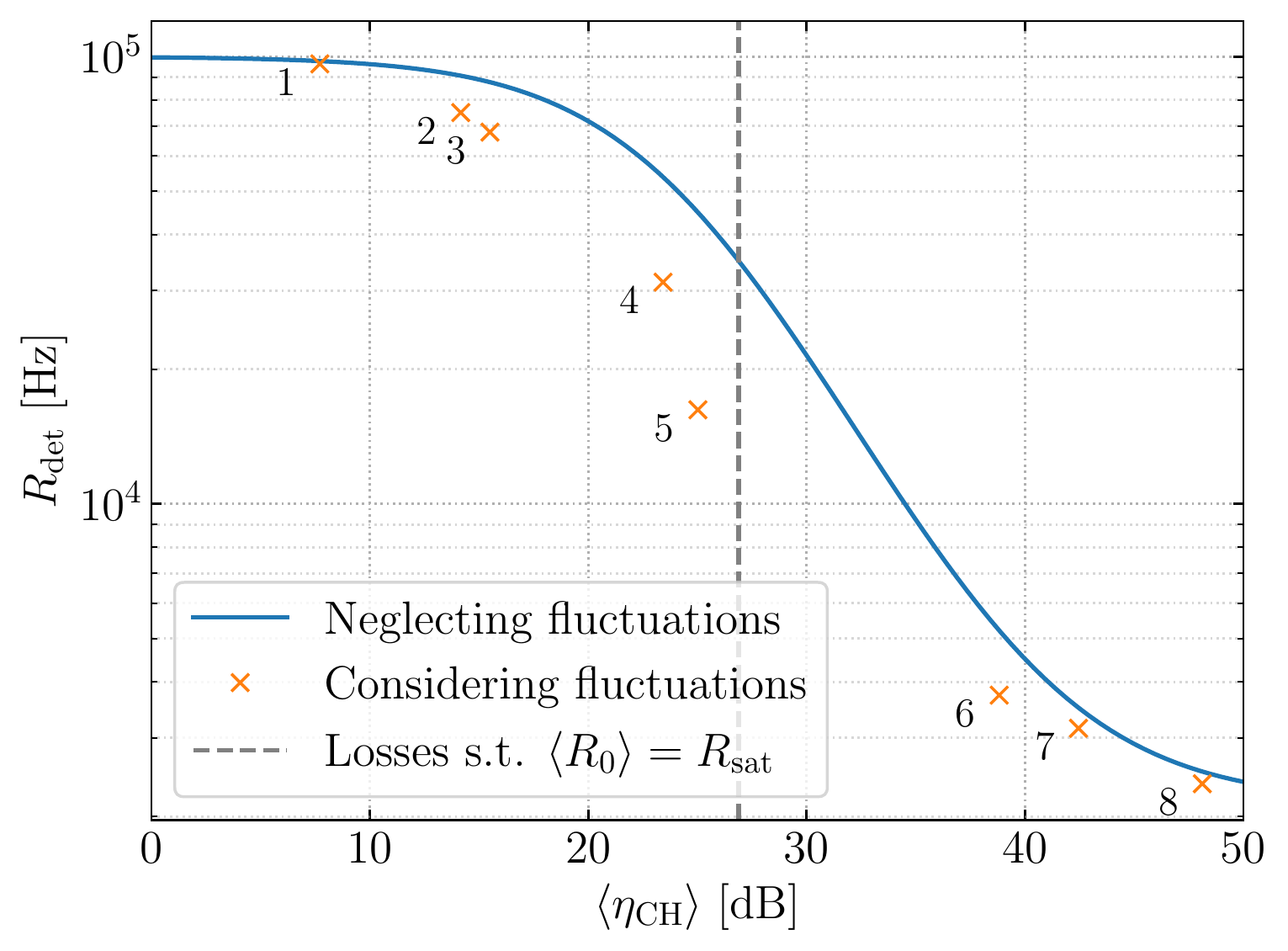}
    \caption{Comparison between a simulation of the raw key rate that neglects fluctuations and the ones which consider them for the eight distributions of Fig.~\ref{fig:joint_distributions} (to which the numbers refer).}
    \label{fig:trend}
\end{figure}

In Fig.~\ref{fig:trend}, we consider an arbitrary QKD scenario which features a 1 GHz source, $T_d= 10~\mu$s, $15\%$ detection efficiency.
We show a simulation of the raw key rate (as a function of the channel efficiency $\eta_{\rm CH}$) which neglects fluctuations, and compare it with the more correct values which consider this effect.
We can see a clear separation between the two methods of estimation, which grows larger when $\langle R_0\rangle\approx R_{\rm sat}$ and for distributions of greater kurtosis.
This shows that performance predictions that consider only the mean value of the channel efficiency can be severely inaccurate.

\section{Contributions to the error rate}
\label{sec:Errors}

Mismatches between Alice and Bob's raw keys influence the performance of QKD in two ways.
First, the more the errors, the more bits must be published to correct them.
Second, they indicate the amount information leaked to an attacker.
In the security scenario in which QKD operates, all errors are attributed to attacks and reduce the length of the final secret key.
Therefore, when simulating a QKD system, several physical sources of error must be considered.

One is intrinsic to the signal: inaccurate quantum state preparation or measurement might cause a mismatch between Alice's encoded bit and Bob's decoded one, even if the carrier photon arrives at the detector.
We quantify this with the coding error, which we define as the conditional probability of a mismatch given that a signal photon is detected.
In principle, the channel can also increase it if it can change the state of the photons, but this does not happen in typical stationary free space systems with polarization encoding, because the medium in which light travels is not birefringent.
The only way to reduce the coding error is to build better quantum state encoders and decoders, and better systems to align them to each other~\cite{Agnesi2019,Avesani2020}.

Then, there is random noise.
A portion of it is caused by single-photon detectors, in the form of dark counts and afterpulses.
The former are random events that happen even in the total absence of light, whereas the latter are spurious signals caused by true ones and are typical of avalanche diodes.
Another portion is introduced by the channel background light, especially in the free-space case that we are studying.

In Sec.~\ref{ss:Background} we quantify this background light and in Sec.~\ref{ss:TemporalFiltering} we study a way to counter noise with temporal filtering.

\subsection{Diffuse atmospheric background}
\label{ss:Background}

\begin{figure}[b!]
\centering
\includegraphics[width=\columnwidth]{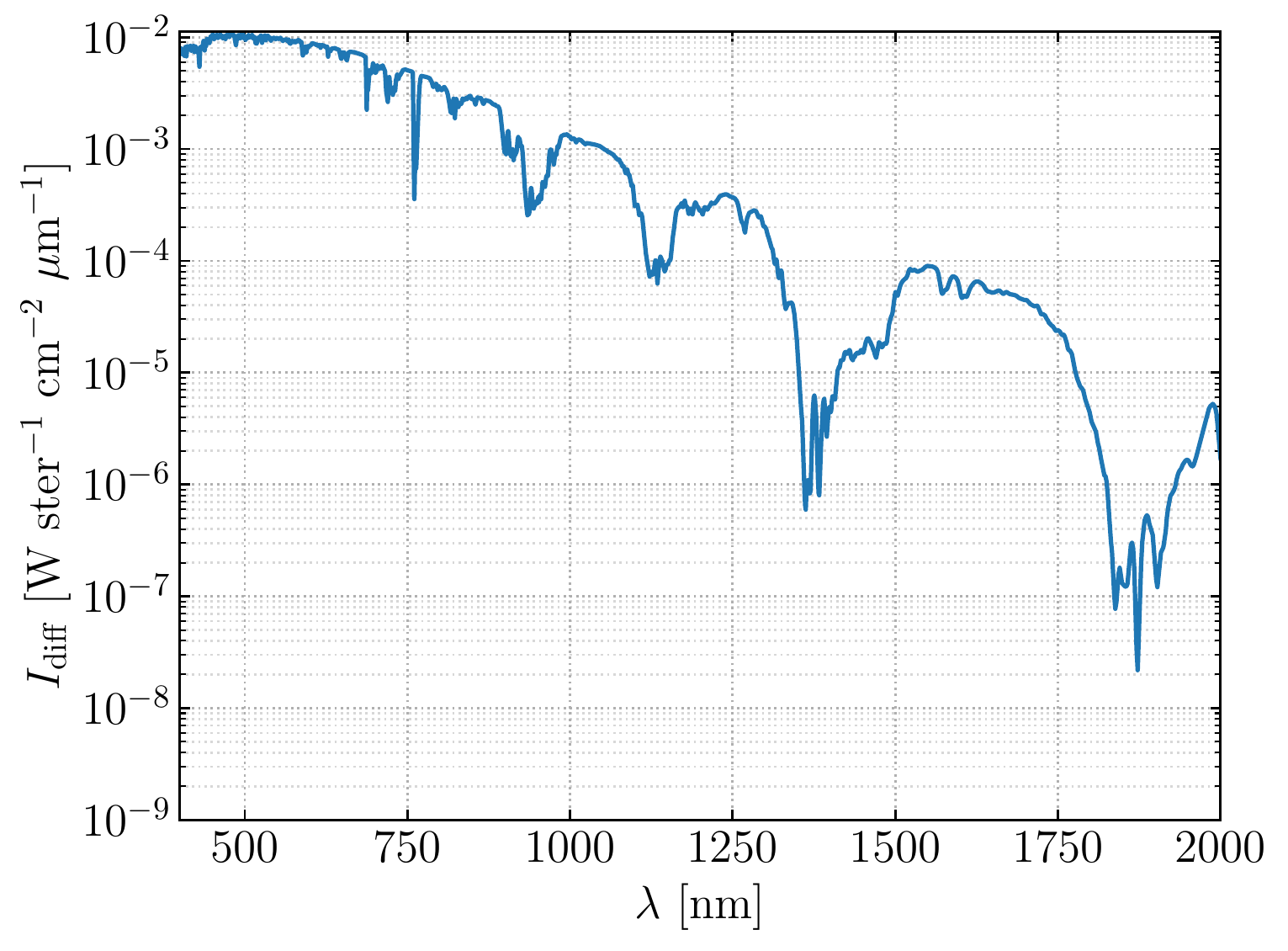}
\caption{Diffuse atmospheric radiance spectrum, horizontal path.}
\label{fig:scatterradiance_lowtran}
\end{figure}

A crucial requirement for the realization of daylight free-space QKD is the successful filtering of the background radiation.
Fig.~\ref{fig:scatterradiance_lowtran} shows the spectrum of the diffuse atmospheric radiance $I_{\rm diff}$, extracted with LOWTRAN.
As we can see, the spectrum peaks at blue wavelengths, and decreases for wavelengths in the infra-red.

This datum considers only the sky brightness and neglects the fact that the transmitter partially blocks the field-of-view (FOV) of the receiver, therefore it is overestimated in realistic scenarios, especially for short links.
Nonetheless, we will use it in the following discussion, which focuses on the impact of background on the performance of QKD.

As a first-order approximation, we can consider the diffuse radiance to be uniform over the receiver FOV.
We can estimate the detection rate of background photons per detection window $\expval{r_{\rm sky}}$ at the quantum signal wavelength as a function of the receiver aperture $D_{\rm Rx}$, solid-angle field of view $\Omega$, and filtering bandwidth $\delta\lambda$ ($h$ is the Planck's constant)
\begin{equation}
\expval{r_{\rm sky}} = \frac{I_{\rm diff}\cdot \pi\left(\frac{D_{\rm Rx}}{2}\right)^2 \cdot \Omega \cdot \delta\lambda }{ hc/\lambda} ~.
\label{eq:rskyGeneral}
\end{equation}
For the typically small field of view characteristic of free-space communication systems, $\Omega$ can be approximated by
\begin{equation}
    \Omega=2\pi\left(1-\cos({\rm FOV})\right)\approx \pi {\rm FOV}^2 ~,
\end{equation}
where ${\rm FOV}$ is the one-dimensional field-of-view.

\begin{figure}
    \centering
    \includegraphics[width=\columnwidth]{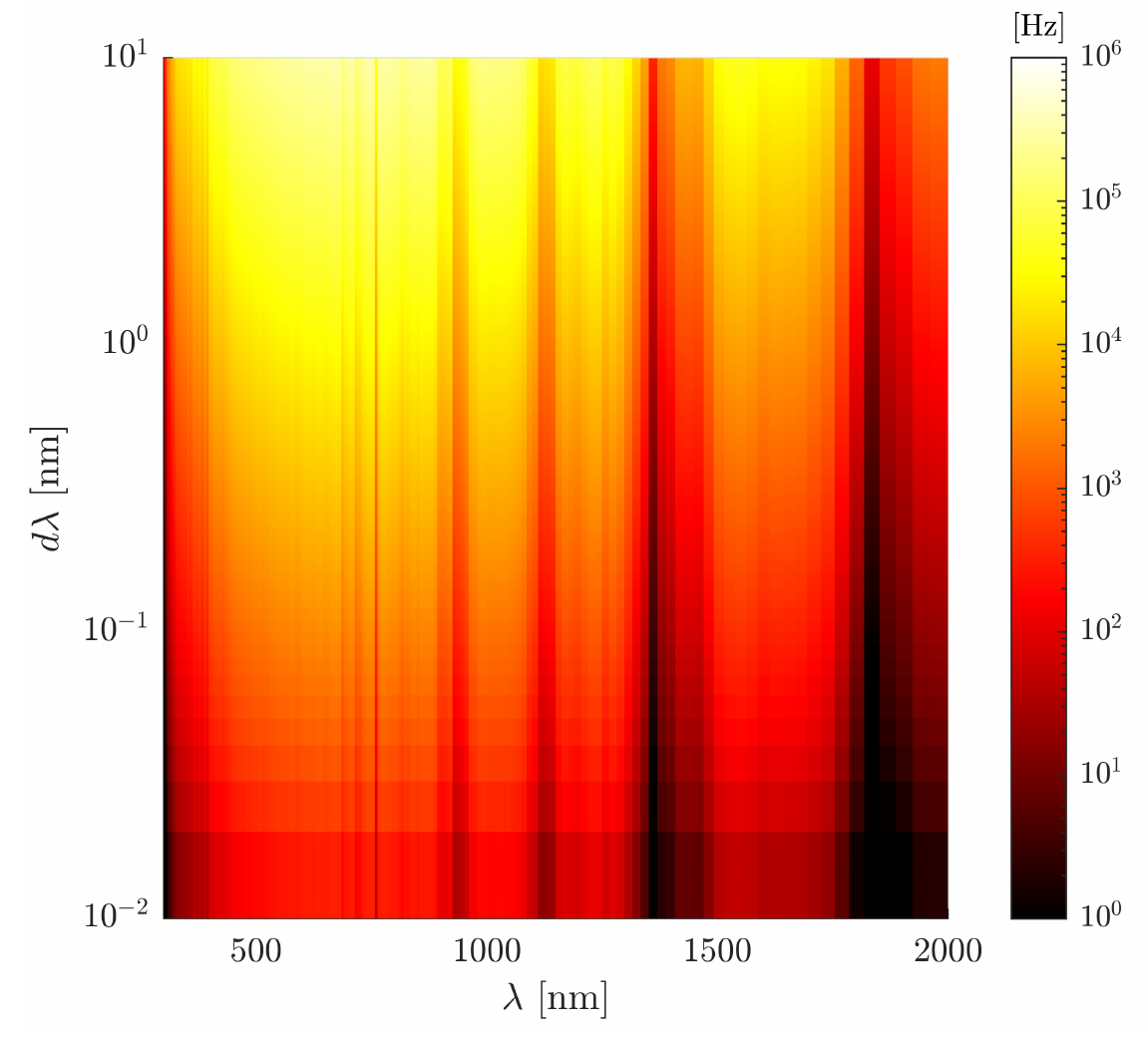}
    \caption{Noise count rate due to the diffuse atmospheric background as a function of qubit wavelength and linewidth of the spectral filter, for the SMF receiver case. Note that the dark bands in the figure also correspond to absorption windows of the atmosphere, see Fig.~\ref{fig:lowtran_absorption_coefficient}.}
    \label{fig:noise_count_smf}
\end{figure}

The FOV of the receiver optical system depends on the optical design:
if the optical fiber, or free-space detector, is placed on an image-plane of the entrance pupil, which is the case for free-space detectors with large active area ($\sim 150~\mu m$) or large-core multi-mode fibers (MMF), the FOV is essentially a free parameter, limited only by the size of the optical elements (lenses, mirrors) used in the optical system, and may be as large as $400~\mu\rm rad$.
In the case of single-mode fibers or free-space detectors with small active area $<10~\mu$m, the optimal choice is to place them on the focal plane of the optical system. In this configuration the field of view is constrained by the size of the active area or fiber mode field diameter (MFD). Since the receiver focal length $f$ is chosen so that the optical system efficiency in Eq.~\eqref{eqn:eta0} is maximized, we have
\begin{equation}
    {\rm MFD} = \beta_{\rm opt}\frac{2}{\pi} \lambda \frac{f}{D_{\rm Rx}} ~,
\end{equation}
where $\beta_{\rm opt}$ is shown in Fig.~\ref{fig:eta0_beta_alpha} and $\beta_{\rm opt}=1.12$ for unobstructed apertures.
If we define the FOV as the pupil incidence angle at which the spot on the focal plane is deflected to a distance equal to half the MFD, then we have that:
\begin{equation}
{\rm FOV} = \frac{\rm MFD}{2 f}  =\frac{\beta_{\rm opt}}{\pi} \frac{\lambda}{D_{\rm Rx}} ~.
\label{eq:FOVSMF}
\end{equation}
Another consequence of this constraint is that the detection rate of diffuse background photons coupled into the system becomes almost independent of the receiver optical system parameters. Indeed, combining Eqs. \eqref{eq:rskyGeneral} and \eqref{eq:FOVSMF}, we find that
\begin{equation}
    \expval{r_{\rm sky}}_{\rm SMF} = \frac{\beta_{\rm opt}^2}{4}\frac{I_{\rm diff}}{hc} \lambda^3  \delta\lambda 
\end{equation}
does not depend on $f$ nor $D_{\rm Rx}$.
Fig.~\ref{fig:noise_count_smf} shows the expected noise count rate for a SMF coupled receiver.

\subsection{Temporal gating}
\label{ss:TemporalFiltering}

Typical DV-QKD systems apply a temporal filter to all detected events, with the purpose of reducing the impact of noise.
Indeed, the latter is uniformly distributed in time whereas signal photons, being emitted at regular intervals, have a predictable time of arrival.
In post-processing, one can apply a $T_{\rm gat}$-wide temporal window centered at this time and discard all events that fall outside of it, thus suppressing noise by a factor $T_{\rm gat}/\tau$, where $\tau$ is the repetition period of the source.
However, there is a tradeoff, because the temporal distribution of the signal events is enlarged by the optical pulse width, by the jitter of the source, of the detectors, and of the time-digitizing hardware.
A small value of $T_{\rm gat}$, while strongly reducing noise, might discard too much of the signal, negatively impacting the final SKR.
Assuming a normal distribution of standard deviation $J$ for the signal time of arrival, the filter reduces the signal detection rate by a factor $\erf\left(\frac{T_{\rm gat}}{J2\sqrt{2}}\right)$.

A numerical study of the tradeoff can guide the choice of $T_{\rm gat}$. The figure of merit to maximize is the final SKR, which includes the contribution of the detection and error rates.
We focus on the ratio $T_{\rm gat}/J$ between it and the standard deviation $J$ of the temporal distribution of the signal (including all the aforementioned jitter contributions).
We can expect the tradeoff to be influenced by (i) the signal-to-noise ratio (SNR) when the noise is gated to a window as wide as the signal ($\pm 3$ times the standard deviation $J$) and (ii) the coding error.

The SNR obtained before any gating is not sufficient to describe the situation, because it does not consider the width of the signal.
Intuitively, for the same ungated SNR, a temporally wider signal favors smaller values of $T_{\rm gat}/J$ to eliminate more noise.
Our definition of the SNR, by considering only the portion of noise that falls under the signal, effectively combines the ungated SNR and the width of the signal in a single parameter.
Other protocol parameters such as decoy intensity levels and probabilities also have an importance, but we focus only on the two quantities above for simplicity.
All the parameters of the model except the SNR and coding error are arbitrarily fixed to a realistic QKD scenario.

\begin{figure}
    \centering
    \includegraphics[width=\columnwidth]{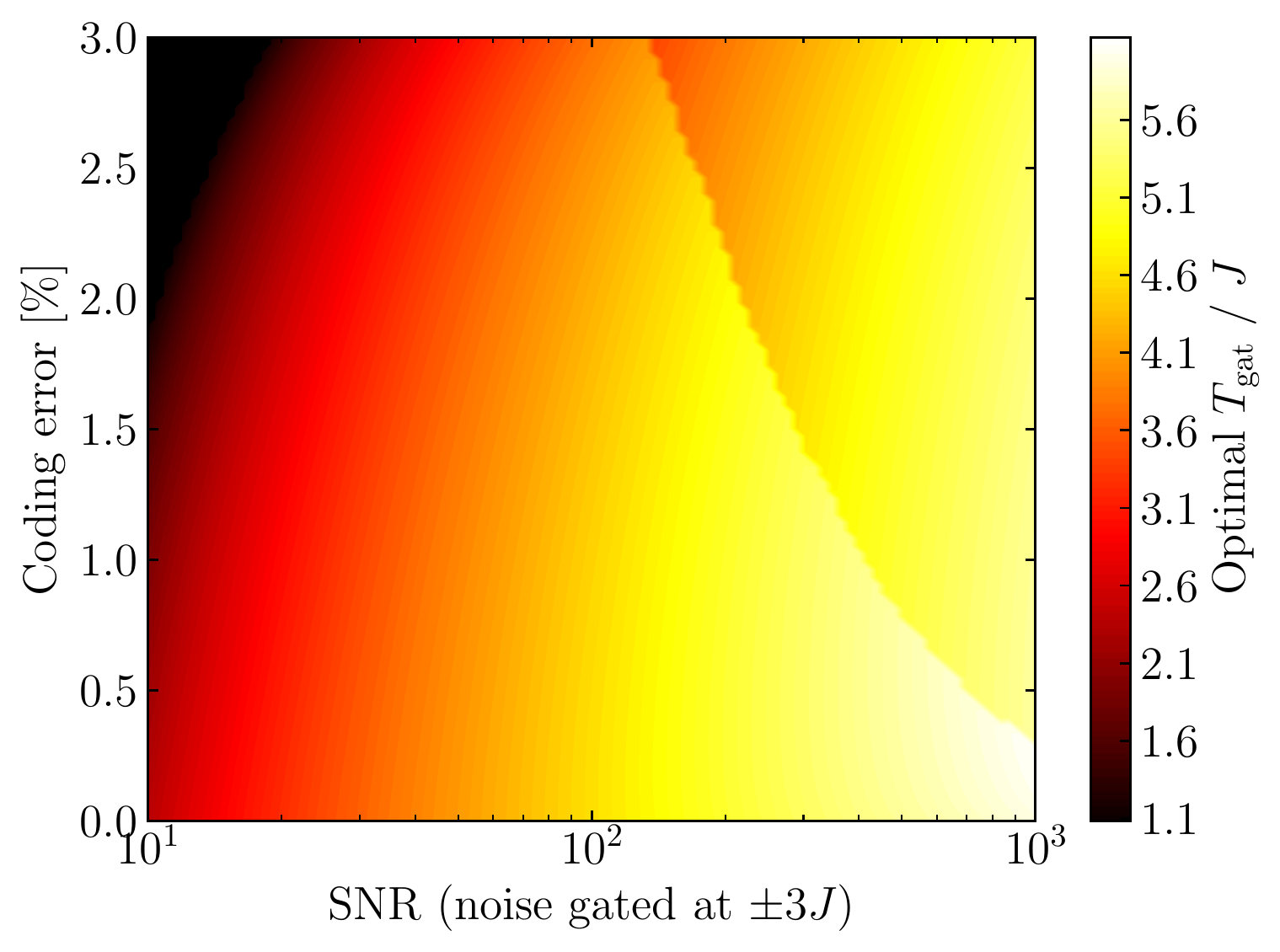}
    \caption{Optimization of $T_{\rm gat}/J$ to maximize the secret key rate, for a grid of values of the SNR (with noise gated to a $\pm 3J$-wide window) and the coding/decoding error. The quantitative details are influenced by the specific scenario used in the simulation, but the behavior of $T_{\rm gat}/J$ is quite general.}
    \label{fig:Gating}
\end{figure}

In Fig.~\ref{fig:Gating}, we can see the results of an optimization with the Nelder-Mead algorithm~\cite{Nelder1965} of the $T_{\rm gat}/J$ ratio to maximize the SKR, for a grid of values of the SNR (noise gated at $\pm 3J$) and coding error.
Predictably, the smaller the SNR, the smaller the gating window should be, in order to discard more noise.
High values of the coding/decoding error decrease the optimal $T_{\rm gat}/J$.
Indeed, although temporal gating alone cannot change the coding/decoding error, the higher the total QBER, the more important it is to reduce it, even at the expense of discarding part of the signal.
This shifts the balance of the tradeoff towards smaller windows, as can be seen in the top-left corner of the figure.

The small ridge in Fig.~\ref{fig:Gating} indicates a sudden jump in the optimal value of $T_{\rm gat}/J$.
This is because there are several terms contributing to the secret key rate (see Eq.~\eqref{eq:SKR}), and each is computed with several methods~\cite{Rusca2018}, choosing the best for every configuration of the parameters. 
This leads to the presence of multiple local maxima: when one of them is promoted to global maximum, overcoming another, the optimal $T_{\rm gat}/J$ changes.
The exact position and entity of the ridge is not universal and depends on the specific scenario that we chose in this simulation, but its presence is to be expected in general when the SKR is optimized.

\section{Estimation of the SKR}
\label{sec:Secret}

The rate of production of the secret key in a QKD experiment is calculated from the detection and error rates through a security analysis which bounds the amount of information leaked to an adversary Eve.
For our choice of protocol, efficient BB84 with decoy states, we follow Ref.~\cite{Rusca2018} and find
\begin{equation}
 \mathrm{SKR} = \frac{1}{t}\left(s_{Z,0} + s_{Z,1}(1 - h(\phi_Z)) - \ell_{\rm EC} -\ell_{\rm c} - \ell_{\rm sec}\right) .
 \label{eq:SKR}
\end{equation}
Here, $s_{Z,0}$ and $s_{Z,1}$ are the lower bounds on the number of vacuum and single-photon detections in the key-generating $Z$ basis, $\phi_{Z}$ is the upper bound on the phase error rate corresponding to single photon pulses, $h(\cdot)$ is the binary entropy, $\ell_{\rm EC}$ and $\ell_{\rm c}$ are the number of bits published during the error correction and confirmation of correctness steps, and $\ell_{\rm sec}=6 \log_2(\frac{19}{\epsilon_{\rm sec}})$, where $\epsilon_{\rm sec}=10^{-9}$ is the secrecy parameter associated to the key.
Finally, $t$ is the duration of the qubit transmission.

Terms $s_{Z,0}$, $s_{Z,1}$, and $\phi_{Z}$ are estimated from detections and errors observed in the experiment and depend on protocol parameters and statistical effects.
In what follows, we study how to optimize these parameters to maximize the SKR and finally estimate it in some exemplary scenarios.
 
\subsection{Finite-key effects}
\label{ss:Finite}

Because actual experiments accumulate only a finite amount of data, the parameters mentioned above cannot be estimated with perfect precision, leaving an opening for potential attackers.
To counter this, QKD uses a broad range of statistical analyses~\cite{Renner2005,Scarani2008,Tomamichel2012}.
Following Ref.~\cite{Rusca2018}, we construct confidence intervals based on the Hoeffding inequality~\cite{Hoeffding1963} around the parameters of interest, and then use the pessimistic extrema of the intervals as our estimates.
This penalizes the performance of the system, but guarantees that the key is secure with a very high probability $1-\epsilon_\mathrm{sec}$.

Fortunately, we can mitigate this cost by optimizing some parameters of the protocol.
These are: the probability $p_Z$ that each of Alice and Bob choose the key basis $Z$ for their preparations and measurements, the probability $p_\mu$ that Alice chooses the stronger intensity level, and the intensities $\mu, \nu$ of each level.
Their optimal values that maximize the SKR change depending on the amount of data (block length) that is used in the statistical analysis.
In our study, we consider two example QKD scenarios: a high-end system  with SNSPDs, a GHz source, and low coding error (scenario A), and a less expensive one with SPADs, a slower source, and a higher coding error.
The most relevant parameters of each scenario are listed in Table~\ref{tab:FiniteScenarios}.

\begin{table}
    \centering
    \begin{tabular}{c|cc}
        \toprule
        Quantity & Scenario A & Scenario B \\ 
        \midrule
        Source repetition rate & 1 GHz & 100 MHz \\
        Detector efficiency & 80 \% & 15 \% \\
        Coding Error & 0.5 \% & 1.5 \% \\
        Dark count rate & 10 Hz & 2 kHz \\
        Dead time & 10 ns & 20 $\mu$s \\
        Afterpulse probability & 0 & 10 \% \\
        Temporal jitter & 10 ps & 200 ps \\
        Additional receiver losses & 3 dB & 3 dB \\
        Wavelength & 1550 nm & 1550 nm \\
        Background photons rate & 5 kHz & 5 kHz \\
        \bottomrule
    \end{tabular}
    \caption{Relevant parameters for the two scenarios considered in the studies of Secs. \ref{ss:Finite} and \ref{ss:SKRTypical}.
    For the former, $\langle \eta_{\rm CH}\rangle \approx -7$ dB, whereas for the latter, the distributions of Fig.~\ref{fig:joint_distributions} are used.}
\label{tab:FiniteScenarios}
\end{table}

\begin{figure}[b]
\centering
\subfloat[Scenario A.]{\includegraphics[width=0.495\textwidth]{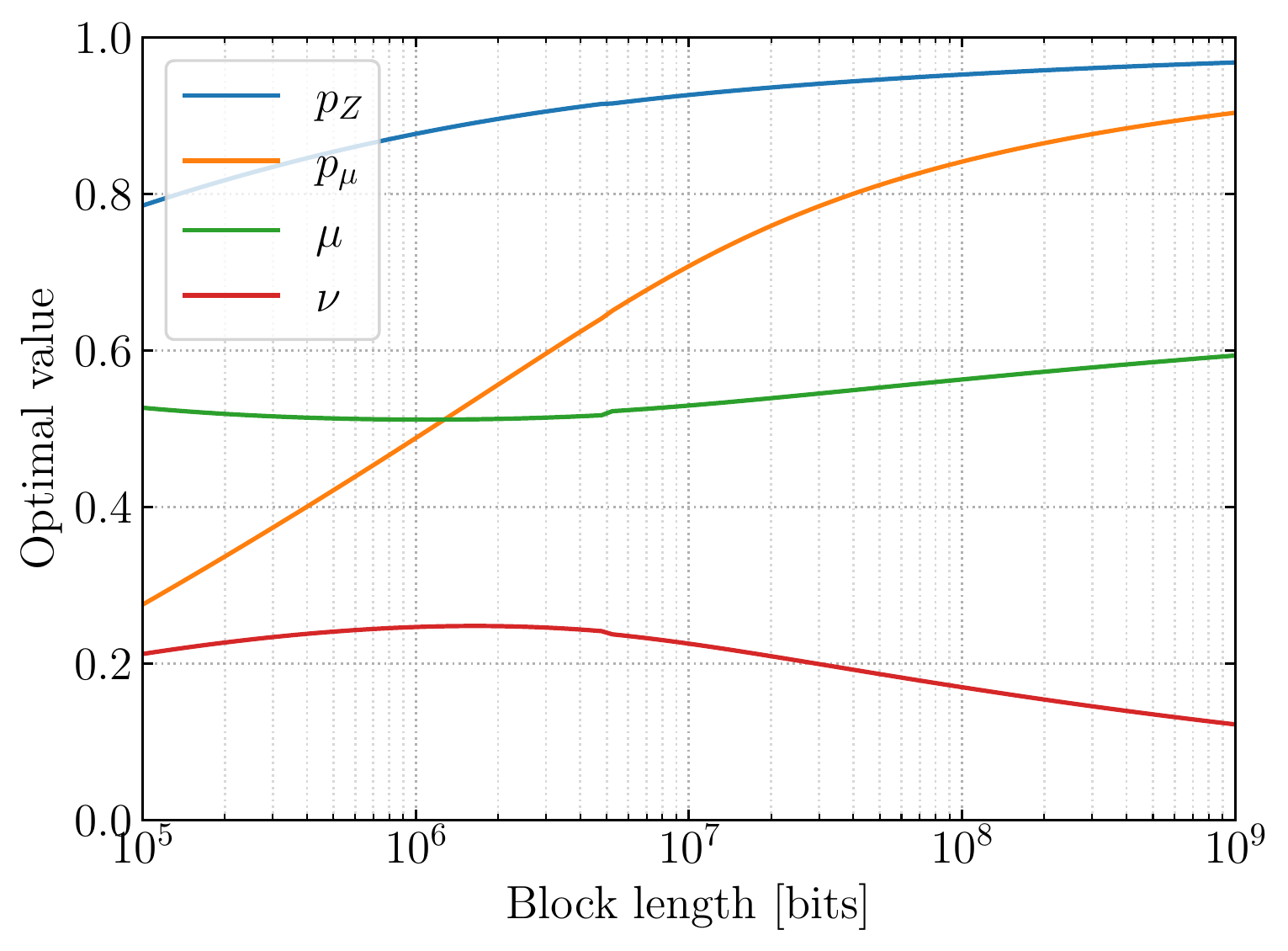}}\hfill
\subfloat[Scenario B.\label{fig:OptimalProtocolParamsB}]{\includegraphics[width=0.495\textwidth]{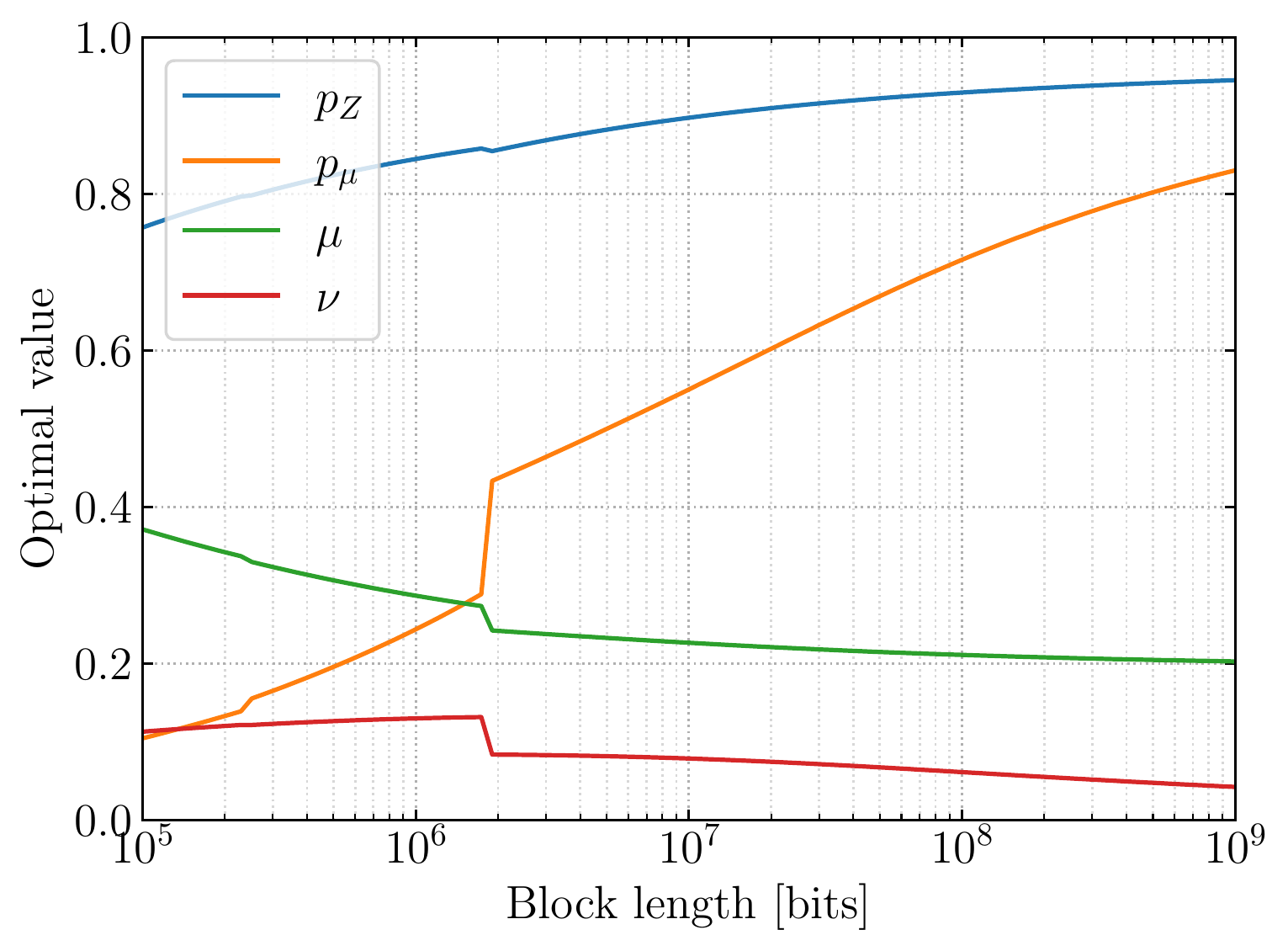}}
\caption{Optimization of the protocol parameters against the block length to maximize the secret key rate.}
\label{fig:OptimalProtocolParams}
\end{figure}

In Fig.~\ref{fig:OptimalProtocolParams} we show the results of the optimization (with a simulated annealing procedure~\cite{Xiang1997}) when the channel is characterized by the distribution of Case 1 (Fig.~\ref{fig:joint_distributions}), with $\langle \eta_{\rm CH}\rangle \approx -7$ dB.
We can see an upward trend for $p_Z$ and $p_\mu$ for growing block length, indeed the larger the total sample size, the easier it is to accumulate the needed statistics in the check-basis basis $X$ (mutually unbiased with respect to $Z$) and intensity level $\nu$ even if they are chosen rarely.
The mean photon numbers $\mu$ and $\nu$ show more stability, with a slight downward trend for $\nu$ in the right-hand side of the plot.
This is justified considering that its limit for infinite block length is zero, as this makes the bounds of the decoy-state method tighter.
For shorter block lengths, $\nu$ must grow to increase the detection rate and accumulate the needed statistics.
The optimal value of $\mu$ is the one that strikes the right balance between a low multi-photon emission probability and a high detection rate.
However, it is also influenced by other factors like afterpulses and detector saturation, which is why it is smaller in scenario B.
For low and decreasing block length, as $p_\mu$ shrinks to increase the statistics accumulated in the low intensity level, $\mu$ grows to keep the signal rate high, and $\nu$ responds by slightly decreasing to tighten the bounds.

The jump which can be seen in Fig.~\ref{fig:OptimalProtocolParamsB} is similar to the ridge of Fig.~\ref{fig:Gating}: a local maximum is promoted to global and the optimal parameters suddenly move.
The exact position and entity of the jump depends strongly on the scenario, but its presence is to be expected in optimizations like these.

\begin{figure}
\centering
\subfloat[Scenario A.]{\includegraphics[width=0.495\textwidth]{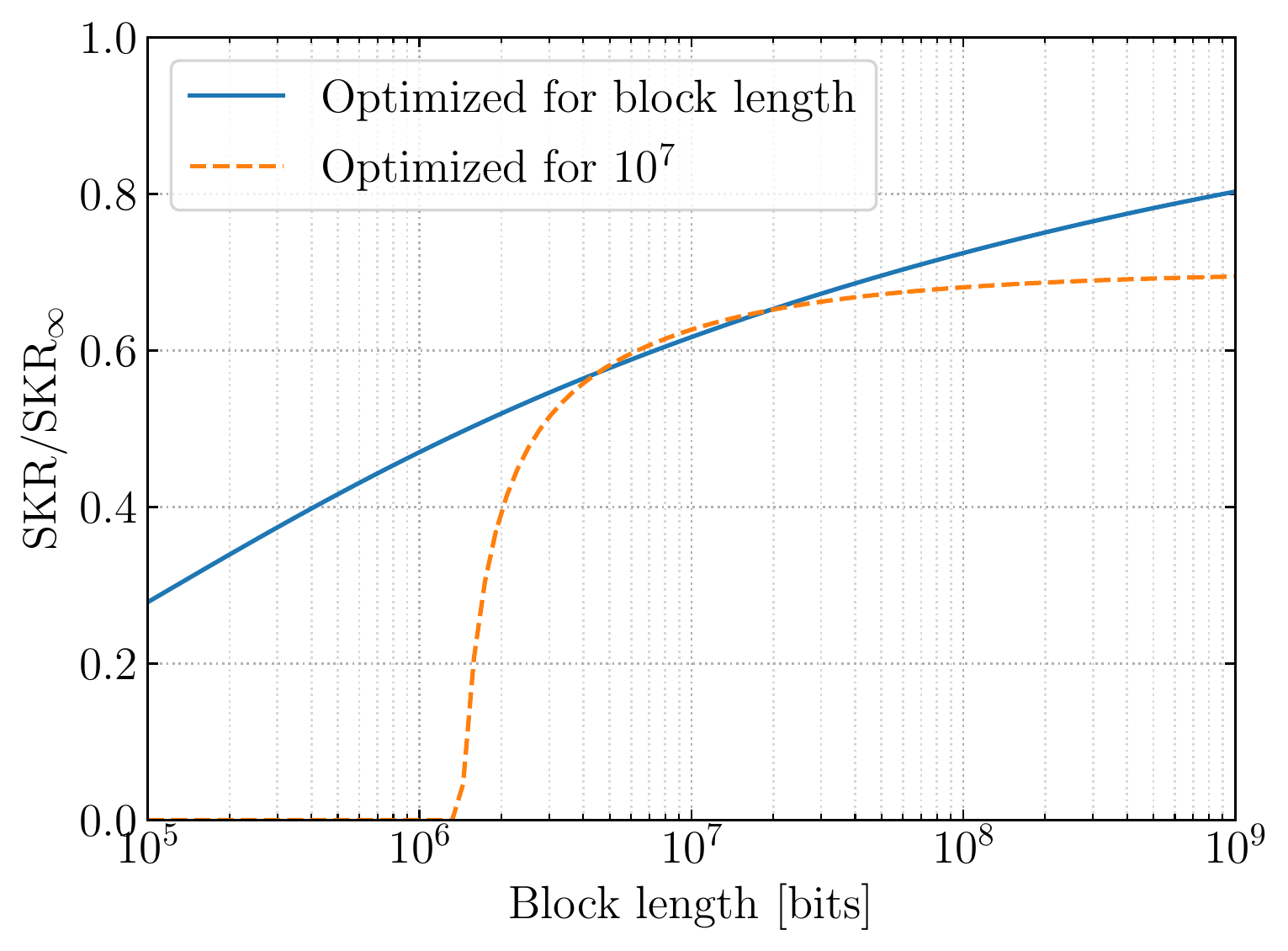}}\hfill
\subfloat[Scenario B.]{\includegraphics[width=0.495\textwidth]{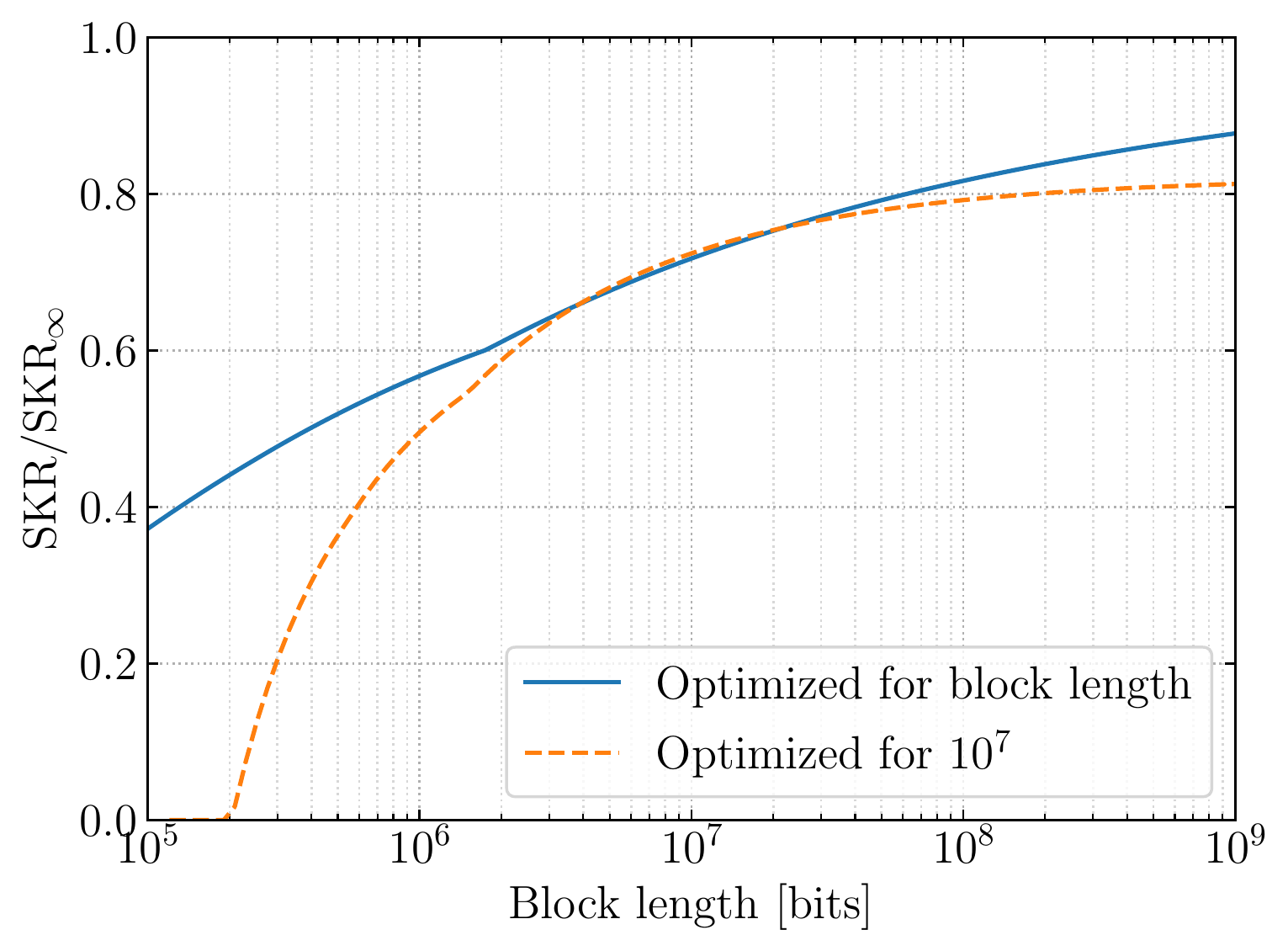}}
\caption{Cost of finite-key effects, represented by the ratio between the SKR and $\mathrm{SKR}_\infty$, which would be obtained by optimizing the protocol parameters for infinite length and by accumulating an infinitely long key block.}
\label{fig:CostFinite}
\end{figure}

In Fig.~\ref{fig:CostFinite} we show the cost of finite-key effects and of using a wrong set of parameters.
We express it as the ratio between the SKR and $\mathrm{SKR}_\infty$, i.e., the SKR that would be obtained by optimizing the above parameters for infinite length and by accumulating an infinitely long key block.
The dashed line corresponds to the choice of optimizing the parameters for a fixed block length of $10^7$ bits.
Generating a key is possible but the SKR quickly drops to zero if block lengths shorter than $10^7$ are used.
This happens more slowly in scenario B because the noisier detectors reduce the impact of tuning signal-related parameters.

The solid line shows what happens if the parameters are optimized for each block length.
The results improve drastically, underlining the importance of optimizing the protocol parameters for the predicted size that will be used operatively.
However, even with a large block length of $10^9$ bits, the SKR is reduced by a sizeable portion with respect to $\mathrm{SKR}_\infty$.
While this precise value depends on the chosen scenarios, the fact that finite-key effects should not be neglected even for large block lengths is general.

\subsection{SKR in typical scenarios}
\label{ss:SKRTypical}

\begin{figure}[b]
    \centering
    \includegraphics[width=\columnwidth]{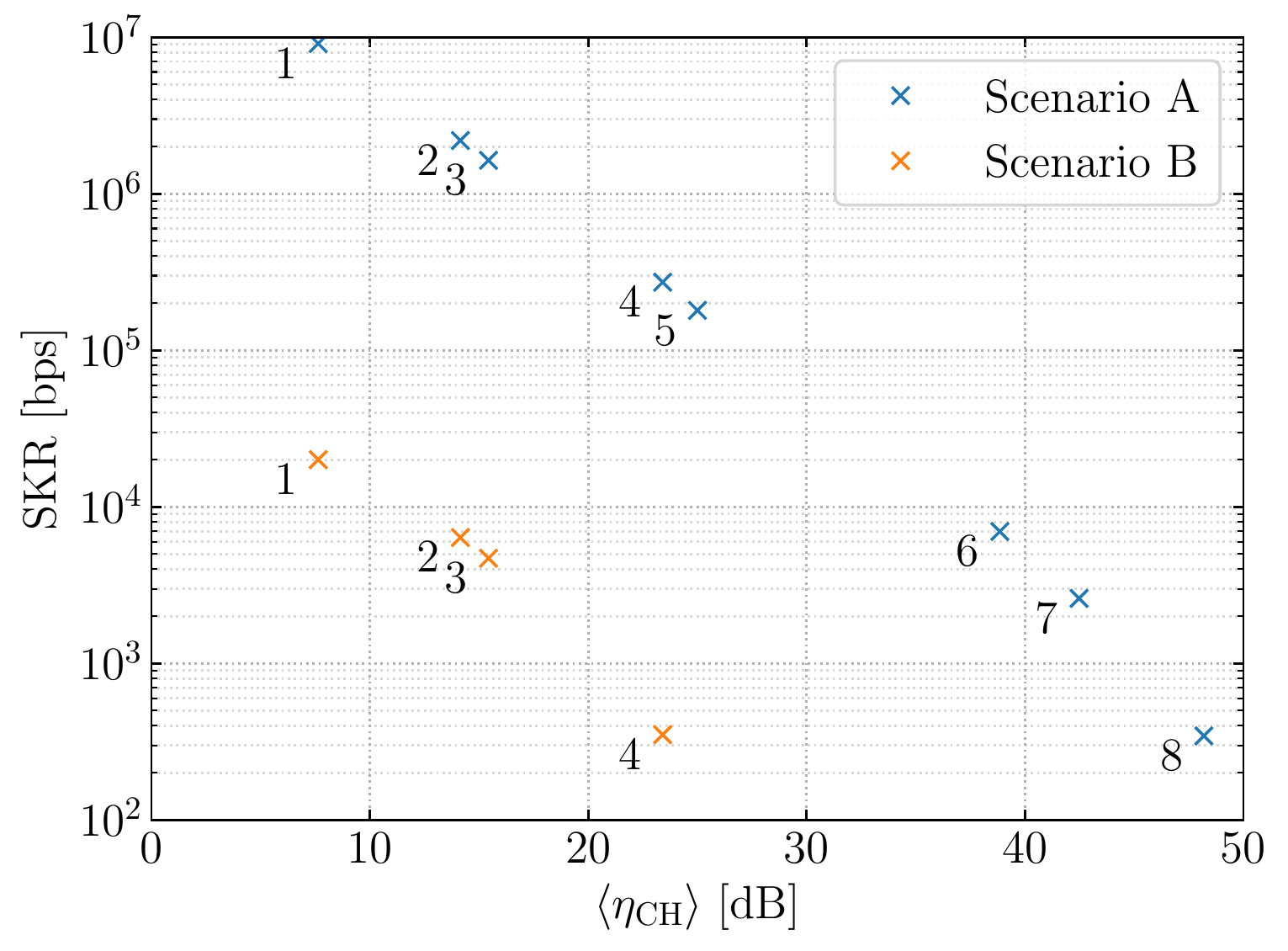}
    \caption{Secret key rate for the two scenarios and the eight cases of Fig.~\ref{fig:joint_distributions} (to which the numbers refer), considering all the effects studied in this work.}
    \label{fig:AllSKR}
\end{figure}

To conclude our analysis and give an example of the capabilities of our full model, we calculate the final SKR of a QKD system considering all the effects we have studied in this work.
We do this for the two scenarios of Table~\ref{tab:FiniteScenarios} and the eight channel efficiency distributions of Fig.~\ref{fig:joint_distributions}.
For each configuration, we use the simulated annealing algorithm to optimize the protocol parameters (the same of Sec. \ref{ss:Finite}) and the temporal gating, for a fixed block length of $10^7$ bits.

The results are shown in Fig.~\ref{fig:AllSKR}.
We can observe the characteristic linear behavior of the SKR with transmittance, and glimpse a drop for strong losses caused by the prevalence of noise.
However, thanks to the breadth of phenomena that our model includes, these values go beyond the simple verification of this typical trend, and are accurate estimates of the performance of the considered QKD systems.
Scenario B is strongly penalized by its slower source, lower detection efficiency, afterpulses and dead time.
Because of this, only the first four distributions yield a positive SKR.

\section{Conclusion and Outlook}
In this work, we studied many of the relevant phenomena that influence the performance of a ground-to-ground QKD BB84 system.
Particular focus was given to the channel model that estimates the efficiency of the link, considering atmospheric absorption, turbulence-induced beam broadening, wandering, scintillation, and the effect of single-mode fiber coupling.
We showed how adaptive optics can reduce losses and suggested ways to optimize the receiver diameter.
We found that calculating only the mean efficiency can sometimes be insufficient, and the entire probability distribution is needed.
This is because of the saturation of single-photon detectors, which may suppress the high tails of distribution, reducing the detection rate more than one would expect by considering only the mean.

We analyzed most of the sources of error in QKD and some mitigation techniques, showing how to find the best temporal gating to filter out noise.
We included also the finite-key effects that reduce the performance because of imperfect parameter estimation.
We highlighted how optimizing the probabilities of basis choice and the properties of the decoy states can alleviate this cost.
Finally, we put everything together to estimate the final secret key rate in some example scenarios.

Our model can be expanded further, for instance to include tracking imprecision in moving links and imperfect quantum state preparation beyond the coding error.
However, it is comprehensive enough to guide the design of QKD systems and underline what problems should be considered.
This can help the implementation and deployment of free-space daylight links in future QKD networks.

\begin{acknowledgments}
    This work was supported by Agenzia Spaziale Italiana, project {\it Q-SecGroundSpace} (Accordo n. 2018-14-HH.0, CUP: E16J16001490001). CloudVeneto is acknowledged for the computational resources.
\end{acknowledgments}
\vfill

\bibliographystyle{apsrev4-2}

\begin{thebibliography}{55}%
\makeatletter
\providecommand \@ifxundefined [1]{%
 \@ifx{#1\undefined}
}%
\providecommand \@ifnum [1]{%
 \ifnum #1\expandafter \@firstoftwo
 \else \expandafter \@secondoftwo
 \fi
}%
\providecommand \@ifx [1]{%
 \ifx #1\expandafter \@firstoftwo
 \else \expandafter \@secondoftwo
 \fi
}%
\providecommand \natexlab [1]{#1}%
\providecommand \enquote  [1]{``#1''}%
\providecommand \bibnamefont  [1]{#1}%
\providecommand \bibfnamefont [1]{#1}%
\providecommand \citenamefont [1]{#1}%
\providecommand \href@noop [0]{\@secondoftwo}%
\providecommand \href [0]{\begingroup \@sanitize@url \@href}%
\providecommand \@href[1]{\@@startlink{#1}\@@href}%
\providecommand \@@href[1]{\endgroup#1\@@endlink}%
\providecommand \@sanitize@url [0]{\catcode `\\12\catcode `\$12\catcode
  `\&12\catcode `\#12\catcode `\^12\catcode `\_12\catcode `\%12\relax}%
\providecommand \@@startlink[1]{}%
\providecommand \@@endlink[0]{}%
\providecommand \url  [0]{\begingroup\@sanitize@url \@url }%
\providecommand \@url [1]{\endgroup\@href {#1}{\urlprefix }}%
\providecommand \urlprefix  [0]{URL }%
\providecommand \Eprint [0]{\href }%
\providecommand \doibase [0]{https://doi.org/}%
\providecommand \selectlanguage [0]{\@gobble}%
\providecommand \bibinfo  [0]{\@secondoftwo}%
\providecommand \bibfield  [0]{\@secondoftwo}%
\providecommand \translation [1]{[#1]}%
\providecommand \BibitemOpen [0]{}%
\providecommand \bibitemStop [0]{}%
\providecommand \bibitemNoStop [0]{.\EOS\space}%
\providecommand \EOS [0]{\spacefactor3000\relax}%
\providecommand \BibitemShut  [1]{\csname bibitem#1\endcsname}%
\let\auto@bib@innerbib\@empty
%</preamble>
\bibitem [{\citenamefont {Gisin}\ \emph {et~al.}(2002)\citenamefont {Gisin},
  \citenamefont {Ribordy}, \citenamefont {Tittel},\ and\ \citenamefont
  {Zbinden}}]{GisinQKD}%
  \BibitemOpen
  \bibfield  {author} {\bibinfo {author} {\bibfnamefont {N.}~\bibnamefont
  {Gisin}}, \bibinfo {author} {\bibfnamefont {G.}~\bibnamefont {Ribordy}},
  \bibinfo {author} {\bibfnamefont {W.}~\bibnamefont {Tittel}},\ and\ \bibinfo
  {author} {\bibfnamefont {H.}~\bibnamefont {Zbinden}},\ }\href
  {https://doi.org/10.1103/RevModPhys.74.145} {\bibfield  {journal} {\bibinfo
  {journal} {Rev. Mod. Phys.}\ }\textbf {\bibinfo {volume} {74}},\ \bibinfo
  {pages} {145} (\bibinfo {year} {2002})}\BibitemShut {NoStop}%
\bibitem [{\citenamefont {Scarani}\ \emph {et~al.}(2009)\citenamefont
  {Scarani}, \citenamefont {Bechmann-Pasquinucci}, \citenamefont {Cerf},
  \citenamefont {Du\ifmmode~\check{s}\else \v{s}\fi{}ek}, \citenamefont
  {L\"utkenhaus},\ and\ \citenamefont {Peev}}]{ScaraniSecurityQKD}%
  \BibitemOpen
  \bibfield  {author} {\bibinfo {author} {\bibfnamefont {V.}~\bibnamefont
  {Scarani}}, \bibinfo {author} {\bibfnamefont {H.}~\bibnamefont
  {Bechmann-Pasquinucci}}, \bibinfo {author} {\bibfnamefont {N.~J.}\
  \bibnamefont {Cerf}}, \bibinfo {author} {\bibfnamefont {M.}~\bibnamefont
  {Du\ifmmode~\check{s}\else \v{s}\fi{}ek}}, \bibinfo {author} {\bibfnamefont
  {N.}~\bibnamefont {L\"utkenhaus}},\ and\ \bibinfo {author} {\bibfnamefont
  {M.}~\bibnamefont {Peev}},\ }\href
  {https://doi.org/10.1103/RevModPhys.81.1301} {\bibfield  {journal} {\bibinfo
  {journal} {Rev. Mod. Phys.}\ }\textbf {\bibinfo {volume} {81}},\ \bibinfo
  {pages} {1301} (\bibinfo {year} {2009})}\BibitemShut {NoStop}%
\bibitem [{\citenamefont {Diamanti}\ \emph {et~al.}(2016)\citenamefont
  {Diamanti}, \citenamefont {Lo}, \citenamefont {Qi},\ and\ \citenamefont
  {Yuan}}]{DiamantiQKD}%
  \BibitemOpen
  \bibfield  {author} {\bibinfo {author} {\bibfnamefont {E.}~\bibnamefont
  {Diamanti}}, \bibinfo {author} {\bibfnamefont {H.-K.}\ \bibnamefont {Lo}},
  \bibinfo {author} {\bibfnamefont {B.}~\bibnamefont {Qi}},\ and\ \bibinfo
  {author} {\bibfnamefont {Z.}~\bibnamefont {Yuan}},\ }\href
  {https://doi.org/10.1038/npjqi.2016.25} {\bibfield  {journal} {\bibinfo
  {journal} {npj Quantum Inf.}\ }\textbf {\bibinfo {volume} {2}},\ \bibinfo
  {pages} {16025} (\bibinfo {year} {2016})}\BibitemShut {NoStop}%
\bibitem [{\citenamefont {Pirandola}\ \emph {et~al.}(2020)\citenamefont
  {Pirandola}, \citenamefont {Andersen}, \citenamefont {Banchi}, \citenamefont
  {Berta}, \citenamefont {Bunandar}, \citenamefont {Colbeck}, \citenamefont
  {Englund}, \citenamefont {Gehring}, \citenamefont {Lupo}, \citenamefont
  {Ottaviani}, \citenamefont {Pereira}, \citenamefont {Razavi}, \citenamefont
  {Shaari}, \citenamefont {Tomamichel}, \citenamefont {Usenko}, \citenamefont
  {Vallone}, \citenamefont {Villoresi},\ and\ \citenamefont
  {Wallden}}]{pir2019advances}%
  \BibitemOpen
  \bibfield  {author} {\bibinfo {author} {\bibfnamefont {S.}~\bibnamefont
  {Pirandola}}, \bibinfo {author} {\bibfnamefont {U.~L.}\ \bibnamefont
  {Andersen}}, \bibinfo {author} {\bibfnamefont {L.}~\bibnamefont {Banchi}},
  \bibinfo {author} {\bibfnamefont {M.}~\bibnamefont {Berta}}, \bibinfo
  {author} {\bibfnamefont {D.}~\bibnamefont {Bunandar}}, \bibinfo {author}
  {\bibfnamefont {R.}~\bibnamefont {Colbeck}}, \bibinfo {author} {\bibfnamefont
  {D.}~\bibnamefont {Englund}}, \bibinfo {author} {\bibfnamefont
  {T.}~\bibnamefont {Gehring}}, \bibinfo {author} {\bibfnamefont
  {C.}~\bibnamefont {Lupo}}, \bibinfo {author} {\bibfnamefont {C.}~\bibnamefont
  {Ottaviani}}, \bibinfo {author} {\bibfnamefont {J.~L.}\ \bibnamefont
  {Pereira}}, \bibinfo {author} {\bibfnamefont {M.}~\bibnamefont {Razavi}},
  \bibinfo {author} {\bibfnamefont {J.~S.}\ \bibnamefont {Shaari}}, \bibinfo
  {author} {\bibfnamefont {M.}~\bibnamefont {Tomamichel}}, \bibinfo {author}
  {\bibfnamefont {V.~C.}\ \bibnamefont {Usenko}}, \bibinfo {author}
  {\bibfnamefont {G.}~\bibnamefont {Vallone}}, \bibinfo {author} {\bibfnamefont
  {P.}~\bibnamefont {Villoresi}},\ and\ \bibinfo {author} {\bibfnamefont
  {P.}~\bibnamefont {Wallden}},\ }\href {https://doi.org/10.1364/AOP.361502}
  {\bibfield  {journal} {\bibinfo  {journal} {Adv. Opt. Photon.}\ }\textbf
  {\bibinfo {volume} {12}},\ \bibinfo {pages} {1012} (\bibinfo {year}
  {2020})}\BibitemShut {NoStop}%
\bibitem [{\citenamefont {Peev}\ \emph {et~al.}(2009)\citenamefont {Peev},
  \citenamefont {Pacher}, \citenamefont {All{\'{e}}aume}, \citenamefont
  {Barreiro}, \citenamefont {Bouda}, \citenamefont {Boxleitner}, \citenamefont
  {Debuisschert}, \citenamefont {Diamanti}, \citenamefont {Dianati},
  \citenamefont {Dynes}, \citenamefont {Fasel}, \citenamefont {Fossier},
  \citenamefont {F\"{u}rst}, \citenamefont {Gautier}, \citenamefont {Gay},
  \citenamefont {Gisin}, \citenamefont {Grangier}, \citenamefont {Happe},
  \citenamefont {Hasani}, \citenamefont {Hentschel}, \citenamefont {H\"{u}bel},
  \citenamefont {Humer}, \citenamefont {L\"{a}nger}, \citenamefont
  {Legr{\'{e}}}, \citenamefont {Lieger}, \citenamefont {Lodewyck},
  \citenamefont {Lor\"{u}nser}, \citenamefont {L\"{u}tkenhaus}, \citenamefont
  {Marhold}, \citenamefont {Matyus}, \citenamefont {Maurhart}, \citenamefont
  {Monat}, \citenamefont {Nauerth}, \citenamefont {Page}, \citenamefont
  {Poppe}, \citenamefont {Querasser}, \citenamefont {Ribordy}, \citenamefont
  {Robyr}, \citenamefont {Salvail}, \citenamefont {Sharpe}, \citenamefont
  {Shields}, \citenamefont {Stucki}, \citenamefont {Suda}, \citenamefont
  {Tamas}, \citenamefont {Themel}, \citenamefont {Thew}, \citenamefont {Thoma},
  \citenamefont {Treiber}, \citenamefont {Trinkler}, \citenamefont
  {Tualle-Brouri}, \citenamefont {Vannel}, \citenamefont {Walenta},
  \citenamefont {Weier}, \citenamefont {Weinfurter}, \citenamefont {Wimberger},
  \citenamefont {Yuan}, \citenamefont {Zbinden},\ and\ \citenamefont
  {Zeilinger}}]{Peev2009}%
  \BibitemOpen
  \bibfield  {author} {\bibinfo {author} {\bibfnamefont {M.}~\bibnamefont
  {Peev}}, \bibinfo {author} {\bibfnamefont {C.}~\bibnamefont {Pacher}},
  \bibinfo {author} {\bibfnamefont {R.}~\bibnamefont {All{\'{e}}aume}},
  \bibinfo {author} {\bibfnamefont {C.}~\bibnamefont {Barreiro}}, \bibinfo
  {author} {\bibfnamefont {J.}~\bibnamefont {Bouda}}, \bibinfo {author}
  {\bibfnamefont {W.}~\bibnamefont {Boxleitner}}, \bibinfo {author}
  {\bibfnamefont {T.}~\bibnamefont {Debuisschert}}, \bibinfo {author}
  {\bibfnamefont {E.}~\bibnamefont {Diamanti}}, \bibinfo {author}
  {\bibfnamefont {M.}~\bibnamefont {Dianati}}, \bibinfo {author} {\bibfnamefont
  {J.~F.}\ \bibnamefont {Dynes}}, \bibinfo {author} {\bibfnamefont
  {S.}~\bibnamefont {Fasel}}, \bibinfo {author} {\bibfnamefont
  {S.}~\bibnamefont {Fossier}}, \bibinfo {author} {\bibfnamefont
  {M.}~\bibnamefont {F\"{u}rst}}, \bibinfo {author} {\bibfnamefont {J.-D.}\
  \bibnamefont {Gautier}}, \bibinfo {author} {\bibfnamefont {O.}~\bibnamefont
  {Gay}}, \bibinfo {author} {\bibfnamefont {N.}~\bibnamefont {Gisin}}, \bibinfo
  {author} {\bibfnamefont {P.}~\bibnamefont {Grangier}}, \bibinfo {author}
  {\bibfnamefont {A.}~\bibnamefont {Happe}}, \bibinfo {author} {\bibfnamefont
  {Y.}~\bibnamefont {Hasani}}, \bibinfo {author} {\bibfnamefont
  {M.}~\bibnamefont {Hentschel}}, \bibinfo {author} {\bibfnamefont
  {H.}~\bibnamefont {H\"{u}bel}}, \bibinfo {author} {\bibfnamefont
  {G.}~\bibnamefont {Humer}}, \bibinfo {author} {\bibfnamefont
  {T.}~\bibnamefont {L\"{a}nger}}, \bibinfo {author} {\bibfnamefont
  {M.}~\bibnamefont {Legr{\'{e}}}}, \bibinfo {author} {\bibfnamefont
  {R.}~\bibnamefont {Lieger}}, \bibinfo {author} {\bibfnamefont
  {J.}~\bibnamefont {Lodewyck}}, \bibinfo {author} {\bibfnamefont
  {T.}~\bibnamefont {Lor\"{u}nser}}, \bibinfo {author} {\bibfnamefont
  {N.}~\bibnamefont {L\"{u}tkenhaus}}, \bibinfo {author} {\bibfnamefont
  {A.}~\bibnamefont {Marhold}}, \bibinfo {author} {\bibfnamefont
  {T.}~\bibnamefont {Matyus}}, \bibinfo {author} {\bibfnamefont
  {O.}~\bibnamefont {Maurhart}}, \bibinfo {author} {\bibfnamefont
  {L.}~\bibnamefont {Monat}}, \bibinfo {author} {\bibfnamefont
  {S.}~\bibnamefont {Nauerth}}, \bibinfo {author} {\bibfnamefont {J.-B.}\
  \bibnamefont {Page}}, \bibinfo {author} {\bibfnamefont {A.}~\bibnamefont
  {Poppe}}, \bibinfo {author} {\bibfnamefont {E.}~\bibnamefont {Querasser}},
  \bibinfo {author} {\bibfnamefont {G.}~\bibnamefont {Ribordy}}, \bibinfo
  {author} {\bibfnamefont {S.}~\bibnamefont {Robyr}}, \bibinfo {author}
  {\bibfnamefont {L.}~\bibnamefont {Salvail}}, \bibinfo {author} {\bibfnamefont
  {A.~W.}\ \bibnamefont {Sharpe}}, \bibinfo {author} {\bibfnamefont {A.~J.}\
  \bibnamefont {Shields}}, \bibinfo {author} {\bibfnamefont {D.}~\bibnamefont
  {Stucki}}, \bibinfo {author} {\bibfnamefont {M.}~\bibnamefont {Suda}},
  \bibinfo {author} {\bibfnamefont {C.}~\bibnamefont {Tamas}}, \bibinfo
  {author} {\bibfnamefont {T.}~\bibnamefont {Themel}}, \bibinfo {author}
  {\bibfnamefont {R.~T.}\ \bibnamefont {Thew}}, \bibinfo {author}
  {\bibfnamefont {Y.}~\bibnamefont {Thoma}}, \bibinfo {author} {\bibfnamefont
  {A.}~\bibnamefont {Treiber}}, \bibinfo {author} {\bibfnamefont
  {P.}~\bibnamefont {Trinkler}}, \bibinfo {author} {\bibfnamefont
  {R.}~\bibnamefont {Tualle-Brouri}}, \bibinfo {author} {\bibfnamefont
  {F.}~\bibnamefont {Vannel}}, \bibinfo {author} {\bibfnamefont
  {N.}~\bibnamefont {Walenta}}, \bibinfo {author} {\bibfnamefont
  {H.}~\bibnamefont {Weier}}, \bibinfo {author} {\bibfnamefont
  {H.}~\bibnamefont {Weinfurter}}, \bibinfo {author} {\bibfnamefont
  {I.}~\bibnamefont {Wimberger}}, \bibinfo {author} {\bibfnamefont {Z.~L.}\
  \bibnamefont {Yuan}}, \bibinfo {author} {\bibfnamefont {H.}~\bibnamefont
  {Zbinden}},\ and\ \bibinfo {author} {\bibfnamefont {A.}~\bibnamefont
  {Zeilinger}},\ }\href {https://doi.org/10.1088/1367-2630/11/7/075001}
  {\bibfield  {journal} {\bibinfo  {journal} {New J. Phys.}\ }\textbf {\bibinfo
  {volume} {11}},\ \bibinfo {pages} {075001} (\bibinfo {year}
  {2009})}\BibitemShut {NoStop}%
\bibitem [{\citenamefont {Sasaki}\ \emph {et~al.}(2011)\citenamefont {Sasaki},
  \citenamefont {Fujiwara}, \citenamefont {Ishizuka}, \citenamefont {Klaus},
  \citenamefont {Wakui}, \citenamefont {Takeoka}, \citenamefont {Miki},
  \citenamefont {Yamashita}, \citenamefont {Wang}, \citenamefont {Tanaka},
  \citenamefont {Yoshino}, \citenamefont {Nambu}, \citenamefont {Takahashi},
  \citenamefont {Tajima}, \citenamefont {Tomita}, \citenamefont {Domeki},
  \citenamefont {Hasegawa}, \citenamefont {Sakai}, \citenamefont {Kobayashi},
  \citenamefont {Asai}, \citenamefont {Shimizu}, \citenamefont {Tokura},
  \citenamefont {Tsurumaru}, \citenamefont {Matsui}, \citenamefont {Honjo},
  \citenamefont {Tamaki}, \citenamefont {Takesue}, \citenamefont {Tokura},
  \citenamefont {Dynes}, \citenamefont {Dixon}, \citenamefont {Sharpe},
  \citenamefont {Yuan}, \citenamefont {Shields}, \citenamefont {Uchikoga},
  \citenamefont {Legr\'{e}}, \citenamefont {Robyr}, \citenamefont {Trinkler},
  \citenamefont {Monat}, \citenamefont {Page}, \citenamefont {Ribordy},
  \citenamefont {Poppe}, \citenamefont {Allacher}, \citenamefont {Maurhart},
  \citenamefont {L\"{a}nger}, \citenamefont {Peev},\ and\ \citenamefont
  {Zeilinger}}]{Sasaki2011}%
  \BibitemOpen
  \bibfield  {author} {\bibinfo {author} {\bibfnamefont {M.}~\bibnamefont
  {Sasaki}}, \bibinfo {author} {\bibfnamefont {M.}~\bibnamefont {Fujiwara}},
  \bibinfo {author} {\bibfnamefont {H.}~\bibnamefont {Ishizuka}}, \bibinfo
  {author} {\bibfnamefont {W.}~\bibnamefont {Klaus}}, \bibinfo {author}
  {\bibfnamefont {K.}~\bibnamefont {Wakui}}, \bibinfo {author} {\bibfnamefont
  {M.}~\bibnamefont {Takeoka}}, \bibinfo {author} {\bibfnamefont
  {S.}~\bibnamefont {Miki}}, \bibinfo {author} {\bibfnamefont {T.}~\bibnamefont
  {Yamashita}}, \bibinfo {author} {\bibfnamefont {Z.}~\bibnamefont {Wang}},
  \bibinfo {author} {\bibfnamefont {A.}~\bibnamefont {Tanaka}}, \bibinfo
  {author} {\bibfnamefont {K.}~\bibnamefont {Yoshino}}, \bibinfo {author}
  {\bibfnamefont {Y.}~\bibnamefont {Nambu}}, \bibinfo {author} {\bibfnamefont
  {S.}~\bibnamefont {Takahashi}}, \bibinfo {author} {\bibfnamefont
  {A.}~\bibnamefont {Tajima}}, \bibinfo {author} {\bibfnamefont
  {A.}~\bibnamefont {Tomita}}, \bibinfo {author} {\bibfnamefont
  {T.}~\bibnamefont {Domeki}}, \bibinfo {author} {\bibfnamefont
  {T.}~\bibnamefont {Hasegawa}}, \bibinfo {author} {\bibfnamefont
  {Y.}~\bibnamefont {Sakai}}, \bibinfo {author} {\bibfnamefont
  {H.}~\bibnamefont {Kobayashi}}, \bibinfo {author} {\bibfnamefont
  {T.}~\bibnamefont {Asai}}, \bibinfo {author} {\bibfnamefont {K.}~\bibnamefont
  {Shimizu}}, \bibinfo {author} {\bibfnamefont {T.}~\bibnamefont {Tokura}},
  \bibinfo {author} {\bibfnamefont {T.}~\bibnamefont {Tsurumaru}}, \bibinfo
  {author} {\bibfnamefont {M.}~\bibnamefont {Matsui}}, \bibinfo {author}
  {\bibfnamefont {T.}~\bibnamefont {Honjo}}, \bibinfo {author} {\bibfnamefont
  {K.}~\bibnamefont {Tamaki}}, \bibinfo {author} {\bibfnamefont
  {H.}~\bibnamefont {Takesue}}, \bibinfo {author} {\bibfnamefont
  {Y.}~\bibnamefont {Tokura}}, \bibinfo {author} {\bibfnamefont {J.~F.}\
  \bibnamefont {Dynes}}, \bibinfo {author} {\bibfnamefont {A.~R.}\ \bibnamefont
  {Dixon}}, \bibinfo {author} {\bibfnamefont {A.~W.}\ \bibnamefont {Sharpe}},
  \bibinfo {author} {\bibfnamefont {Z.~L.}\ \bibnamefont {Yuan}}, \bibinfo
  {author} {\bibfnamefont {A.~J.}\ \bibnamefont {Shields}}, \bibinfo {author}
  {\bibfnamefont {S.}~\bibnamefont {Uchikoga}}, \bibinfo {author}
  {\bibfnamefont {M.}~\bibnamefont {Legr\'{e}}}, \bibinfo {author}
  {\bibfnamefont {S.}~\bibnamefont {Robyr}}, \bibinfo {author} {\bibfnamefont
  {P.}~\bibnamefont {Trinkler}}, \bibinfo {author} {\bibfnamefont
  {L.}~\bibnamefont {Monat}}, \bibinfo {author} {\bibfnamefont {J.-B.}\
  \bibnamefont {Page}}, \bibinfo {author} {\bibfnamefont {G.}~\bibnamefont
  {Ribordy}}, \bibinfo {author} {\bibfnamefont {A.}~\bibnamefont {Poppe}},
  \bibinfo {author} {\bibfnamefont {A.}~\bibnamefont {Allacher}}, \bibinfo
  {author} {\bibfnamefont {O.}~\bibnamefont {Maurhart}}, \bibinfo {author}
  {\bibfnamefont {T.}~\bibnamefont {L\"{a}nger}}, \bibinfo {author}
  {\bibfnamefont {M.}~\bibnamefont {Peev}},\ and\ \bibinfo {author}
  {\bibfnamefont {A.}~\bibnamefont {Zeilinger}},\ }\href
  {https://doi.org/10.1364/OE.19.010387} {\bibfield  {journal} {\bibinfo
  {journal} {Opt. Express}\ }\textbf {\bibinfo {volume} {19}},\ \bibinfo
  {pages} {10387} (\bibinfo {year} {2011})}\BibitemShut {NoStop}%
\bibitem [{\citenamefont {Kimble}(2008)}]{Kimble08}%
  \BibitemOpen
  \bibfield  {author} {\bibinfo {author} {\bibfnamefont {H.~J.}\ \bibnamefont
  {Kimble}},\ }\href {https://doi.org/10.1038/nature07127} {\bibfield
  {journal} {\bibinfo  {journal} {Nature}\ }\textbf {\bibinfo {volume} {453}},\
  \bibinfo {pages} {1023} (\bibinfo {year} {2008})}\BibitemShut {NoStop}%
\bibitem [{\citenamefont {Pirandola}\ and\ \citenamefont
  {Braunstein}(2016)}]{Pirandolacomment}%
  \BibitemOpen
  \bibfield  {author} {\bibinfo {author} {\bibfnamefont {S.}~\bibnamefont
  {Pirandola}}\ and\ \bibinfo {author} {\bibfnamefont {S.~L.}\ \bibnamefont
  {Braunstein}},\ }\href {https://doi.org/10.1038/532169a} {\bibfield
  {journal} {\bibinfo  {journal} {Nature}\ }\textbf {\bibinfo {volume} {532}},\
  \bibinfo {pages} {169} (\bibinfo {year} {2016})}\BibitemShut {NoStop}%
\bibitem [{\citenamefont {Wehner}\ \emph {et~al.}(2018)\citenamefont {Wehner},
  \citenamefont {Elkouss},\ and\ \citenamefont {Hanson}}]{Wehnereaam9288}%
  \BibitemOpen
  \bibfield  {author} {\bibinfo {author} {\bibfnamefont {S.}~\bibnamefont
  {Wehner}}, \bibinfo {author} {\bibfnamefont {D.}~\bibnamefont {Elkouss}},\
  and\ \bibinfo {author} {\bibfnamefont {R.}~\bibnamefont {Hanson}},\ }\href
  {https://doi.org/10.1126/science.aam9288} {\bibfield  {journal} {\bibinfo
  {journal} {Science}\ }\textbf {\bibinfo {volume} {362}},\ \bibinfo {pages}
  {eaam9288} (\bibinfo {year} {2018})}\BibitemShut {NoStop}%
\bibitem [{\citenamefont {Liao}\ \emph
  {et~al.}(2017{\natexlab{a}})\citenamefont {Liao}, \citenamefont {Cai},
  \citenamefont {Liu}, \citenamefont {Zhang}, \citenamefont {Li}, \citenamefont
  {Ren}, \citenamefont {Yin}, \citenamefont {Shen}, \citenamefont {Cao},
  \citenamefont {Li}, \citenamefont {Li}, \citenamefont {Chen}, \citenamefont
  {Sun}, \citenamefont {Jia}, \citenamefont {Wu}, \citenamefont {Jiang},
  \citenamefont {Wang}, \citenamefont {Huang}, \citenamefont {Wang},
  \citenamefont {Zhou}, \citenamefont {Deng}, \citenamefont {Xi}, \citenamefont
  {Ma}, \citenamefont {Hu}, \citenamefont {Zhang}, \citenamefont {Chen},
  \citenamefont {Liu}, \citenamefont {Wang}, \citenamefont {Zhu}, \citenamefont
  {Lu}, \citenamefont {Shu}, \citenamefont {Peng}, \citenamefont {Wang},\ and\
  \citenamefont {Pan}}]{Micius_Liao2017}%
  \BibitemOpen
  \bibfield  {author} {\bibinfo {author} {\bibfnamefont {S.-K.}\ \bibnamefont
  {Liao}}, \bibinfo {author} {\bibfnamefont {W.-Q.}\ \bibnamefont {Cai}},
  \bibinfo {author} {\bibfnamefont {W.-Y.}\ \bibnamefont {Liu}}, \bibinfo
  {author} {\bibfnamefont {L.}~\bibnamefont {Zhang}}, \bibinfo {author}
  {\bibfnamefont {Y.}~\bibnamefont {Li}}, \bibinfo {author} {\bibfnamefont
  {J.-G.}\ \bibnamefont {Ren}}, \bibinfo {author} {\bibfnamefont
  {J.}~\bibnamefont {Yin}}, \bibinfo {author} {\bibfnamefont {Q.}~\bibnamefont
  {Shen}}, \bibinfo {author} {\bibfnamefont {Y.}~\bibnamefont {Cao}}, \bibinfo
  {author} {\bibfnamefont {Z.-P.}\ \bibnamefont {Li}}, \bibinfo {author}
  {\bibfnamefont {F.-Z.}\ \bibnamefont {Li}}, \bibinfo {author} {\bibfnamefont
  {X.-W.}\ \bibnamefont {Chen}}, \bibinfo {author} {\bibfnamefont {L.-H.}\
  \bibnamefont {Sun}}, \bibinfo {author} {\bibfnamefont {J.-J.}\ \bibnamefont
  {Jia}}, \bibinfo {author} {\bibfnamefont {J.-C.}\ \bibnamefont {Wu}},
  \bibinfo {author} {\bibfnamefont {X.-J.}\ \bibnamefont {Jiang}}, \bibinfo
  {author} {\bibfnamefont {J.-F.}\ \bibnamefont {Wang}}, \bibinfo {author}
  {\bibfnamefont {Y.-M.}\ \bibnamefont {Huang}}, \bibinfo {author}
  {\bibfnamefont {Q.}~\bibnamefont {Wang}}, \bibinfo {author} {\bibfnamefont
  {Y.-L.}\ \bibnamefont {Zhou}}, \bibinfo {author} {\bibfnamefont
  {L.}~\bibnamefont {Deng}}, \bibinfo {author} {\bibfnamefont {T.}~\bibnamefont
  {Xi}}, \bibinfo {author} {\bibfnamefont {L.}~\bibnamefont {Ma}}, \bibinfo
  {author} {\bibfnamefont {T.}~\bibnamefont {Hu}}, \bibinfo {author}
  {\bibfnamefont {Q.}~\bibnamefont {Zhang}}, \bibinfo {author} {\bibfnamefont
  {Y.-A.}\ \bibnamefont {Chen}}, \bibinfo {author} {\bibfnamefont {N.-L.}\
  \bibnamefont {Liu}}, \bibinfo {author} {\bibfnamefont {X.-B.}\ \bibnamefont
  {Wang}}, \bibinfo {author} {\bibfnamefont {Z.-C.}\ \bibnamefont {Zhu}},
  \bibinfo {author} {\bibfnamefont {C.-Y.}\ \bibnamefont {Lu}}, \bibinfo
  {author} {\bibfnamefont {R.}~\bibnamefont {Shu}}, \bibinfo {author}
  {\bibfnamefont {C.-Z.}\ \bibnamefont {Peng}}, \bibinfo {author}
  {\bibfnamefont {J.-Y.}\ \bibnamefont {Wang}},\ and\ \bibinfo {author}
  {\bibfnamefont {J.-W.}\ \bibnamefont {Pan}},\ }\href
  {https://doi.org/10.1038/nature23655} {\bibfield  {journal} {\bibinfo
  {journal} {Nature}\ }\textbf {\bibinfo {volume} {549}},\ \bibinfo {pages}
  {43} (\bibinfo {year} {2017}{\natexlab{a}})}\BibitemShut {NoStop}%
\bibitem [{\citenamefont {Yin}\ \emph {et~al.}(2017)\citenamefont {Yin},
  \citenamefont {Cao}, \citenamefont {Li}, \citenamefont {Ren}, \citenamefont
  {Liao}, \citenamefont {Zhang}, \citenamefont {Cai}, \citenamefont {Liu},
  \citenamefont {Li}, \citenamefont {Dai}, \citenamefont {Li}, \citenamefont
  {Huang}, \citenamefont {Deng}, \citenamefont {Li}, \citenamefont {Zhang},
  \citenamefont {Liu}, \citenamefont {Chen}, \citenamefont {Lu}, \citenamefont
  {Shu}, \citenamefont {Peng}, \citenamefont {Wang},\ and\ \citenamefont
  {Pan}}]{Micius_BBM92_Yin2017}%
  \BibitemOpen
  \bibfield  {author} {\bibinfo {author} {\bibfnamefont {J.}~\bibnamefont
  {Yin}}, \bibinfo {author} {\bibfnamefont {Y.}~\bibnamefont {Cao}}, \bibinfo
  {author} {\bibfnamefont {Y.-H.}\ \bibnamefont {Li}}, \bibinfo {author}
  {\bibfnamefont {J.-G.}\ \bibnamefont {Ren}}, \bibinfo {author} {\bibfnamefont
  {S.-K.}\ \bibnamefont {Liao}}, \bibinfo {author} {\bibfnamefont
  {L.}~\bibnamefont {Zhang}}, \bibinfo {author} {\bibfnamefont {W.-Q.}\
  \bibnamefont {Cai}}, \bibinfo {author} {\bibfnamefont {W.-Y.}\ \bibnamefont
  {Liu}}, \bibinfo {author} {\bibfnamefont {B.}~\bibnamefont {Li}}, \bibinfo
  {author} {\bibfnamefont {H.}~\bibnamefont {Dai}}, \bibinfo {author}
  {\bibfnamefont {M.}~\bibnamefont {Li}}, \bibinfo {author} {\bibfnamefont
  {Y.-M.}\ \bibnamefont {Huang}}, \bibinfo {author} {\bibfnamefont
  {L.}~\bibnamefont {Deng}}, \bibinfo {author} {\bibfnamefont {L.}~\bibnamefont
  {Li}}, \bibinfo {author} {\bibfnamefont {Q.}~\bibnamefont {Zhang}}, \bibinfo
  {author} {\bibfnamefont {N.-L.}\ \bibnamefont {Liu}}, \bibinfo {author}
  {\bibfnamefont {Y.-A.}\ \bibnamefont {Chen}}, \bibinfo {author}
  {\bibfnamefont {C.-Y.}\ \bibnamefont {Lu}}, \bibinfo {author} {\bibfnamefont
  {R.}~\bibnamefont {Shu}}, \bibinfo {author} {\bibfnamefont {C.-Z.}\
  \bibnamefont {Peng}}, \bibinfo {author} {\bibfnamefont {J.-Y.}\ \bibnamefont
  {Wang}},\ and\ \bibinfo {author} {\bibfnamefont {J.-W.}\ \bibnamefont
  {Pan}},\ }\href {https://doi.org/10.1103/PhysRevLett.119.200501} {\bibfield
  {journal} {\bibinfo  {journal} {Phys. Rev. Lett.}\ }\textbf {\bibinfo
  {volume} {119}},\ \bibinfo {pages} {200501} (\bibinfo {year}
  {2017})}\BibitemShut {NoStop}%
\bibitem [{\citenamefont {Bedington}\ \emph {et~al.}(2017)\citenamefont
  {Bedington}, \citenamefont {Arrazola},\ and\ \citenamefont
  {Ling}}]{Bedington2017}%
  \BibitemOpen
  \bibfield  {author} {\bibinfo {author} {\bibfnamefont {R.}~\bibnamefont
  {Bedington}}, \bibinfo {author} {\bibfnamefont {J.~M.}\ \bibnamefont
  {Arrazola}},\ and\ \bibinfo {author} {\bibfnamefont {A.}~\bibnamefont
  {Ling}},\ }\href {https://doi.org/10.1038/s41534-017-0031-5} {\bibfield
  {journal} {\bibinfo  {journal} {npj Quantum Inf.}\ }\textbf {\bibinfo
  {volume} {3}},\ \bibinfo {pages} {30} (\bibinfo {year} {2017})}\BibitemShut
  {NoStop}%
\bibitem [{\citenamefont {Agnesi}\ \emph {et~al.}(2018)\citenamefont {Agnesi},
  \citenamefont {Vedovato}, \citenamefont {Schiavon}, \citenamefont {Dequal},
  \citenamefont {Calderaro}, \citenamefont {Tomasin}, \citenamefont {Marangon},
  \citenamefont {Stanco}, \citenamefont {Luceri}, \citenamefont {Bianco},
  \citenamefont {Vallone},\ and\ \citenamefont {Villoresi}}]{Agnesi2018}%
  \BibitemOpen
  \bibfield  {author} {\bibinfo {author} {\bibfnamefont {C.}~\bibnamefont
  {Agnesi}}, \bibinfo {author} {\bibfnamefont {F.}~\bibnamefont {Vedovato}},
  \bibinfo {author} {\bibfnamefont {M.}~\bibnamefont {Schiavon}}, \bibinfo
  {author} {\bibfnamefont {D.}~\bibnamefont {Dequal}}, \bibinfo {author}
  {\bibfnamefont {L.}~\bibnamefont {Calderaro}}, \bibinfo {author}
  {\bibfnamefont {M.}~\bibnamefont {Tomasin}}, \bibinfo {author} {\bibfnamefont
  {D.~G.}\ \bibnamefont {Marangon}}, \bibinfo {author} {\bibfnamefont
  {A.}~\bibnamefont {Stanco}}, \bibinfo {author} {\bibfnamefont
  {V.}~\bibnamefont {Luceri}}, \bibinfo {author} {\bibfnamefont
  {G.}~\bibnamefont {Bianco}}, \bibinfo {author} {\bibfnamefont
  {G.}~\bibnamefont {Vallone}},\ and\ \bibinfo {author} {\bibfnamefont
  {P.}~\bibnamefont {Villoresi}},\ }\href
  {https://doi.org/10.1098/rsta.2017.0461} {\bibfield  {journal} {\bibinfo
  {journal} {Philos. Trans. Royal Soc. A}\ }\textbf {\bibinfo {volume} {376}},\
  \bibinfo {pages} {20170461} (\bibinfo {year} {2018})}\BibitemShut {NoStop}%
\bibitem [{\citenamefont {Khan}\ \emph {et~al.}(2018)\citenamefont {Khan},
  \citenamefont {Heim}, \citenamefont {Neuzner},\ and\ \citenamefont
  {Marquardt}}]{Khan2018}%
  \BibitemOpen
  \bibfield  {author} {\bibinfo {author} {\bibfnamefont {I.}~\bibnamefont
  {Khan}}, \bibinfo {author} {\bibfnamefont {B.}~\bibnamefont {Heim}}, \bibinfo
  {author} {\bibfnamefont {A.}~\bibnamefont {Neuzner}},\ and\ \bibinfo {author}
  {\bibfnamefont {C.}~\bibnamefont {Marquardt}},\ }\href
  {https://doi.org/10.1364/opn.29.2.000026} {\bibfield  {journal} {\bibinfo
  {journal} {Opt. Photon. News}\ }\textbf {\bibinfo {volume} {29}},\ \bibinfo
  {pages} {26} (\bibinfo {year} {2018})}\BibitemShut {NoStop}%
\bibitem [{\citenamefont {Boaron}\ \emph {et~al.}(2018)\citenamefont {Boaron},
  \citenamefont {Boso}, \citenamefont {Rusca}, \citenamefont {Vulliez},
  \citenamefont {Autebert}, \citenamefont {Caloz}, \citenamefont {Perrenoud},
  \citenamefont {Gras}, \citenamefont {Bussi{\`{e}}res}, \citenamefont {Li},
  \citenamefont {Nolan}, \citenamefont {Martin},\ and\ \citenamefont
  {Zbinden}}]{BoaronRecord}%
  \BibitemOpen
  \bibfield  {author} {\bibinfo {author} {\bibfnamefont {A.}~\bibnamefont
  {Boaron}}, \bibinfo {author} {\bibfnamefont {G.}~\bibnamefont {Boso}},
  \bibinfo {author} {\bibfnamefont {D.}~\bibnamefont {Rusca}}, \bibinfo
  {author} {\bibfnamefont {C.}~\bibnamefont {Vulliez}}, \bibinfo {author}
  {\bibfnamefont {C.}~\bibnamefont {Autebert}}, \bibinfo {author}
  {\bibfnamefont {M.}~\bibnamefont {Caloz}}, \bibinfo {author} {\bibfnamefont
  {M.}~\bibnamefont {Perrenoud}}, \bibinfo {author} {\bibfnamefont
  {G.}~\bibnamefont {Gras}}, \bibinfo {author} {\bibfnamefont {F.}~\bibnamefont
  {Bussi{\`{e}}res}}, \bibinfo {author} {\bibfnamefont {M.-J.}\ \bibnamefont
  {Li}}, \bibinfo {author} {\bibfnamefont {D.}~\bibnamefont {Nolan}}, \bibinfo
  {author} {\bibfnamefont {A.}~\bibnamefont {Martin}},\ and\ \bibinfo {author}
  {\bibfnamefont {H.}~\bibnamefont {Zbinden}},\ }\href
  {https://doi.org/10.1103/physrevlett.121.190502} {\bibfield  {journal}
  {\bibinfo  {journal} {Phys. Rev. Lett.}\ }\textbf {\bibinfo {volume} {121}},\
  \bibinfo {pages} {190502} (\bibinfo {year} {2018})}\BibitemShut {NoStop}%
\bibitem [{\citenamefont {Yoshino}\ \emph {et~al.}(2013)\citenamefont
  {Yoshino}, \citenamefont {Ochi}, \citenamefont {Fujiwara}, \citenamefont
  {Sasaki},\ and\ \citenamefont {Tajima}}]{Yoshino2013}%
  \BibitemOpen
  \bibfield  {author} {\bibinfo {author} {\bibfnamefont {K.-I.}\ \bibnamefont
  {Yoshino}}, \bibinfo {author} {\bibfnamefont {T.}~\bibnamefont {Ochi}},
  \bibinfo {author} {\bibfnamefont {M.}~\bibnamefont {Fujiwara}}, \bibinfo
  {author} {\bibfnamefont {M.}~\bibnamefont {Sasaki}},\ and\ \bibinfo {author}
  {\bibfnamefont {A.}~\bibnamefont {Tajima}},\ }\href
  {https://doi.org/10.1364/OE.21.031395} {\bibfield  {journal} {\bibinfo
  {journal} {Opt. Express}\ }\textbf {\bibinfo {volume} {21}},\ \bibinfo
  {pages} {31395} (\bibinfo {year} {2013})}\BibitemShut {NoStop}%
\bibitem [{\citenamefont {Islam}\ \emph {et~al.}(2017)\citenamefont {Islam},
  \citenamefont {Lim}, \citenamefont {Cahall}, \citenamefont {Kim},\ and\
  \citenamefont {Gauthier}}]{Islam2017}%
  \BibitemOpen
  \bibfield  {author} {\bibinfo {author} {\bibfnamefont {N.~T.}\ \bibnamefont
  {Islam}}, \bibinfo {author} {\bibfnamefont {C.~C.~W.}\ \bibnamefont {Lim}},
  \bibinfo {author} {\bibfnamefont {C.}~\bibnamefont {Cahall}}, \bibinfo
  {author} {\bibfnamefont {J.}~\bibnamefont {Kim}},\ and\ \bibinfo {author}
  {\bibfnamefont {D.~J.}\ \bibnamefont {Gauthier}},\ }\href
  {https://doi.org/10.1126/sciadv.1701491} {\bibfield  {journal} {\bibinfo
  {journal} {Sci. Adv.}\ }\textbf {\bibinfo {volume} {3}},\ \bibinfo {pages}
  {e1701491} (\bibinfo {year} {2017})}\BibitemShut {NoStop}%
\bibitem [{\citenamefont {{Yuan}}\ \emph {et~al.}(2018)\citenamefont {{Yuan}},
  \citenamefont {{Plews}}, \citenamefont {{Takahashi}}, \citenamefont {{Doi}},
  \citenamefont {{Tam}}, \citenamefont {{Sharpe}}, \citenamefont {{Dixon}},
  \citenamefont {{Lavelle}}, \citenamefont {{Dynes}}, \citenamefont
  {{Murakami}}, \citenamefont {{Kujiraoka}}, \citenamefont {{Lucamarini}},
  \citenamefont {{Tanizawa}}, \citenamefont {{Sato}},\ and\ \citenamefont
  {{Shields}}}]{Yuan2018}%
  \BibitemOpen
  \bibfield  {author} {\bibinfo {author} {\bibfnamefont {Z.}~\bibnamefont
  {{Yuan}}}, \bibinfo {author} {\bibfnamefont {A.}~\bibnamefont {{Plews}}},
  \bibinfo {author} {\bibfnamefont {R.}~\bibnamefont {{Takahashi}}}, \bibinfo
  {author} {\bibfnamefont {K.}~\bibnamefont {{Doi}}}, \bibinfo {author}
  {\bibfnamefont {W.}~\bibnamefont {{Tam}}}, \bibinfo {author} {\bibfnamefont
  {A.~W.}\ \bibnamefont {{Sharpe}}}, \bibinfo {author} {\bibfnamefont {A.~R.}\
  \bibnamefont {{Dixon}}}, \bibinfo {author} {\bibfnamefont {E.}~\bibnamefont
  {{Lavelle}}}, \bibinfo {author} {\bibfnamefont {J.~F.}\ \bibnamefont
  {{Dynes}}}, \bibinfo {author} {\bibfnamefont {A.}~\bibnamefont {{Murakami}}},
  \bibinfo {author} {\bibfnamefont {M.}~\bibnamefont {{Kujiraoka}}}, \bibinfo
  {author} {\bibfnamefont {M.}~\bibnamefont {{Lucamarini}}}, \bibinfo {author}
  {\bibfnamefont {Y.}~\bibnamefont {{Tanizawa}}}, \bibinfo {author}
  {\bibfnamefont {H.}~\bibnamefont {{Sato}}},\ and\ \bibinfo {author}
  {\bibfnamefont {A.~J.}\ \bibnamefont {{Shields}}},\ }\href
  {https://doi.org/10.1109/JLT.2018.2843136} {\bibfield  {journal} {\bibinfo
  {journal} {J. Light. Technol.}\ }\textbf {\bibinfo {volume} {36}},\ \bibinfo
  {pages} {3427} (\bibinfo {year} {2018})}\BibitemShut {NoStop}%
\bibitem [{\citenamefont {Minder}\ \emph {et~al.}(2019)\citenamefont {Minder},
  \citenamefont {Pittaluga}, \citenamefont {Roberts}, \citenamefont
  {Lucamarini}, \citenamefont {Dynes}, \citenamefont {Yuan},\ and\
  \citenamefont {Shields}}]{MinderPittaluga2019}%
  \BibitemOpen
  \bibfield  {author} {\bibinfo {author} {\bibfnamefont {M.}~\bibnamefont
  {Minder}}, \bibinfo {author} {\bibfnamefont {M.}~\bibnamefont {Pittaluga}},
  \bibinfo {author} {\bibfnamefont {G.~L.}\ \bibnamefont {Roberts}}, \bibinfo
  {author} {\bibfnamefont {M.}~\bibnamefont {Lucamarini}}, \bibinfo {author}
  {\bibfnamefont {J.~F.}\ \bibnamefont {Dynes}}, \bibinfo {author}
  {\bibfnamefont {Z.~L.}\ \bibnamefont {Yuan}},\ and\ \bibinfo {author}
  {\bibfnamefont {A.~J.}\ \bibnamefont {Shields}},\ }\href
  {https://doi.org/10.1038/s41566-019-0377-7} {\bibfield  {journal} {\bibinfo
  {journal} {Nature Photonics}\ }\textbf {\bibinfo {volume} {13}},\ \bibinfo
  {pages} {334} (\bibinfo {year} {2019})}\BibitemShut {NoStop}%
\bibitem [{\citenamefont {Agnesi}\ \emph {et~al.}(2020)\citenamefont {Agnesi},
  \citenamefont {Avesani}, \citenamefont {Calderaro}, \citenamefont {Stanco},
  \citenamefont {Foletto}, \citenamefont {Zahidy}, \citenamefont {Scriminich},
  \citenamefont {Vedovato}, \citenamefont {Vallone},\ and\ \citenamefont
  {Villoresi}}]{Optica_Agnesi2019}%
  \BibitemOpen
  \bibfield  {author} {\bibinfo {author} {\bibfnamefont {C.}~\bibnamefont
  {Agnesi}}, \bibinfo {author} {\bibfnamefont {M.}~\bibnamefont {Avesani}},
  \bibinfo {author} {\bibfnamefont {L.}~\bibnamefont {Calderaro}}, \bibinfo
  {author} {\bibfnamefont {A.}~\bibnamefont {Stanco}}, \bibinfo {author}
  {\bibfnamefont {G.}~\bibnamefont {Foletto}}, \bibinfo {author} {\bibfnamefont
  {M.}~\bibnamefont {Zahidy}}, \bibinfo {author} {\bibfnamefont
  {A.}~\bibnamefont {Scriminich}}, \bibinfo {author} {\bibfnamefont
  {F.}~\bibnamefont {Vedovato}}, \bibinfo {author} {\bibfnamefont
  {G.}~\bibnamefont {Vallone}},\ and\ \bibinfo {author} {\bibfnamefont
  {P.}~\bibnamefont {Villoresi}},\ }\href
  {https://doi.org/10.1364/OPTICA.381013} {\bibfield  {journal} {\bibinfo
  {journal} {Optica}\ }\textbf {\bibinfo {volume} {7}},\ \bibinfo {pages} {284}
  (\bibinfo {year} {2020})}\BibitemShut {NoStop}%
\bibitem [{\citenamefont {Liu}\ \emph {et~al.}(2019)\citenamefont {Liu},
  \citenamefont {Yu}, \citenamefont {Zhang}, \citenamefont {Guan},
  \citenamefont {Chen}, \citenamefont {Zhang}, \citenamefont {Hu},
  \citenamefont {Li}, \citenamefont {Jiang}, \citenamefont {Lin}, \citenamefont
  {Chen}, \citenamefont {You}, \citenamefont {Wang}, \citenamefont {Wang},
  \citenamefont {Zhang},\ and\ \citenamefont
  {Pan}}]{experimental_twin_field_Liu2019}%
  \BibitemOpen
  \bibfield  {author} {\bibinfo {author} {\bibfnamefont {Y.}~\bibnamefont
  {Liu}}, \bibinfo {author} {\bibfnamefont {Z.-W.}\ \bibnamefont {Yu}},
  \bibinfo {author} {\bibfnamefont {W.}~\bibnamefont {Zhang}}, \bibinfo
  {author} {\bibfnamefont {J.-Y.}\ \bibnamefont {Guan}}, \bibinfo {author}
  {\bibfnamefont {J.-P.}\ \bibnamefont {Chen}}, \bibinfo {author}
  {\bibfnamefont {C.}~\bibnamefont {Zhang}}, \bibinfo {author} {\bibfnamefont
  {X.-L.}\ \bibnamefont {Hu}}, \bibinfo {author} {\bibfnamefont
  {H.}~\bibnamefont {Li}}, \bibinfo {author} {\bibfnamefont {C.}~\bibnamefont
  {Jiang}}, \bibinfo {author} {\bibfnamefont {J.}~\bibnamefont {Lin}}, \bibinfo
  {author} {\bibfnamefont {T.-Y.}\ \bibnamefont {Chen}}, \bibinfo {author}
  {\bibfnamefont {L.}~\bibnamefont {You}}, \bibinfo {author} {\bibfnamefont
  {Z.}~\bibnamefont {Wang}}, \bibinfo {author} {\bibfnamefont {X.-B.}\
  \bibnamefont {Wang}}, \bibinfo {author} {\bibfnamefont {Q.}~\bibnamefont
  {Zhang}},\ and\ \bibinfo {author} {\bibfnamefont {J.-W.}\ \bibnamefont
  {Pan}},\ }\href {https://doi.org/10.1103/PhysRevLett.123.100505} {\bibfield
  {journal} {\bibinfo  {journal} {Phys. Rev. Lett.}\ }\textbf {\bibinfo
  {volume} {123}},\ \bibinfo {pages} {100505} (\bibinfo {year}
  {2019})}\BibitemShut {NoStop}%
\bibitem [{\citenamefont {Avesani}\ \emph
  {et~al.}(2021{\natexlab{a}})\citenamefont {Avesani}, \citenamefont
  {Calderaro}, \citenamefont {Foletto}, \citenamefont {Agnesi}, \citenamefont
  {Picciariello}, \citenamefont {Santagiustina}, \citenamefont {Scriminich},
  \citenamefont {Stanco}, \citenamefont {Vedovato}, \citenamefont {Zahidy},
  \citenamefont {Vallone},\ and\ \citenamefont
  {Villoresi}}]{centro_di_calcolo_Avesani2021}%
  \BibitemOpen
  \bibfield  {author} {\bibinfo {author} {\bibfnamefont {M.}~\bibnamefont
  {Avesani}}, \bibinfo {author} {\bibfnamefont {L.}~\bibnamefont {Calderaro}},
  \bibinfo {author} {\bibfnamefont {G.}~\bibnamefont {Foletto}}, \bibinfo
  {author} {\bibfnamefont {C.}~\bibnamefont {Agnesi}}, \bibinfo {author}
  {\bibfnamefont {F.}~\bibnamefont {Picciariello}}, \bibinfo {author}
  {\bibfnamefont {F.~B.~L.}\ \bibnamefont {Santagiustina}}, \bibinfo {author}
  {\bibfnamefont {A.}~\bibnamefont {Scriminich}}, \bibinfo {author}
  {\bibfnamefont {A.}~\bibnamefont {Stanco}}, \bibinfo {author} {\bibfnamefont
  {F.}~\bibnamefont {Vedovato}}, \bibinfo {author} {\bibfnamefont
  {M.}~\bibnamefont {Zahidy}}, \bibinfo {author} {\bibfnamefont
  {G.}~\bibnamefont {Vallone}},\ and\ \bibinfo {author} {\bibfnamefont
  {P.}~\bibnamefont {Villoresi}},\ }\href {https://doi.org/10.1364/OL.422890}
  {\bibfield  {journal} {\bibinfo  {journal} {Opt. Lett.}\ }\textbf {\bibinfo
  {volume} {46}},\ \bibinfo {pages} {2848} (\bibinfo {year}
  {2021}{\natexlab{a}})}\BibitemShut {NoStop}%
\bibitem [{\citenamefont {Liao}\ \emph
  {et~al.}(2017{\natexlab{b}})\citenamefont {Liao}, \citenamefont {Yong},
  \citenamefont {Liu}, \citenamefont {Shentu}, \citenamefont {Li},
  \citenamefont {Lin}, \citenamefont {Dai}, \citenamefont {Zhao}, \citenamefont
  {Li}, \citenamefont {Guan}, \citenamefont {Chen}, \citenamefont {Gong},
  \citenamefont {Li}, \citenamefont {Lin}, \citenamefont {Pan}, \citenamefont
  {Pelc}, \citenamefont {Fejer}, \citenamefont {Zhang}, \citenamefont {Liu},
  \citenamefont {Yin}, \citenamefont {Ren}, \citenamefont {Wang}, \citenamefont
  {Zhang}, \citenamefont {Peng},\ and\ \citenamefont
  {Pan}}]{Liao2017_daylight}%
  \BibitemOpen
  \bibfield  {author} {\bibinfo {author} {\bibfnamefont {S.-K.}\ \bibnamefont
  {Liao}}, \bibinfo {author} {\bibfnamefont {H.-L.}\ \bibnamefont {Yong}},
  \bibinfo {author} {\bibfnamefont {C.}~\bibnamefont {Liu}}, \bibinfo {author}
  {\bibfnamefont {G.-L.}\ \bibnamefont {Shentu}}, \bibinfo {author}
  {\bibfnamefont {D.-D.}\ \bibnamefont {Li}}, \bibinfo {author} {\bibfnamefont
  {J.}~\bibnamefont {Lin}}, \bibinfo {author} {\bibfnamefont {H.}~\bibnamefont
  {Dai}}, \bibinfo {author} {\bibfnamefont {S.-Q.}\ \bibnamefont {Zhao}},
  \bibinfo {author} {\bibfnamefont {B.}~\bibnamefont {Li}}, \bibinfo {author}
  {\bibfnamefont {J.-Y.}\ \bibnamefont {Guan}}, \bibinfo {author}
  {\bibfnamefont {W.}~\bibnamefont {Chen}}, \bibinfo {author} {\bibfnamefont
  {Y.-H.}\ \bibnamefont {Gong}}, \bibinfo {author} {\bibfnamefont
  {Y.}~\bibnamefont {Li}}, \bibinfo {author} {\bibfnamefont {Z.-H.}\
  \bibnamefont {Lin}}, \bibinfo {author} {\bibfnamefont {G.-S.}\ \bibnamefont
  {Pan}}, \bibinfo {author} {\bibfnamefont {J.~S.}\ \bibnamefont {Pelc}},
  \bibinfo {author} {\bibfnamefont {M.~M.}\ \bibnamefont {Fejer}}, \bibinfo
  {author} {\bibfnamefont {W.-Z.}\ \bibnamefont {Zhang}}, \bibinfo {author}
  {\bibfnamefont {W.-Y.}\ \bibnamefont {Liu}}, \bibinfo {author} {\bibfnamefont
  {J.}~\bibnamefont {Yin}}, \bibinfo {author} {\bibfnamefont {J.-G.}\
  \bibnamefont {Ren}}, \bibinfo {author} {\bibfnamefont {X.-B.}\ \bibnamefont
  {Wang}}, \bibinfo {author} {\bibfnamefont {Q.}~\bibnamefont {Zhang}},
  \bibinfo {author} {\bibfnamefont {C.-Z.}\ \bibnamefont {Peng}},\ and\
  \bibinfo {author} {\bibfnamefont {J.-W.}\ \bibnamefont {Pan}},\ }\href
  {https://doi.org/10.1038/nphoton.2017.116} {\bibfield  {journal} {\bibinfo
  {journal} {Nat. Photonics}\ }\textbf {\bibinfo {volume} {11}},\ \bibinfo
  {pages} {509} (\bibinfo {year} {2017}{\natexlab{b}})}\BibitemShut {NoStop}%
\bibitem [{\citenamefont {Gong}\ \emph {et~al.}(2018)\citenamefont {Gong},
  \citenamefont {Yang}, \citenamefont {Yong}, \citenamefont {Guan},
  \citenamefont {Shentu}, \citenamefont {Liu}, \citenamefont {Li},
  \citenamefont {Cao}, \citenamefont {Yin}, \citenamefont {Liao}, \citenamefont
  {Ren}, \citenamefont {Zhang}, \citenamefont {Peng},\ and\ \citenamefont
  {Pan}}]{Gong2018}%
  \BibitemOpen
  \bibfield  {author} {\bibinfo {author} {\bibfnamefont {Y.-H.}\ \bibnamefont
  {Gong}}, \bibinfo {author} {\bibfnamefont {K.-X.}\ \bibnamefont {Yang}},
  \bibinfo {author} {\bibfnamefont {H.-L.}\ \bibnamefont {Yong}}, \bibinfo
  {author} {\bibfnamefont {J.-Y.}\ \bibnamefont {Guan}}, \bibinfo {author}
  {\bibfnamefont {G.-L.}\ \bibnamefont {Shentu}}, \bibinfo {author}
  {\bibfnamefont {C.}~\bibnamefont {Liu}}, \bibinfo {author} {\bibfnamefont
  {F.-Z.}\ \bibnamefont {Li}}, \bibinfo {author} {\bibfnamefont
  {Y.}~\bibnamefont {Cao}}, \bibinfo {author} {\bibfnamefont {J.}~\bibnamefont
  {Yin}}, \bibinfo {author} {\bibfnamefont {S.-K.}\ \bibnamefont {Liao}},
  \bibinfo {author} {\bibfnamefont {J.-G.}\ \bibnamefont {Ren}}, \bibinfo
  {author} {\bibfnamefont {Q.}~\bibnamefont {Zhang}}, \bibinfo {author}
  {\bibfnamefont {C.-Z.}\ \bibnamefont {Peng}},\ and\ \bibinfo {author}
  {\bibfnamefont {J.-W.}\ \bibnamefont {Pan}},\ }\href
  {https://doi.org/10.1364/oe.26.018897} {\bibfield  {journal} {\bibinfo
  {journal} {Opt. Express}\ }\textbf {\bibinfo {volume} {26}},\ \bibinfo
  {pages} {18897} (\bibinfo {year} {2018})}\BibitemShut {NoStop}%
\bibitem [{\citenamefont {Avesani}\ \emph
  {et~al.}(2021{\natexlab{b}})\citenamefont {Avesani}, \citenamefont
  {Calderaro}, \citenamefont {Schiavon}, \citenamefont {Stanco}, \citenamefont
  {Agnesi}, \citenamefont {Santamato}, \citenamefont {Zahidy}, \citenamefont
  {Scriminich}, \citenamefont {Foletto}, \citenamefont {Contestabile},
  \citenamefont {Chiesa}, \citenamefont {Rotta}, \citenamefont {Artiglia},
  \citenamefont {Montanaro}, \citenamefont {Romagnoli}, \citenamefont
  {Sorianello}, \citenamefont {Vedovato}, \citenamefont {Vallone},\ and\
  \citenamefont {Villoresi}}]{qcosone_Avesani2021}%
  \BibitemOpen
  \bibfield  {author} {\bibinfo {author} {\bibfnamefont {M.}~\bibnamefont
  {Avesani}}, \bibinfo {author} {\bibfnamefont {L.}~\bibnamefont {Calderaro}},
  \bibinfo {author} {\bibfnamefont {M.}~\bibnamefont {Schiavon}}, \bibinfo
  {author} {\bibfnamefont {A.}~\bibnamefont {Stanco}}, \bibinfo {author}
  {\bibfnamefont {C.}~\bibnamefont {Agnesi}}, \bibinfo {author} {\bibfnamefont
  {A.}~\bibnamefont {Santamato}}, \bibinfo {author} {\bibfnamefont
  {M.}~\bibnamefont {Zahidy}}, \bibinfo {author} {\bibfnamefont
  {A.}~\bibnamefont {Scriminich}}, \bibinfo {author} {\bibfnamefont
  {G.}~\bibnamefont {Foletto}}, \bibinfo {author} {\bibfnamefont
  {G.}~\bibnamefont {Contestabile}}, \bibinfo {author} {\bibfnamefont
  {M.}~\bibnamefont {Chiesa}}, \bibinfo {author} {\bibfnamefont
  {D.}~\bibnamefont {Rotta}}, \bibinfo {author} {\bibfnamefont
  {M.}~\bibnamefont {Artiglia}}, \bibinfo {author} {\bibfnamefont
  {A.}~\bibnamefont {Montanaro}}, \bibinfo {author} {\bibfnamefont
  {M.}~\bibnamefont {Romagnoli}}, \bibinfo {author} {\bibfnamefont
  {V.}~\bibnamefont {Sorianello}}, \bibinfo {author} {\bibfnamefont
  {F.}~\bibnamefont {Vedovato}}, \bibinfo {author} {\bibfnamefont
  {G.}~\bibnamefont {Vallone}},\ and\ \bibinfo {author} {\bibfnamefont
  {P.}~\bibnamefont {Villoresi}},\ }\href
  {https://doi.org/10.1038/s41534-021-00421-2} {\bibfield  {journal} {\bibinfo
  {journal} {npj Quantum Information}\ }\textbf {\bibinfo {volume} {7}},\
  \bibinfo {pages} {93} (\bibinfo {year} {2021}{\natexlab{b}})}\BibitemShut
  {NoStop}%
\bibitem [{\citenamefont {Jian}\ \emph {et~al.}(2014)\citenamefont {Jian},
  \citenamefont {Ke}, \citenamefont {Chao}, \citenamefont {Peng}, \citenamefont
  {Dagang},\ and\ \citenamefont {Zhoushi}}]{Jian2014}%
  \BibitemOpen
  \bibfield  {author} {\bibinfo {author} {\bibfnamefont {H.}~\bibnamefont
  {Jian}}, \bibinfo {author} {\bibfnamefont {D.}~\bibnamefont {Ke}}, \bibinfo
  {author} {\bibfnamefont {L.}~\bibnamefont {Chao}}, \bibinfo {author}
  {\bibfnamefont {Z.}~\bibnamefont {Peng}}, \bibinfo {author} {\bibfnamefont
  {J.}~\bibnamefont {Dagang}},\ and\ \bibinfo {author} {\bibfnamefont
  {Y.}~\bibnamefont {Zhoushi}},\ }\href {https://doi.org/10.1364/OE.22.016000}
  {\bibfield  {journal} {\bibinfo  {journal} {Opt. Express}\ }\textbf {\bibinfo
  {volume} {22}},\ \bibinfo {pages} {16000} (\bibinfo {year}
  {2014})}\BibitemShut {NoStop}%
\bibitem [{\citenamefont {Rusca}\ \emph {et~al.}(2018)\citenamefont {Rusca},
  \citenamefont {Boaron}, \citenamefont {Gr{\"{u}}nenfelder}, \citenamefont
  {Martin},\ and\ \citenamefont {Zbinden}}]{Rusca2018}%
  \BibitemOpen
  \bibfield  {author} {\bibinfo {author} {\bibfnamefont {D.}~\bibnamefont
  {Rusca}}, \bibinfo {author} {\bibfnamefont {A.}~\bibnamefont {Boaron}},
  \bibinfo {author} {\bibfnamefont {F.}~\bibnamefont {Gr{\"{u}}nenfelder}},
  \bibinfo {author} {\bibfnamefont {A.}~\bibnamefont {Martin}},\ and\ \bibinfo
  {author} {\bibfnamefont {H.}~\bibnamefont {Zbinden}},\ }\href
  {https://doi.org/10.1063/1.5023340} {\bibfield  {journal} {\bibinfo
  {journal} {Applied Physics Letters}\ }\textbf {\bibinfo {volume} {112}},\
  \bibinfo {pages} {171104} (\bibinfo {year} {2018})},\ \Eprint
  {https://arxiv.org/abs/1801.03443} {arXiv:1801.03443} \BibitemShut {NoStop}%
\bibitem [{\citenamefont {Vasylyev}\ \emph {et~al.}(2016)\citenamefont
  {Vasylyev}, \citenamefont {Semenov},\ and\ \citenamefont
  {Vogel}}]{Vasylyev2016}%
  \BibitemOpen
  \bibfield  {author} {\bibinfo {author} {\bibfnamefont {D.}~\bibnamefont
  {Vasylyev}}, \bibinfo {author} {\bibfnamefont {A.~A.}\ \bibnamefont
  {Semenov}},\ and\ \bibinfo {author} {\bibfnamefont {W.}~\bibnamefont
  {Vogel}},\ }\href {https://doi.org/10.1103/PhysRevLett.117.090501} {\bibfield
   {journal} {\bibinfo  {journal} {Phys. Rev. Lett.}\ }\textbf {\bibinfo
  {volume} {117}},\ \bibinfo {pages} {090501} (\bibinfo {year}
  {2016})}\BibitemShut {NoStop}%
\bibitem [{\citenamefont {Vasylyev}\ \emph {et~al.}(2018)\citenamefont
  {Vasylyev}, \citenamefont {Vogel},\ and\ \citenamefont
  {Semenov}}]{Vasylyev2018}%
  \BibitemOpen
  \bibfield  {author} {\bibinfo {author} {\bibfnamefont {D.}~\bibnamefont
  {Vasylyev}}, \bibinfo {author} {\bibfnamefont {W.}~\bibnamefont {Vogel}},\
  and\ \bibinfo {author} {\bibfnamefont {A.~A.}\ \bibnamefont {Semenov}},\
  }\href {https://doi.org/10.1103/PhysRevA.97.063852} {\bibfield  {journal}
  {\bibinfo  {journal} {Phys. Rev. A}\ }\textbf {\bibinfo {volume} {97}},\
  \bibinfo {pages} {063852} (\bibinfo {year} {2018})}\BibitemShut {NoStop}%
\bibitem [{\citenamefont {Pirandola}(2021{\natexlab{a}})}]{Pirandola2021free}%
  \BibitemOpen
  \bibfield  {author} {\bibinfo {author} {\bibfnamefont {S.}~\bibnamefont
  {Pirandola}},\ }\href {https://doi.org/10.1103/PhysRevResearch.3.013279}
  {\bibfield  {journal} {\bibinfo  {journal} {Phys. Rev. Research}\ }\textbf
  {\bibinfo {volume} {3}},\ \bibinfo {pages} {013279} (\bibinfo {year}
  {2021}{\natexlab{a}})}\BibitemShut {NoStop}%
\bibitem [{\citenamefont {Pirandola}(2021{\natexlab{b}})}]{Pirandola2021sat}%
  \BibitemOpen
  \bibfield  {author} {\bibinfo {author} {\bibfnamefont {S.}~\bibnamefont
  {Pirandola}},\ }\href {https://doi.org/10.1103/PhysRevResearch.3.023130}
  {\bibfield  {journal} {\bibinfo  {journal} {Phys. Rev. Research}\ }\textbf
  {\bibinfo {volume} {3}},\ \bibinfo {pages} {023130} (\bibinfo {year}
  {2021}{\natexlab{b}})}\BibitemShut {NoStop}%
\bibitem [{\citenamefont {Canuet}\ \emph {et~al.}(2018)\citenamefont {Canuet},
  \citenamefont {V\'{e}drenne}, \citenamefont {Conan}, \citenamefont {Petit},
  \citenamefont {Artaud}, \citenamefont {Rissons},\ and\ \citenamefont
  {Lacan}}]{Canuet2018}%
  \BibitemOpen
  \bibfield  {author} {\bibinfo {author} {\bibfnamefont {L.}~\bibnamefont
  {Canuet}}, \bibinfo {author} {\bibfnamefont {N.}~\bibnamefont
  {V\'{e}drenne}}, \bibinfo {author} {\bibfnamefont {J.}~\bibnamefont {Conan}},
  \bibinfo {author} {\bibfnamefont {C.}~\bibnamefont {Petit}}, \bibinfo
  {author} {\bibfnamefont {G.}~\bibnamefont {Artaud}}, \bibinfo {author}
  {\bibfnamefont {A.}~\bibnamefont {Rissons}},\ and\ \bibinfo {author}
  {\bibfnamefont {J.}~\bibnamefont {Lacan}},\ }\href
  {https://doi.org/10.1364/JOSAA.35.000148} {\bibfield  {journal} {\bibinfo
  {journal} {J. Opt. Soc. Am. A}\ }\textbf {\bibinfo {volume} {35}},\ \bibinfo
  {pages} {148} (\bibinfo {year} {2018})}\BibitemShut {NoStop}%
\bibitem [{\citenamefont {Kneizys}\ \emph {et~al.}(1988)\citenamefont
  {Kneizys}, \citenamefont {Shettle}, \citenamefont {Abreu}, \citenamefont
  {Chetwynd},\ and\ \citenamefont {Anderson}}]{LOWTRAN}%
  \BibitemOpen
  \bibfield  {author} {\bibinfo {author} {\bibfnamefont {F.}~\bibnamefont
  {Kneizys}}, \bibinfo {author} {\bibfnamefont {E.}~\bibnamefont {Shettle}},
  \bibinfo {author} {\bibfnamefont {L.}~\bibnamefont {Abreu}}, \bibinfo
  {author} {\bibfnamefont {J.}~\bibnamefont {Chetwynd}},\ and\ \bibinfo
  {author} {\bibfnamefont {G.}~\bibnamefont {Anderson}},\ }\href@noop {}
  {\bibinfo {title} {User guide to lowtran 7}} (\bibinfo {year}
  {1988})\BibitemShut {NoStop}%
\bibitem [{\citenamefont {Andrews}\ and\ \citenamefont
  {Phillips}(2005)}]{Andrews_book}%
  \BibitemOpen
  \bibfield  {author} {\bibinfo {author} {\bibfnamefont {L.~C.}\ \bibnamefont
  {Andrews}}\ and\ \bibinfo {author} {\bibfnamefont {R.~L.}\ \bibnamefont
  {Phillips}},\ }\href@noop {} {\emph {\bibinfo {title} {Laser beam propagation
  through random media}}},\ \bibinfo {edition} {2nd}\ ed.\ (\bibinfo
  {publisher} {SPIE Press},\ \bibinfo {year} {2005})\BibitemShut {NoStop}%
\bibitem [{\citenamefont {Ricklin}\ and\ \citenamefont
  {Davidson}(2002)}]{Ricklin2002}%
  \BibitemOpen
  \bibfield  {author} {\bibinfo {author} {\bibfnamefont {J.}~\bibnamefont
  {Ricklin}}\ and\ \bibinfo {author} {\bibfnamefont {F.~M.}\ \bibnamefont
  {Davidson}},\ }\href {https://doi.org/10.1364/JOSAA.19.001794} {\bibfield
  {journal} {\bibinfo  {journal} {J. Opt. Soc. Am. A}\ }\textbf {\bibinfo
  {volume} {19}},\ \bibinfo {pages} {1794} (\bibinfo {year}
  {2002})}\BibitemShut {NoStop}%
\bibitem [{\citenamefont {Ricklin}\ and\ \citenamefont
  {Davidson}(2003)}]{Ricklin2003}%
  \BibitemOpen
  \bibfield  {author} {\bibinfo {author} {\bibfnamefont {J.}~\bibnamefont
  {Ricklin}}\ and\ \bibinfo {author} {\bibfnamefont {F.~M.}\ \bibnamefont
  {Davidson}},\ }\href {https://doi.org/10.1364/JOSAA.20.000856} {\bibfield
  {journal} {\bibinfo  {journal} {J. Opt. Soc. Am. A}\ }\textbf {\bibinfo
  {volume} {20}},\ \bibinfo {pages} {856} (\bibinfo {year} {2003})}\BibitemShut
  {NoStop}%
\bibitem [{\citenamefont {Fried}(1966)}]{Fried1966}%
  \BibitemOpen
  \bibfield  {author} {\bibinfo {author} {\bibfnamefont {D.~L.}\ \bibnamefont
  {Fried}},\ }\href {https://doi.org/10.1364/josa.56.001372} {\bibfield
  {journal} {\bibinfo  {journal} {J. Opt. Soc. Am.}\ }\textbf {\bibinfo
  {volume} {56}},\ \bibinfo {pages} {1372} (\bibinfo {year}
  {1966})}\BibitemShut {NoStop}%
\bibitem [{\citenamefont {Ruilier}\ and\ \citenamefont
  {Cassaing}(2001)}]{Ruilier2001}%
  \BibitemOpen
  \bibfield  {author} {\bibinfo {author} {\bibfnamefont {C.}~\bibnamefont
  {Ruilier}}\ and\ \bibinfo {author} {\bibfnamefont {F.}~\bibnamefont
  {Cassaing}},\ }\href {https://doi.org/10.1364/JOSAA.18.000143} {\bibfield
  {journal} {\bibinfo  {journal} {J. Opt. Soc. Am. A}\ }\textbf {\bibinfo
  {volume} {18}},\ \bibinfo {pages} {143} (\bibinfo {year} {2001})}\BibitemShut
  {NoStop}%
\bibitem [{\citenamefont {Fried}(1965)}]{Fried1965}%
  \BibitemOpen
  \bibfield  {author} {\bibinfo {author} {\bibfnamefont {D.~L.}\ \bibnamefont
  {Fried}},\ }\href {https://doi.org/10.1364/JOSA.55.001427} {\bibfield
  {journal} {\bibinfo  {journal} {J. Opt. Soc. Am.}\ }\textbf {\bibinfo
  {volume} {55}},\ \bibinfo {pages} {1427} (\bibinfo {year}
  {1965})}\BibitemShut {NoStop}%
\bibitem [{\citenamefont {Noll}(1976)}]{Noll1976}%
  \BibitemOpen
  \bibfield  {author} {\bibinfo {author} {\bibfnamefont {R.~J.}\ \bibnamefont
  {Noll}},\ }\href {https://doi.org/10.1364/JOSA.66.000207} {\bibfield
  {journal} {\bibinfo  {journal} {J. Opt. Soc. Am.}\ }\textbf {\bibinfo
  {volume} {66}},\ \bibinfo {pages} {207} (\bibinfo {year} {1976})}\BibitemShut
  {NoStop}%
\bibitem [{\citenamefont {Boreman}\ and\ \citenamefont
  {Dainty}(1996)}]{Boreman1996}%
  \BibitemOpen
  \bibfield  {author} {\bibinfo {author} {\bibfnamefont {G.~D.}\ \bibnamefont
  {Boreman}}\ and\ \bibinfo {author} {\bibfnamefont {C.}~\bibnamefont
  {Dainty}},\ }\href {https://doi.org/10.1364/JOSAA.13.000517} {\bibfield
  {journal} {\bibinfo  {journal} {J. Opt. Soc. Am. A}\ }\textbf {\bibinfo
  {volume} {13}},\ \bibinfo {pages} {517} (\bibinfo {year} {1996})}\BibitemShut
  {NoStop}%
\bibitem [{\citenamefont {Ma}\ \emph {et~al.}(2015)\citenamefont {Ma},
  \citenamefont {Ma}, \citenamefont {Yang},\ and\ \citenamefont
  {Ran}}]{Ma2015}%
  \BibitemOpen
  \bibfield  {author} {\bibinfo {author} {\bibfnamefont {J.}~\bibnamefont
  {Ma}}, \bibinfo {author} {\bibfnamefont {L.}~\bibnamefont {Ma}}, \bibinfo
  {author} {\bibfnamefont {Q.}~\bibnamefont {Yang}},\ and\ \bibinfo {author}
  {\bibfnamefont {Q.}~\bibnamefont {Ran}},\ }\href
  {https://doi.org/10.1364/AO.54.009287} {\bibfield  {journal} {\bibinfo
  {journal} {Appl. Opt.}\ }\textbf {\bibinfo {volume} {54}},\ \bibinfo {pages}
  {9287} (\bibinfo {year} {2015})}\BibitemShut {NoStop}%
\bibitem [{\citenamefont {Malacara}(2007)}]{malacara_book}%
  \BibitemOpen
  \bibfield  {author} {\bibinfo {author} {\bibfnamefont {D.}~\bibnamefont
  {Malacara}},\ }\href@noop {} {\emph {\bibinfo {title} {Optical Shop
  Testing}}},\ \bibinfo {edition} {3rd}\ ed.\ (\bibinfo  {publisher} {Wiley},\
  \bibinfo {year} {2007})\BibitemShut {NoStop}%
\bibitem [{\citenamefont {Roddier}(2004)}]{Roddier_AO}%
  \BibitemOpen
  \bibfield  {author} {\bibinfo {author} {\bibfnamefont {F.}~\bibnamefont
  {Roddier}},\ }\href@noop {} {\emph {\bibinfo {title} {Adaptive Optics in
  Astronomy}}},\ \bibinfo {edition} {2nd}\ ed.\ (\bibinfo  {publisher} {CUP},\
  \bibinfo {year} {2004})\BibitemShut {NoStop}%
\bibitem [{\citenamefont {Conan}\ \emph {et~al.}(1995)\citenamefont {Conan},
  \citenamefont {Rousset},\ and\ \citenamefont {Madec}}]{Conan1995}%
  \BibitemOpen
  \bibfield  {author} {\bibinfo {author} {\bibfnamefont {J.-M.}\ \bibnamefont
  {Conan}}, \bibinfo {author} {\bibfnamefont {G.}~\bibnamefont {Rousset}},\
  and\ \bibinfo {author} {\bibfnamefont {P.-Y.}\ \bibnamefont {Madec}},\ }\href
  {https://doi.org/10.1364/JOSAA.12.001559} {\bibfield  {journal} {\bibinfo
  {journal} {J. Opt. Soc. Am. A}\ }\textbf {\bibinfo {volume} {12}},\ \bibinfo
  {pages} {1559} (\bibinfo {year} {1995})}\BibitemShut {NoStop}%
\bibitem [{\citenamefont {Gil-Pelaez}(1951)}]{gil_pelaez}%
  \BibitemOpen
  \bibfield  {author} {\bibinfo {author} {\bibfnamefont {J.}~\bibnamefont
  {Gil-Pelaez}},\ }\href {https://doi.org/10.1093/biomet/38.3-4.481} {\bibfield
   {journal} {\bibinfo  {journal} {Biometrika}\ }\textbf {\bibinfo {volume}
  {38}},\ \bibinfo {pages} {481} (\bibinfo {year} {1951})},\ \Eprint
  {https://arxiv.org/abs/https://academic.oup.com/biomet/article-pdf/38/3-4/481/718851/38-3-4-481.pdf}
  {https://academic.oup.com/biomet/article-pdf/38/3-4/481/718851/38-3-4-481.pdf}
  \BibitemShut {NoStop}%
\bibitem [{\citenamefont {{W. Müller}}(1974)}]{Muller1974}%
  \BibitemOpen
  \bibfield  {author} {\bibinfo {author} {\bibfnamefont {J.}~\bibnamefont {{W.
  Müller}}},\ }\href
  {https://doi.org/https://doi.org/10.1016/0029-554X(74)90283-3} {\bibfield
  {journal} {\bibinfo  {journal} {Nuclear Instruments and Methods}\ }\textbf
  {\bibinfo {volume} {117}},\ \bibinfo {pages} {401} (\bibinfo {year}
  {1974})}\BibitemShut {NoStop}%
\bibitem [{\citenamefont {Agnesi}\ \emph {et~al.}(2019)\citenamefont {Agnesi},
  \citenamefont {Avesani}, \citenamefont {Stanco}, \citenamefont {Villoresi},\
  and\ \citenamefont {Vallone}}]{Agnesi2019}%
  \BibitemOpen
  \bibfield  {author} {\bibinfo {author} {\bibfnamefont {C.}~\bibnamefont
  {Agnesi}}, \bibinfo {author} {\bibfnamefont {M.}~\bibnamefont {Avesani}},
  \bibinfo {author} {\bibfnamefont {A.}~\bibnamefont {Stanco}}, \bibinfo
  {author} {\bibfnamefont {P.}~\bibnamefont {Villoresi}},\ and\ \bibinfo
  {author} {\bibfnamefont {G.}~\bibnamefont {Vallone}},\ }\href
  {https://doi.org/10.1364/OL.44.002398} {\bibfield  {journal} {\bibinfo
  {journal} {Optics Letters}\ }\textbf {\bibinfo {volume} {44}},\ \bibinfo
  {pages} {2398} (\bibinfo {year} {2019})}\BibitemShut {NoStop}%
\bibitem [{\citenamefont {Avesani}\ \emph {et~al.}(2020)\citenamefont
  {Avesani}, \citenamefont {Agnesi}, \citenamefont {Stanco}, \citenamefont
  {Vallone},\ and\ \citenamefont {Villoresi}}]{Avesani2020}%
  \BibitemOpen
  \bibfield  {author} {\bibinfo {author} {\bibfnamefont {M.}~\bibnamefont
  {Avesani}}, \bibinfo {author} {\bibfnamefont {C.}~\bibnamefont {Agnesi}},
  \bibinfo {author} {\bibfnamefont {A.}~\bibnamefont {Stanco}}, \bibinfo
  {author} {\bibfnamefont {G.}~\bibnamefont {Vallone}},\ and\ \bibinfo {author}
  {\bibfnamefont {P.}~\bibnamefont {Villoresi}},\ }\href
  {https://doi.org/10.1364/OL.396412} {\bibfield  {journal} {\bibinfo
  {journal} {Optics Letters}\ }\textbf {\bibinfo {volume} {45}},\ \bibinfo
  {pages} {4706} (\bibinfo {year} {2020})},\ \Eprint
  {https://arxiv.org/abs/2004.11877} {arXiv:2004.11877} \BibitemShut {NoStop}%
\bibitem [{\citenamefont {Nelder}\ and\ \citenamefont
  {Mead}(1965)}]{Nelder1965}%
  \BibitemOpen
  \bibfield  {author} {\bibinfo {author} {\bibfnamefont {J.~A.}\ \bibnamefont
  {Nelder}}\ and\ \bibinfo {author} {\bibfnamefont {R.}~\bibnamefont {Mead}},\
  }\href {https://doi.org/10.1093/comjnl/7.4.308} {\bibfield  {journal}
  {\bibinfo  {journal} {The Computer Journal}\ }\textbf {\bibinfo {volume}
  {7}},\ \bibinfo {pages} {308} (\bibinfo {year} {1965})}\BibitemShut {NoStop}%
\bibitem [{\citenamefont {Renner}(2005)}]{Renner2005}%
  \BibitemOpen
  \bibfield  {author} {\bibinfo {author} {\bibfnamefont {R.}~\bibnamefont
  {Renner}},\ }\emph {\bibinfo {title} {{Security of Quantum Key
  Distribution}}},\ \href
  {https://doi.org/https://doi.org/10.3929/ethz-a-005115027} {Ph.D. thesis}
  (\bibinfo {year} {2005}),\ \Eprint {https://arxiv.org/abs/0512258}
  {arXiv:0512258 [quant-ph]} \BibitemShut {NoStop}%
\bibitem [{\citenamefont {Scarani}\ and\ \citenamefont
  {Renner}(2008)}]{Scarani2008}%
  \BibitemOpen
  \bibfield  {author} {\bibinfo {author} {\bibfnamefont {V.}~\bibnamefont
  {Scarani}}\ and\ \bibinfo {author} {\bibfnamefont {R.}~\bibnamefont
  {Renner}},\ }\href {https://doi.org/10.1103/PhysRevLett.100.200501}
  {\bibfield  {journal} {\bibinfo  {journal} {Physical Review Letters}\
  }\textbf {\bibinfo {volume} {100}},\ \bibinfo {pages} {200501} (\bibinfo
  {year} {2008})},\ \Eprint {https://arxiv.org/abs/0708.0709} {arXiv:0708.0709}
  \BibitemShut {NoStop}%
\bibitem [{\citenamefont {Tomamichel}\ \emph {et~al.}(2012)\citenamefont
  {Tomamichel}, \citenamefont {Lim}, \citenamefont {Gisin},\ and\ \citenamefont
  {Renner}}]{Tomamichel2012}%
  \BibitemOpen
  \bibfield  {author} {\bibinfo {author} {\bibfnamefont {M.}~\bibnamefont
  {Tomamichel}}, \bibinfo {author} {\bibfnamefont {C.~C.~W.}\ \bibnamefont
  {Lim}}, \bibinfo {author} {\bibfnamefont {N.}~\bibnamefont {Gisin}},\ and\
  \bibinfo {author} {\bibfnamefont {R.}~\bibnamefont {Renner}},\ }\href
  {https://doi.org/10.1038/ncomms1631} {\bibfield  {journal} {\bibinfo
  {journal} {Nature Communications}\ }\textbf {\bibinfo {volume} {3}},\
  \bibinfo {pages} {634} (\bibinfo {year} {2012})},\ \Eprint
  {https://arxiv.org/abs/1103.4130} {arXiv:1103.4130} \BibitemShut {NoStop}%
\bibitem [{\citenamefont {Hoeffding}(1963)}]{Hoeffding1963}%
  \BibitemOpen
  \bibfield  {author} {\bibinfo {author} {\bibfnamefont {W.}~\bibnamefont
  {Hoeffding}},\ }\href {https://doi.org/10.1080/01621459.1963.10500830}
  {\bibfield  {journal} {\bibinfo  {journal} {Journal of the American
  Statistical Association}\ }\textbf {\bibinfo {volume} {58}},\ \bibinfo
  {pages} {13} (\bibinfo {year} {1963})}\BibitemShut {NoStop}%
\bibitem [{\citenamefont {Xiang}\ \emph {et~al.}(1997)\citenamefont {Xiang},
  \citenamefont {Sun}, \citenamefont {Fan},\ and\ \citenamefont
  {Gong}}]{Xiang1997}%
  \BibitemOpen
  \bibfield  {author} {\bibinfo {author} {\bibfnamefont {Y.}~\bibnamefont
  {Xiang}}, \bibinfo {author} {\bibfnamefont {D.}~\bibnamefont {Sun}}, \bibinfo
  {author} {\bibfnamefont {W.}~\bibnamefont {Fan}},\ and\ \bibinfo {author}
  {\bibfnamefont {X.}~\bibnamefont {Gong}},\ }\href
  {https://doi.org/10.1016/S0375-9601(97)00474-X} {\bibfield  {journal}
  {\bibinfo  {journal} {Physics Letters A}\ }\textbf {\bibinfo {volume}
  {233}},\ \bibinfo {pages} {216} (\bibinfo {year} {1997})}\BibitemShut
  {NoStop}%
\end{thebibliography}

\end{document}